\documentclass[aps,prd,eqsecnum,11pt,showpacs,preprintnumbers,nofootinbib,superscriptaddress]{revtex4-1}
\usepackage{geometry}                % See geometry.pdf to learn the layout options. There are lots.
\usepackage[pdftex]{graphicx}
\usepackage{float}
\usepackage{amssymb,amsmath,amsopn,bm,dsfont}
\DeclareMathAlphabet{\matheul}{U}{eus}{m}{n}
\allowdisplaybreaks

\usepackage[colorlinks,pdfstartview=FitH]{hyperref}
\hypersetup{linkcolor=blue,citecolor=blue,filecolor=black,urlcolor=blue}

\newcommand{\gobble}[1]{}

\newcommand{\RH}{R_{{\mkern-1mu}_\textit{H}}}
\newcommand{\wH}{\omega_{{\mkern-1mu}_\textit{H}}}
\newcommand{\SBI}{S}

\renewcommand\a{\alpha}
\renewcommand\b{\beta}
\renewcommand\d{\delta}
\renewcommand\k{\kappa}
\renewcommand\l{\lambda}
\renewcommand\r{\rho}

\renewcommand\t{\tau}
\newcommand\y{\upsilon}
\renewcommand\j{\psi}
\renewcommand\th{\theta}

\newcommand\e{\epsilon}
\newcommand\g{\gamma}
\newcommand\z{\zeta}
\newcommand\m{\mu}
\newcommand\n{\nu}
\newcommand\x{\xi}

\newcommand\h{\eta}

\newcommand\f{\phi}
\newcommand\w{\omega}
\newcommand\ve{\varepsilon}

\newcommand\vf{\varphi}

\renewcommand\L{\Lambda}

\renewcommand\S{\Sigma}
\newcommand\Th{\Theta}

\newcommand\G{\Gamma}
\newcommand\F{\Phi}

\newcommand\W{\Omega}
\newcommand{\cC}{{\cal C}}

\newcommand{\cJ}{{\cal J}}
\newcommand{\cK}{{\cal K}}

\newcommand{\cM}{{\cal M}}
\newcommand{\cO}{{\cal O}}
\newcommand{\cS}{{\cal S}}

\newcommand{\cR}{{\cal R}}
\newcommand{\cG}{{\cal G}}

\newcommand{\pa}{\partial}

\newcommand{\nn}{\nonumber \\}
\newcommand{\na}{\nabla}
\newcommand{\sdfrac}[2]{\mbox{\small$\displaystyle\frac{#1}{#2}$}}

\newcommand{\bn}{{\boldsymbol n}}
\newcommand{\bN}{{\boldsymbol N}}
\newcommand{\bl}{{\boldsymbol \ell}}
\newcommand{\by}{{\boldsymbol \y}}
\newcommand{\be}{\boldsymbol e}

\newcommand{\bKt}{{\boldsymbol K_{(t)}}}
\newcommand{\bKf}{{\boldsymbol K_{(\f)}}}
\newcommand{\bK}{{\boldsymbol K}}

\newcommand{\hth}{\hat{\th}}
\newcommand{\bhth}{{\mathbf{\hat{\boldsymbol\th}}}}
\newcommand{\hvf}{{\hat{\phantom{\rule{0pt}{5.5pt}}\vf}}}
\newcommand{\bhvf}{{\mathbf{\hat{\phantom{\rule{0pt}{6pt}}\boldsymbol\vf}}}}

\newcommand{\prim}{\prime\,}

\def\nbox#1#2{\vcenter{\hrule \hbox{\vrule height#2in
\kern#1in \vrule} \hrule}}
\def\sq{\,\raise.5pt\hbox{$\nbox{.10}{.10}$}\,}
\def\sqb{\,\raise.5pt\hbox{$\overline{\nbox{.09}{.09}}$}\,}

\allowdisplaybreaks

\begin{document}

\preprint{LA-UR-20-30222}

\title{Surface stress tensor and junction conditions on a rotating null horizon}
\author{Philip Beltracchi}
\email{philipbeltracchi@gmail.com}
\affiliation{Department of Physics and Astronomy, University of Utah\\Salt Lake City, Utah 84112, USA}

\author{Paolo Gondolo}
\email{paolo.gondolo@utah.edu}
\affiliation{Department of Physics and Astronomy, University of Utah\\Salt Lake City, Utah 84112, USA}
\affiliation{Department of Physics, Tokyo Institute of Technology, 2-12-1 Ookayama, Meguro-ku, Tokyo 152-8551, Japan}
\affiliation{Kavli Institute for the Physics and Mathematics of the Universe, The University of Tokyo, Kashiwa, Chiba 277-8583, Japan}

\author{Emil Mottola}
\email{mottola.emil@gmail.com, emottola@unm.edu}
\affiliation{Theoretical Division, T-2, MS B283, \\Los Alamos National Laboratory\\ Los Alamos, New Mexico 87545, USA}
\affiliation{Department of Physics and Astronomy, University of New Mexico\\
Albuquerque, New Mexico 87131, USA}

\begin{abstract}
The general form of the surface stress tensor of an infinitesimally thin shell located on a rotating null horizon is derived, when different interior 
and exterior geometries are joined there. Although the induced metric on the surface must be the same approached from either side, 
the first derivatives of the metric need not be. Such discontinuities lead to a Dirac $\d$-distribution in the Einstein tensor localized 
on the horizon. For a general stationary axisymmetric geometry the surface stress tensor can be expressed in terms of two geometric invariants 
that characterize the surface, namely the discontinuities $[\k]$ and $[\cJ]$ of the surface gravity $\k$ and angular momentum density $\cJ$. 
The Komar energy and angular momentum are given in coordinates adapted to the Killing symmetries, and the surface contributions to each 
determined in terms of $[\k]$ and $[\cJ]$. Guided by these, a simple modification of the original Israel junction conditions is verified directly from 
the Einstein tensor density to give the correct finite result for the surface stress, when the normal $\bn$ to the surface is allowed to tend continuously 
to a null vector. The relation to Israel's original junction conditions, which fail on null surfaces, is given. The modified junction conditions 
are suitable to the matching of a rotating ``black hole” exterior to any interior geometry joined at the Kerr null horizon surface, even when the 
surface normal is itself discontinuous and the Barrab\`es-Israel formalism is also inapplicable. This joining on a rotating null horizon is purely 
of the matter shell type and does not contain a propagating gravitational shock wave.

\end{abstract} 
\maketitle
\vspace{-1cm}

\section{Introduction}

Singular surfaces in general relativity (GR) have a long history, beginning with Lanczos and Sen just a few years after the introduction of GR 
itself~\cite{Lanczos:1922, Sen:1924,Lanczos:1924,OBrSyn:1952,Lichnerowicz:1955,Synge:1957,Synge:1960,Dautcourt:1964}.
A singular surface arises when one tries to join or ``glue” the geometries of two different regions of spacetime $\cM_+$ and $\cM_-$
at their mutual boundary $\S =\! \cM_+ \cup\, \cM_-$, where the metric and coordinates of the two regions coincide, but the metric derivatives 
in general do not. The discontinuities in the metric derivatives lead to a $\d$-function distributional surface stress tensor 
localized on the infinitesimal idealized boundary layer on $\S$ \cite{Israel:1966a,Taub:1980,Clarke:1987,Barrabes:1989,BarrabesIsrael:1991,Poisson:2002}.

Singular surfaces in GR arise in describing the dynamics of first order phase transitions in the universe, by the nucleation and growth of bubbles of a new phase 
of matter within the old metastable phase \cite{BerKuzTka:1983,BerKuzTka:1987,BlauGueGuth:1987,AurKisManSpa:1987}. 
The boundary layer or surface of the bubble separating the phases of different properties and geometries can be approximated as an infinitely thin bubble wall 
or shell. In the approach of Israel \cite{Israel:1966a}, the surface stress-energy tensor of this bubble wall is related to the discontinuity of the extrinsic curvature 
of the three-dimensional hypersurface when embedded in the four-dimensional spacetime on either side of it.  In the application to nucleating bubbles and their 
dynamical evolution, the normal vector $\bn$ to the hypersurface is spacelike and can be normalized to unity, while the two-surface follows a timelike trajectory. The 
Israel formalism \cite{Israel:1966a} is quite satisfactory to handle this case, as well as the converse situation where the normal vector to the hypersurface is 
everywhere timelike.

When the hypersurface is lightlike, the normal vector becomes null: $\bn\cdot \bn =0$. This leads to difficulties in the formalism, at least as originally presented 
in \cite{Israel:1966a}, since the extrinsic curvature itself vanishes. Barrab\`es and Israel (BI) subsequently reformulated the junction conditions and surface 
stress energy tensor on null hypersurfaces by introducing a second ``transverse” null  vector $\bN$, and a ``transverse” or ``oblique” extrinsic curvature defined in terms 
of $\bN$, rather than the normal $\bn$ \cite{Barrabes:1989,BarrabesIsrael:1991,Poisson:2002}. In applying the BI formulation both the transverse null vector and 
induced metric are taken to be continuous and nonsingular on the null hypersurface. As we discuss in this paper, these regularity conditions on the choice of 
transversal $\bN$ may not be satisfied in an important case of physical interest, namely the joining of interior and exterior black 
hole (BH) solutions at their horizons, particularly when expressed in coordinates that may be singular there.

The conventional assumption is that coordinate singularities on a BH horizon can be eliminated by a suitable change of coordinates, so that
the horizon is a mathematical boundary only, and once removed, the interior geometry is found by ``analytic continuation” from the exterior \cite{MTW}.
Implicit in this mathematical continuation is the physical assumption that there are no surface stresses on the horizon, which would invalidate analytic 
continuation. Motivation for considering alternative possibilities for BH interiors arises from the unphysical interior geometries that result from mathematical
analytic continuation. In addition to the curvature singularity of infinite density present already in the Schwarzschild case, more generally analytic continuation 
of the vacuum BH Kerr solution, leads to acausal, closed timelike curves in the interior of a rotating BH with angular momentum \cite{HawkingEllis:1973}. 

When quantum effects are considered, additional difficulties arise, most notably the apparent nonconservation of probability and uncertain origin of 
the enormous BH entropy apparently implied by the Hawking effect \cite{Hawking:1976}. These problems are generally classified under the heading 
of the `black hole information paradox' \cite{tHooft:1995}, and have been the source of speculation for more than four decades, with a wide range of views 
on possible resolution \cite{UnruhWald:2017}, some quite radical \cite{Mathur:2015}.  At its root the information paradox arises because of the existence 
of the BH horizon which acts as a one-way causal boundary between two different regions of spacetime, and a sink through which matter can pass,
but from which it cannot reemerge, except possibly only after very long times, and in nothing like its original form.

If on the other hand, metric discontinuities and a nonvanishing surface stress tensor are present on the horizon, the interior cannot be
obtained from the exterior geometry by analytic continuation. In that case the interior need not be singular or have unphysical or acausal features
of any kind, and consistency with quantum unitary evolution can be maintained. Such physical surface stresses are perfectly consistent with
the equivalence principle, and can be described within classical general relativity. An explicit example of this is the nonrotating gravitational 
condensate star, where the Schwarzschild exterior of a nonrotating ``black hole” is matched to a static region of de Sitter space with a positive vacuum 
energy but negative pressure, $p= -\r$. The resulting nonsingular interior ``gravastar” solution can be obtained from Schwarzschild's classical constant 
density interior solution by a limiting process \cite{MazurMottola:2015}, which evades the Buchdahl bound \cite{Buchdahl:1959}, by having an anisotropic 
surface stress tensor localized on the Schwarzschild radius $R_{_H} = 2GM/c^2$. This classical solution is free of any conflict with quantum or statistical 
mechanics, loss of unitarity, or information paradox \cite{MazurMottola:2001,MazurMottola:2004}. 

A similar proposal based on an analogy to quantum phase transitions was suggested in \cite{ChapHohlLaughSant:2001}. A de Sitter interior could develop from 
a phase transition where the quantum vacuum energy changes abruptly~\cite{Mottola:2010}. Independently of the microscopic quantum origin of the phase transition 
boundary layer itself, the interior de Sitter and exterior Schwarzschild classical smooth geometries must be ``glued” together at their mutual horizon boundaries, of infinitesimal 
thickness in classical GR at $\RH = 2GM/c^2$. The discontinuity $[\k]$ in the surface gravities there results in a positive surface tension 
$\t_s = [\k]/8 \pi G = (8\pi G \RH)^{-1}$ \cite{MazurMottola:2015}. 

If such a boundary layer with a nonvanishing surface tensor replaces the classical BH horizon, the existence and physical characteristics of 
the surface layer may be probed by gravitational wave observations of binary neutron star and BH mergers \cite{LIGO:2019}. The possibility of 
observable signatures of a gravastar horizon surface has been attracting increasing interest, in the new era of gravitational wave and 
multi-messenger astronomy \cite{CardosoPani:2019}. Since {\it rotating} ``BHs” are ubiquitous, it is important to determine the properties of the 
surface stress tensor and junction conditions for a null horizon with nonzero angular momentum within classical GR. 
This is the principal motivation and main result of the present work.

Discontinuous boundary layers in GR have obvious parallels to those in electrodynamics where a surface layer of charge density leads to a 
discontinuity in the electric field vector, by Gauss' law. The boundary conditions and discontinuity of the electric field vector are determined
by purely classical considerations, in first approximation independently of the details of the (quantum) matter surface charge layer. In classical
GR the analog of Gauss' law for a stationary axisymmetric geometry is furnished by the Komar mass and angular momentum, in terms of 
the Killing vectors of time translation and azimuthal rotational symmetry. The local differential form of these quantities can be used to relate 
the discontinuities in certain functions of metric derivatives to the surface stress tensor on the horizon where these discontinuities occur. 
Elucidating the physical meaning of possible metric derivative discontinuities on a rotating horizon in terms of geometric invariants and their 
contribution to the mass and angular momentum of the configuration is a second purpose of this study.

In investigating these issues, and guided by the mass and angular momentum expressions, we find that there is a very simple and natural 
refinement of the original Israel junction conditions and surface stress tensor in terms of the discontinuity of an extrinsic curvature 
of the null hypersurface, which is free of the difficulties that motivated the Barrab\`es-Israel (BI) approach \cite{BarrabesIsrael:1991}, and 
dispenses with the need of the transversal $\bN$ and oblique extrinsic curvature altogether. Presenting this new ``old” approach to null 
singular hypersurfaces and verifying it explicitly in the case of stationary, axisymmetric horizons is the third main purpose of this work. 
Whereas Israel's original approach \cite{Israel:1966a} fails for lightlike hypersurfaces, the modified junction conditions work equally well for 
timelike and lightlike hypersurfaces, in which the latter can be viewed as a continuous limit of the former. The reason for that
and the relationship between the two is explained. Finally by studying the Weyl tensor we are also able to characterize the junction 
on a rotating horizon as free of any gravitational shock wave component.

The paper is organized as follows. In Sec.~\ref{Sec:AxiStat} the metric, coordinates and properties of the general stationary, 
axisymmetric geometry are introduced. In Sec.~\ref{Sec:NRotat} the surface stress tensor for a spherical null horizon is reviewed and derived 
directly from the Einstein tensor density that contains the Dirac $\d$-function surface contribution to the energy-momentum-stress tensor. 
This serves to introduce the method we use in Sec.~\ref{Sec:Rotate} to derive the singular stress tensor of the general axisymmetric, stationary 
horizon surface directly from the Einstein tensor density, whose explicit form is given. The result, Eqs.~(\ref{surfTrot}), depends upon just two 
invariant quantities that characterize the surface discontinuities: the surface gravity $\k$ and angular momentum density $\cJ$. Both of these 
appear in the Komar mass and angular momentum which are discussed in Sec.~\ref{Sec:Komar} in both their integral and local differential forms. 
The latter is used to relate total derivatives of certain local quantities to components of the surface stress tensor of the rotating horizon. 
In Secs.~\ref{Sec:Junc}-\ref{Sec:Krot} we show that a relatively small but important modification of the original Israel formalism yields the 
correct junction conditions for a stationary axisymmetric null hypersurface in terms of the discontinuities of a modified extrinsic curvature $\bK^i_{\ j}$, 
by a continuous limiting procedure of the normal $\bn\cdot \bn\!=\! \ve\!\to\! 0$, as the hypersurface and its normal become null, and compare it
to the Barrab\`es-Israel (BI) approach. Sec.~\ref{Sec:Concl} contains a Summary and our Conclusions. Units in which the speed of light $c\!=\!1$ are
used hereafter in the paper.

There are three Appendices. The first contains a compact review of the tetrad formalism, surface and volume integration of forms, and Stokes' 
theorem applied to the axisymmetric geometry used in this paper. Appendix \ref{Sec:Weyl} contains the explicit expressions for the Weyl tensor
of the general axisymmetric, stationary geometry and metric, and its singular surface contributions, showing that the horizon discontinuities
are necessarily associated with a thin shell matter distribution, and not a lightlike gravitational shock wave \cite{BarrabesHogan:2003}. Appendix \ref{Sec:BIrot} 
provides an explicit example of (mis)application of the Barrab\`es-Israel formalism, in order to illustrate the pitfalls one may encounter when 
the normal $\bn$ or volume integration measure are themselves singular or discontinuous on the null hypersurface, as is the case 
for BH horizons in familiar coordinates. 

\vspace{-1mm}
\section{Axisymmetric Stationary Geometry and Metric}
\label{Sec:AxiStat}
\vspace{-1mm}

The condition of stationarity and axial symmetry is invariantly defined by the geometry admitting two
independent Killing vectors, denoted $\bKt$ and $\bKf$ 
\begin{align}
\bKt= K^\m_{(t)} \frac{\pa}{\pa x^\m} = \frac{\pa}{\pa t} \,, \qquad {\rm and}\qquad \bKf =K^\m_{(\f)} \frac{\pa}{\pa x^\m}= \frac{\pa}{\pa \f}
\label{Killvecs}
\end{align}
corresponding to time translation and rotations around the fixed rotation axis respectively. Each of these satisfies the Killing equation
\vspace{-5mm}
\begin{align}
\na_\m K_{\n}+ \na_\n K_{\m} =0 
\label{Killeq}
\end{align}
in component form. Since the two Killing symmetries commute \cite{Carter:1970}, two coordinates $(t,\f)$ can be defined as in (\ref{Killvecs}), 
corresponding to the vector fields generating the symmetries, so that the metric is independent of both $t$ and $\f$. In the case of
asymptotically flat spacetimes, the normalization of $\bKt$ can be fixed by the condition $\bKt\cdot \bKt\!\to \!-1$ in the asymptotic region, while 
the normalization of the spacelike $\bKf$ is set by a regularity condition on the symmetry axis \cite{Stephani:2003tm,Gourgoulhon:2010}, 
which fixes the periodic range of the azimuthal angle coordinate $\f \in [0,2\pi]$. 

There are several forms for the general axisymmetric and stationary metric line element in the literature 
\cite{Lewis:1932,Papapetrou:1966,HartleSharp:1967,Bardeen:1970,FriedmanChandra:1972,Chandrasekhar:1984}. A minimum
of four functions of the remaining two coordinates are necessary to specify the general such metric. It is most convenient to work with the line element
\begin{align}
ds^2 = -e^{2\n} dt^2 + e^{2\j} \big(d \f - \w\, dt\big)^2 + e^{2 \a} dr^2 + e^{2 \b} d\th^2
\label{axisymstat}
\end{align}
which contains five functions, $\n, \j, \a, \b, \w$ of the remaining two coordinates, which are (rather arbitrarily) labeled here as $(r, \th)$.
There remains the freedom to define a coordinate condition on the functions $\a,\b$ to reduce these five functions to the minimal 
four functions of two variables $(r,\th)$. 

For reference we list here the nonzero components of the metric in (\ref{axisymstat}) and its inverse
\begin{subequations}
\begin{align}
g_{tt} &= -e^{2\n} + \w^2\,e^{2\j} 	&g^{tt} &=-e^{-2\n}\\
g_{t\f} &= g_{\f t} = -\w \,e^{2\j}		&g^{t\f}&= g^{\f t} = -\w \,e^{-2\n}\\
g_{\f\f} &= e^{2\j} 				&g^{\f\f}& = e^{-2\j} -\w^2\,e^{-2\n} \\
g_{rr} & = e^{2\a} 				&g^{rr} &= e^{-2\a}\\
g_{\th\th} &= e^{2\b} 				&g^{\th\th} &= e^{-2\b}
\end{align}
\label{metric}\end{subequations}
for the stationary axisymmetric geometry. In these $(t,r,\th,\f)$ coordinates, adapted to the symmetries, the two Killing vectors (\ref{Killvecs})
have contravariant components 
\vspace{-1mm}
\begin{subequations}
\begin{align}
&K_{(t)}^\m = \d^\m_{\ t} = (1,0,0,0) \\
&K_{(\f)}^\m = \d^\m_{\ \f} = (0,0,0,1)
\end{align}
\label{Kill2}\end{subequations}
and their inner products
\vspace{-6mm}
\begin{subequations}
\begin{align}
&\bKt\cdot \bKt =g_{tt} = -e^{2\n} + \w^2 e^{2\j}  \\
&\bKf\cdot \bKf = g_{\f\f} = e^{2\j} \label{Kphi}\\
&\bKt\cdot \bKf = g_{t\f} = - \w \,e^{2\j}
\end{align}
\end{subequations}
are coordinate invariant scalars. Hence $e^{2\n}$, $e^{2\j}$, and $\w$ are separately coordinate invariant scalars. The function $\w(r,\th) = -g_{t\f}/g_{\f\f}$ is 
the rotational frequency of the Lens-Thirring dragging of inertial frames. The condition $e^{2\j}=0$ where (\ref{Kphi}) vanishes (with other metric components nonvanishing) 
defines the axis of rotational symmetry. The particular linear combination 
\vspace{-1mm}
\begin{align}
\bl \equiv \bKt+ \w \bKf 
\end{align}
plays a special role, since $\bKf \cdot \bl = 0$ implies that particles with instantaneous four-velocity proportional to $\bl$ (when not null) define the zero 
angular momentum observers: ZAMOs \cite{Bardeen:1970,ThorneMembrane}. Geometrically, $\bl, \pa/\pa r,\pa/\pa \th,\pa/\pa \f$ are the dual basis to the orthogonal basis 
$dt, dr, d \th, d\f$ appearing in the line element (\ref{axisymstat}). Since
\vspace{-2mm}
\begin{align}
\bl^2 = \big(\bKt + \w \bKf\big) \cdot \big(\bKt + \w \bKf\big) = -e^{2\n}
\label{ellsq}
\end{align}
there are no timelike particle world lines if $e^{2\n} \le 0$, and $e^{2\n} = 0$ is the condition for a marginally trapped surface and event horizon. If $\w= \wH$ is constant 
on the horizon, then $\bl_{_{\!H}}$ is also a Killing vector there and the horizon is a Killing horizon. 

Unlike the $(t,\f)$ coordinates adapted to the symmetry, the geometry and labeling of the other two coordinates $(r,\th)$ is somewhat arbitrary and may be chosen 
in different ways. In fact, the general form of the line element (\ref{axisymstat}) is preserved under any coordinate transformation
$(r,\th)\to (\tilde r,\tilde \th)$ satisfying
\begin{align}
g_{_{\tilde r\tilde \th}} = e^{2\a} \left(\frac{\pa r}{\pa \tilde r}\right) \left(\frac{\pa r}{\pa \tilde \th}\right) 
+ e^{2\b} \left(\frac{\pa \th}{\pa \tilde r}\right) \left(\frac{\pa \th}{\pa \tilde \th}\right) = 0
\label{coordtrans1}
\end{align}
with $r= r(\tilde r,\tilde \th), \th = \th (\tilde r,\tilde \th)$. These two functions are constrained by just one functional condition (\ref{coordtrans1}). The remaining 
freedom allows the specification of one additional condition to reduce the five functions of (\ref{axisymstat}) to the minimal four. Such a coordinate condition  
could be imposed on any combination of the diagonal components $g_{_{rr}} ,g_{_{\th\th}}$. In this paper we shall assume that the horizon where (\ref{ellsq}) 
vanishes defines at fixed $t$ a closed two-surface, and use the coordinate freedom in choosing the $(r,\th)$ coordinates so that this closed two-surface is a 
surface of constant $r\!=\!\RH$ in addition to constant $t$. In other words, without loss of generality, the $r$ coordinate is adapted to the specification of the horizon surface.

\vspace{-1mm}
\section{The Stress Tensor of a Spherical Horizon Surface}
\label{Sec:NRotat}
\vspace{-1mm}

Prior to the case of a rotating horizon and to explain the general method, it is instructive to consider the simpler case of vanishing angular momentum, 
where the geometry is both static and spherically symmetric. The general static, spherically symmetric line element can be expressed in the form
\begin{align}
ds^2\big\vert_{\rm sph} = -f(r)\, dt^2 + \frac{dr^2}{h(r)\!} + r^2 \big(d\th^2 + \sin^2\!\th \, d\f^2\big)
\label{sphstat}
\end{align}
in terms of just two functions $f(r), h(r)$ of $r$ only. This is clearly a special case of (\ref{axisymstat}) with $\w=0$, and
\vspace{-6mm}
\begin{subequations}
\begin{align}
e^{2\n}\big\vert_{\rm sph}  &= f(r)\\
e^{2\a}\big\vert_{\rm sph} &= \frac{1}{h(r)\!}\, = \left(1- \sdfrac{2 G m(r)}{r}\right)^{\!-1}  \\
e^{2\b}\big\vert_{\rm sph}&= r^2\\
e^{2\j}\big\vert_{\rm sph} &= r^2 \sin^2\!\th
\end{align}
\end{subequations}
where $h(r)$ is also expressed in terms of the Misner-Sharp mass function $m(r)$ \cite{MisnerSharp:1964}. The function $\n$ is the gravitational potential 
which approaches $-GM/r$ in the asymptotically flat region, where in the absence of matter $m(r) = M$ and $f/h$ are constants. This is the case for the 
Schwarzschild vacuum BH solution for which $f\!=\!h\!= \!1-2GM/r$. The ratio $f/h$ is constant also if only the weaker condition $-T^t_{\ t} + T^r_{\ r} \equiv \r + p = 0$ 
is satisfied, as for example with a cosmological constant $\L$ source.

The direct method of determining the form that a singular stress tensor may take on an infinitely thin three-dimensional hypersurface is to examine possible 
Dirac $\d(\F)$ contributions to the Einstein tensor for a hypersurface specified by a condition $\F(x^\m)=0$, where $\F\!=\!r-R$ for a hypersurface at fixed $r\!=\!R$.
Einstein's equations may then be used to find the singular surface stress tensor with $\d(\F)$ support, together with the proper integration measure that makes 
this stress tensor meaningful as a distribution when integrated over a small interval in $r$ enclosing the surface.

An horizon occurs at a zero of $e^{2\n}$, which for the line element (\ref{sphstat}) means an $r\!=\!\RH$ where $f(\RH)\!=\!0$.  The induced metric on the horizon 
\begin{align}
\lim_{R\to \RH}ds^2\Big\vert_{r=R} \!\! \equiv d\S_{_H}^2 = \RH^2 \left(d\th^2 + \sin^2\!\th\,d\f^2\right)
\label{dSEF} 
\end{align}
is just the round $\mathbb{S}^2$ metric. The limit of fixed $r = R \to R_{_H}$ is a requisite prescription when the coordinates (\ref{sphstat}) become singular on the event
horizon at $r = R_{_H}$, as for the Schwarzschild vacuum BH solution at $r = R_{_H} = 2GM$. 

The four-volume factor (\ref{detg})
\vspace{-5mm}
\begin{align}
\sqrt{-g} = \sqrt{\sdfrac{f}{h}} \ r^2 \sin \th
\label{sqrtg}
\end{align} 
is finite at the horizon provided that as $r=R\to \RH$ where $f(\RH)\!=\!0$, $h(\RH)\to 0$ also, so that the ratio $f/h$ tends to a finite limit on both sides of the horizon. 
Note that $f(\RH)\!=\!h(\RH)=0$ means that $f$ and $h$ are separately continuous at the horizon, although the ratio $f/h$ and hence $\sqrt{-g}$ need not be
be continuous, or tend to the same limit when approached from either side.

Now consider the case that the interior geometry $r < \RH$ is joined or ``glued” to the null horizon at $r=\RH$ with the same induced metric (\ref{dSEF}) 
but in general discontinuous derivatives. That is, consider a static, spherically symmetric metric of the form (\ref{sphstat}), with continuous and piecewise differentiable 
functions
\vspace{-6mm}
\begin{subequations}
\begin{align}
&f(r) =f_+(r) \, \Th (r-\RH) + f_-(r)\, \Th (\RH-r)\\
&h(r)= h_+(r)\,  \Th (r-\RH) + h_-(r)\, \Th (\RH-r) 
\end{align}\label{gravNrot}
\end{subequations}
where $\Th(x)$ is the Heaviside step function, and 
\begin{align}
f_+(\RH) = f_-(\RH)\,,\qquad h_+(\RH) = h_-(\RH)\,.
\label{matchfh}
\end{align}
The general rule of differentiation of a piecewise differentiable function defined by
\begin{align}
F(x) = F_+(x) \,  \Th (x) + F_-(x) \,  \Th (-x)
\label{genF}
\end{align}
is
\begin{align}
\frac{dF}{dx} = \frac{dF_+(x)}{dx} \,  \Th (x) + \frac{dF_-(x)}{dx} \,  \Th (-x) + \big[ F \big] \,  \d(x)
\label{gendF}
\end{align}
where $\d (x) =d \Th(x)/dx = - d \Th(-x)/dx$ is the Dirac distribution and
\begin{align}
\big[F\big] \equiv \lim_{\ x\to 0^+}\!F_+(x) - \lim_{\ x\to 0^-}\!F_-(x)
\label{discF}
\end{align}
is the discontinuity of $F$ at $x\!=\!0$. Since $f$ and $h$ are piecewise differentiable and continuous, {\it i.e.} they satisfy (\ref{matchfh}), their first derivatives 
$f'=df/dr$ and $h'=dh/dr$ do not contain a $\d$-function term. However since their first derivatives are allowed to be discontinuous, applying (\ref{gendF}) again to $f'$ or $h'$ shows 
that the second derivatives $f'', h''$ do contain $\d$-functions proportional to $[f']$ or $[h']$.

In the Schwarzschild coordinates (\ref{sphstat}), the Einstein tensor components are 
\begin{subequations}
\begin{align}
G_{\ t}^t &= \frac{1}{r} \frac{dh}{dr} + \frac{1}{r^2}\, \big(h-1\big)\label{Gtt}\\
G_{\ r}^r &= \frac{h}{r f}\frac{d f}{dr}  + \frac{1}{r^2}\, \big(h -1\big)\label{Grr}\\
G_{\ \th}^\th =  G_{\ \f}^\f &= \frac{h}{2f} \frac{d^2f}{dr^2} + \frac{h}{4f} \frac{df}{dr} \left(\frac{1}{h} \frac{dh}{dr} - \frac{1}{f} \frac{df}{dr}\right)  
+ \frac{h}{2r}\left( \frac{1}{h} \frac{dh}{dr} + \frac{1}{f} \frac{df}{dr}\right)\label{Gthth}
\end{align}
\label{Gsphsym}\end{subequations}
with one covariant and one contravariant index. Viewed as a $4 \times 4$ matrix $G^\m_{\ \n}$ can be diagonalized, with eigenvalues that are invariant under local 
frame rotations, and free of singularities associated with singular coordinates of the metric tensor. Note that the second derivative of the metric function 
$f$ appears only in (\ref{Gthth}), so that singular $\d$-function stresses localized on the surface can appear only in the angular components 
$G^\th_{\ \th} \!=\! G^\f_{\ \f}$. However, due to divergent factors $1/f$ and $1/h$ multiplied by first derivatives of $f$ or $h$ in (\ref{Gthth}), 
and the possibility that $h/f$ is allowed to be discontinuous at the horizon, care is required, and (\ref{genF})-(\ref{discF}) cannot yet be 
applied to (\ref{Gthth}) for null horizons.

In handling these terms the key observation is that a $\d$-distribution must be defined with respect to an integration measure, and the integration over a small interval in $r$ 
enclosing the surface at fixed $\F\! =\! r-\RH\! =\! 0$ involves the three-volume element (\ref{3surft}), which contains a factor of $\sqrt{-g}= r^2 \sin\th \sqrt{f/h}$.
Since this is the proper measure against which the $\d$-function distribution is to be integrated, it is the tensor {\it densities} $\sqrt{-g}\, G_{\ \n}^\m$ not $G_{\ \n}^\m$ 
that are required for the calculation of the singular $\d$-function stress tensor on the null hypersurface at $r=\RH$. Since the $r^2\sin\th$ factor 
of $\sqrt{-g}$ is finite and continuous on the horizon, but the derivatives $f'$ and $h'$ and the ratio $f/h$ may be discontinuous, consider therefore
\begin{align}
\sqrt{\frac{f}{h}}\,G_{\ \th}^\th = \sqrt{\frac{f}{h}}\,G_{\ \f}^\f = 
\frac{1}{2}\frac{d}{dr} \left(\sqrt{\frac{h}{f}} \frac{df}{dr}\right) + \frac{1}{2r}\left(\sqrt{\frac{f}{h}} \frac{dh}{dr} + \sqrt{\frac{h}{f}} \frac{df}{dr}\right)
\label{Gththdens}
\end{align}
and note the important fact that the potentially singular $f'/f, h'/h$ terms in (\ref{Gthth}) involving the first derivatives of $f$ and $h$ have now combined into a total $r$ 
derivative in the first term on the right side of (\ref{Gththdens}). It is clear then that the last term of (\ref{Gththdens}), although discontinuous, does not contain 
any $\d$-function contributions to the surface stress tensor at $r=\RH$.

The general rule (\ref{genF})-(\ref{discF}) may now be applied to the discontinuous function $\sqrt{h/f}\, f'$, the derivative of which in 
the first term of (\ref{Gththdens}) gives
\begin{align}
\sqrt{\frac{f}{h}}\,G_{\ \th}^\th &= \sqrt{\frac{f}{h}}\,G_{\ \f}^\f =  \left[\frac{1}{2} \sqrt{\frac{h}{f}}\,\frac{df}{dr}\right] \d(r-\RH)  + \dots 
\label{discGthth}
\end{align}
where the ellipsis denotes nonsingular terms containing no Dirac $\d$-functions. Thus the total discontinuity (\ref{discGthth}) of the tensor 
density and singular surface $\d$-function contribution in this quantity is well-defined even if the metric is singular and $f/h$ is discontinuous 
at the horizon. Indeed the relevant quantity whose discontinuity must be computed is the surface gravity
\begin{align}
\k(r) = \frac{1}{2} \sqrt{\frac{h}{f}}\, \frac{df}{dr} 
\label{surfgrav}
\end{align}
which is the relativistic version of the force per unit mass or instantaneous acceleration exerted on a particle fixed at $r$, projected onto the normal to the surface 
at that fixed $r$.

From Einstein's equations, the result (\ref{discGthth}) implies that there is a distributional stress tensor on the horizon surface given by~\cite{MazurMottola:2015}
\begin{align}
^{(\S )}T^A_{\ B}\, \sqrt{\frac{f}{h}} = \cS^A_{\ B} \, \d(r-\RH)\,,\qquad  \cS^A_{\ B}= \frac{[\k]}{8\pi G} \, \d^A_{\ B}  \,,\qquad A,B = \th\,,\f
\label{surfTsph}
\end{align}
which is a well-defined distribution when integrated against the continuous measure $\int dr\, d\th \,d\f\, r^2 \sin\th$ at fixed $t$. This distributional stress tensor $\cS^{A}_{\ B}$ is {\it not} equal to the surface energy tensors in Israel~\cite{Israel:1966a} and Barrab\`es and Israel~\cite{BarrabesIsrael:1991}; their relations are discussed at the end of Sec.~\ref{Sec:BInonrot}.

In the case of the spherical gravastar solution of~\cite{MazurMottola:2015}, the de Sitter interior has
\begin{align}
f_-(r) = \frac{1}{4}\, h_-(r) =  \frac{1}{4}\left(1- \sdfrac{r^2}{\RH^2}\right)\,,\qquad 0\le r \le \RH
\label{fhdeS}
\end{align}
and the Schwarzschild exterior has
\begin{align}
f_+(r) = h_+(r)  = 1- \sdfrac{\RH}{r}\,,\qquad \RH\le r
\label{fhSchw}
\end{align}
so that $f/h$ is discontinuous in this case, but the surface gravities are equal and opposite in sign on opposite sides of the surface at $r= \RH= 2GM$.
Their discontinuity from (\ref{surfgrav}) and (\ref{fhdeS})-(\ref{fhSchw}) is
\begin{align}
\big[\k\big] = \frac{1}{2} \bigg\{ \sdfrac{1}{\RH\!\!} - 2 \Big(\!-\!\sdfrac{1}{2\RH\!\!}\,\Big)\bigg\}= \frac{1}{\RH\!\!} = \frac{1}{2GM}
\label{kapdisc0}
\end{align}
and the surface stress tensor (\ref{surfTsph}) of a spherical gravastar is given by (\ref{surfTsph}) with
\begin{align}
\cS^A_{\ B} = \t_s \, \d^A_{\ B} \,,\qquad A,B= \th\,,\f
\label{Snonrot}
\end{align}
where
\vspace{-6mm}
\begin{align}
\t_s =  \frac{[\k]}{8\pi G} = \frac{1}{16\pi G^2 M\!\!}
\label{taunonrot}
\end{align}
is the surface tension tangential to the spherical horizon surface. The equal and opposite surface gravities at the horizon indicate that there are equal 
and opposite radial forces (ingoing from outside, outgoing from inside) that balance in a static configuration with no net force. These give rise to a positive 
surface tension, and a surface energy differential $dE_s = \t_s \, dA$ that opposes increasing the area $A$ of the surface, indicating stability to small
perturbations~\cite{MazurMottola:2015}. The $\cS^\f_{\ \f}$ and $\cS^\th_{\ \th}$ components in Eq.~\eqref{Snonrot} are the only nonzero components of the surface stress-energy tensor, and represent an isotropic tension on the surface equal to the surface tension $\t_s$. In particular, the energy density term $\cS^{t}_{\ t}$ and the shear on the surface vanish.

Obtaining the finite coordinate invariant result (\ref{Snonrot})-(\ref{taunonrot}) did not require a transformation from Schwarzschild coordinates (\ref{sphstat}) 
to nonsingular coordinates on the horizon, but follows directly from the Einstein tensor density (\ref{Gththdens}). If one does transform to
advanced or retarded Eddington-Finkelstein coordinates 
\vspace{-6mm}
\begin{align}
dv = dt + \frac{dr}{\sqrt{fh}} \qquad {\rm or} \qquad du = dt - \frac{dr}{\sqrt{fh}} 
\end{align}
which are regular on the future or past horizon respectively, it is clear that this coordinate transformation does not affect the $\th, \f$ coordinates,
so that the components of the Einstein tensor density (\ref{Gththdens}), and its discontinuity and $\d$-function distribution (\ref{discGthth}) with support on the 
surface are unchanged. It is also clear therefore that exactly the same surface tensor distributions are produced by this matching of (\ref{fhdeS})
to (\ref{fhSchw}) on {\it both} the future and past horizons in a time-symmetric manner.

The considerations of this section can be generalized to any spherically symmetric spacetime in the interior region, continuously matched to the 
exterior at the horizon or indeed at any $R$, by specifying $f$ and $h$ in each region, and recomputing the discontinuities in the derivatives 
of (\ref{kapdisc0}). In this way the $\d$-function contributions to the surface stress tensor in (\ref{surfTsph}) are related directly to those in the Einstein tensor 
density (\ref{discGthth}) by Einstein's equations. If this matching were performed at a surface where both $f$ and $h$ are nonvanishing and continuous, the 
multiplication of the Einstein tensor by the $\sqrt{f/h}$ factor from $\sqrt{-g}$ is not essential to obtaining a well-defined distribution. However, the fact that the second 
derivatives of the metric functions combine with singular first derivative terms to form the total $d/dr$ derivatives in the density (\ref{Gththdens}) is essential
for null horizon matching where both $f$ and $h$ vanish and $\sqrt{-g}$ can be discontinuous. That this combination occurs is a result of the Killing symmetries 
and related to the integral forms of the Komar mass-energy and angular momentum with the proper integration measure, as demonstrated for the general stationary
axisymmetric geometry in the coordinates adapted to the symmetries in Sec.~\ref{Sec:Komar}.

\vspace{-1mm}
\section{The Stress Tensor of a Rotating Horizon}
\label{Sec:Rotate}
\vspace{-1mm}

In this section we follow the same direct method of determining the stress tensor $\cS^i_{\ j}$ of a singular null hypersurface as in Sec. \ref{Sec:NRotat},
now for the more general case of nonzero angular momentum. We will compute the Einstein tensor density $\sqrt{-g} \, G^{\m}_{~\n}$, retaining only those 
terms that can give rise to a Dirac $\d$-function distribution on the horizon, obtained by differentiation of discontinuous functions, according to the
general rule (\ref{gendF}). The coefficients of the $\d$-function yield the surface stress-energy tensor localized on the horizon by Einstein's equations.

Since at a null horizon $\bl\cdot \bl= - e^{2\n}$ in (\ref{ellsq}) vanishes, the induced metric on the rotating horizon is
\begin{align}
ds^2\Big\vert_{r=\RH} = d\S_{\!_H}^2 = e^{2\j_{\!_H}} \big(d \f - \wH dt\big)^2 + e^{2 \b_{\!_H}} d\th^2
\label{surfmet}
\end{align}
and we require matching of the interior and exterior metrics on the horizon surface so that the metric functions $\b\!=\!\b_{_{\!H}}$, $\j\!=\!\j_{_{\!H}}$, $\w\!=\!\wH$, and their 
derivatives with respect to $\th$ are continuous at the horizon. Hence only their second derivatives with respect to $r$ may give rise to a Dirac $\d$-function on the horizon.
For the potentially singular $\a$ and $\n$ functions, we must expect some combinations of second and first derivatives with respect to $r$ to contribute $\d$-functions 
on the horizon, analogous to (\ref{Gththdens}).

The Einstein tensor $G^\m_{\ \n}$ for the general stationary axisymmetric metric (\ref{axisymstat}) has the following nonvanishing components, 
{\it cf.} \cite{FriedmanChandra:1972,Chandrasekhar:1984}
\begin{subequations}
\begin{align}
&\hspace{-3mm} G^t_{~t}\!= \! e^{-2 \a}\Big\{\b_{rr}\!+\!\j_{rr} + \big(\b_r\!+\!\j_r\big) \big(\j_r \!-\!\a_r \big)\!+\!\b_r^2\Big\}
+ \sdfrac{1}{2} e^{-2 \a -2 \n +2 \j } \Big\{ \w  \w_{rr} \!+\! \w \w_r \big(\b_r \!- \!\a_r \!-\!\n_r \!+\! 3 \j_r\big)\!+ \!\sdfrac{\w_r^2}{2}\Big\}\nn
&\hspace{-6mm} +\! e^{-2\b} \Big\{\a_{\th \th } \!+\! \j_{\th \th } \!+\! \big(\a_{\th } \!+\! \j_\th\big)\big(\j_{\th }\!-\!\b_{\th }\big)\!+\!\a_{\th }^2\Big\} 
+ \sdfrac{1}{2} e^{-2 \b -2 \n + 2\j } \Big\{\w  \w_{\th \th } \!+\! \w\w_\th  \big(\a_\th \!-\!\b_{\th }\!-\!\n_{\th }\!+\!3 \j_{\th }\big)\!+\!\sdfrac{\w_\th^2}{2}\Big\}\\
&\hspace{-4mm} G^t_{\f}\!=\!-\sdfrac{1}{2} e^{-2 \a -2 \n + 2\j } \Big\{\w_{rr}+ \w_r \big(\b_r\!-\!\a_r\!-\!\n_r\!+\!3 \j_r\big)\!\Big\} 
-\sdfrac{1}{2} e^{-2 \b -2 \n + 2\j }  \Big\{\w_{\th \th } + \w_{\th } \big(\a_{\th }\!-\!\b_{\th } \!-\!\n_{\th }\!+\!3 \j_{\th}\big)\!\Big\}\\
&G^\f_{~t}= \sdfrac{1}{2} e^{-2 \a }  \Big\{\w_{rr}\!+\!2 \w(\j_{rr} \!-\!\n_{rr})  \!+\!\w_r \left(\b_r\!-\! \a_r \!-\!\n_r
\!+\! 3 \j_r\right)  \!+\! 2\w \left(\j_r\!-\!\n_r\right) \left(\b_r\!-\!\a_r\!+\!\n_r\!+\!\j_r\right)\!\Big\}\nn
&\hspace{-9mm} \!+\!\sdfrac{\w}{2}e^{-2\a -2 \n + 2\j }\Big\{\!\w \w_{rr} \!+\! 2 \w_r^2 \!+\! \w\w_r(\b_r \!-\! \a_r \!-\!\n_r\!+\! 3 \j_r )\!\Big\}
\!+\!\sdfrac{\w}{2}e^{-2\b -2 \n + 2\j }\Big\{\!\w \w_{\th\th} \!+\! 2 \w_\th^2 \!+\! \w\w_\th (\a_\th\!-\! \b_\th \!-\!\n_\th\!+\! 3 \j_\th)\!\Big\}\nn
&\hspace{5mm} + \sdfrac{1}{2} e^{-2 \b} \Big\{\w_{\th \th } \!+\! 2 \w (\j_{\th \th }\!-\! \n_{\th \th }) 
\!+\!\w_{\th } \left(\a_{\th }\!-\!\b_{\th }\!-\!\n_{\th }\!+\!3 \j_{\th }\right)
\!+\!2 \w  \left(\j_{\th }\!-\!\n_{\th }\right) \left(\a_{\th } \!-\! \b_{\th }\!+\!\n_{\th }\!+\!\j_{\th }\right)\Big\}\\
&\hspace{-6mm}G^\f_{~\f}\!=\!e^{-2 \a } \Big\{\b_{rr}\!+\!\n_{rr} \!+\! (\b_r\!-\!\a_r)(\b_r\!+\!\n_r)\!+\!\n_r^2\Big\}
\!-\!\sdfrac{1}{2} e^{-2 \a -2 \n + 2\j } \Big\{ \w  \w_{rr} \!+\! \w \w_r  (\b_r\!-\!\a_r\!-\! \n_r \!+\! 3 \j_r)\!+\! \sdfrac{3 \w_r^2}{2} \Big\} \nn
&\hspace{-5mm} \!+ e^{-2 \b } \Big\{\a_{\th \th } \!+\!\n_{\th \th } \!+\! (\a_\th\!-\! \b_\th)(\a_\th\!+\! \n_\th)  \!+\! \n_\th^2 \Big\}
\!-\!\sdfrac{1}{2}e^{-2 \b -2 \n + 2\j }\Big\{\w  \w_{\th \th } \!+\! \w\w_\th (\a_\th\!-\!\b_\th \!-\!\n_\th \!+\! 3 \j_\th) \!+\! \sdfrac{3 \w^2_{\th }}{2}\Big\}\\
&G^r_{~r}=e^{-2\a } \big(\b_r\n_r\!+\!\b_r \j_r\!+\!\n_r \j_r\big) +\sdfrac{\w_r^2}{4} \, e^{-2 \a -2 \n + 2\j }  - \sdfrac{\w_{\th }^2}{4} \, e^{-2 \b -2 \n + 2\j }\nn
& \hspace{1.5cm}+  e^{-2 \b } \Big\{ \n_{\th \th }+\j_{\th \th } + (\j_\th \!-\! \b_\th)(\n_{\th }\!+\!\j_{\th})+\n_{\th }^2\Big\}.  \label{Grr_rot}\\
&G^\th_{~\th} = e^{-2 \a } \Big\{\n_{rr}+\j_{rr} + (\j_r\!-\!\a_r )\left(\n_r\!+\!\j_r\right)\!+\!\n_r^2\Big\} -\sdfrac{\w_r^2}{4} \, e^{-2 \a -2 \n + 2\j } 
+\sdfrac{\w_{\th }^2}{4} \, e^{-2 \b -2 \n + 2\j }\nn
& \hspace{1.5cm}+  e^{-2 \b } \big(\a_{\th }\n_{\th }+\a_\th\j_{\th } +\n_{\th } \j_{\th }\big)
\label{Gththrot}\\
&G^r_{~\th}=e^{-2 \a } \Big\{\!-\n_{r\th}\!-\!\j_{r\th} \!+\! \a_{\th } (\n_r\!+\!\j_r)\!+\!\n_{\th } (\b_r\!-\!\n_r)\!+\!\j_{\th } (\b_r\!-\!\j_r)\Big\}
+ \sdfrac{\w_{\th }\, \w_r }{2} \,e^{-2 \a -2 \n + 2\j } \nn
&\hspace{1.5cm} = e^{-2\a+ 2 \b } \,G^\th_{~r}
\end{align}
\label{Einstens}\end{subequations}
where a subscript on a metric function denotes a partial derivative, {\it e.g.} $\w_r\equiv\pa\w/\pa r$, and all components not listed vanish identically by the stationarity 
and axisymmetry of (\ref{axisymstat}).  

As in the nonrotating case considered previously in Sec.~\ref{Sec:NRotat}, the Dirac $\d$-distribution is defined with respect to the three-surface integration measure (\ref{3surft}) 
involving $\sqrt{-g}= \exp(\a + \b + \j + \n)$,  {\it cf.}\,(\ref{detg}). Since $\b$ and $ \j$ are continuous on the horizon but $\a$ and $\n$ need not be, we multiply (\ref{Einstens}) by the 
potentially discontinuous factor $\exp(\a + \n)$ of $\sqrt{-g}$, and retain only terms that may contain a Dirac $\d$-function distribution on the horizon as $R\!\to\,\RH$.

Since terms containing (either first or second) derivatives with respect to $\th$, while possibly discontinuous at the horizon, cannot produce Dirac $\d$-distributions on the surface 
of constant $r$ due to the continuity of $\b$, $\j$, and $\w$ there, we focus on the $r$-derivative terms in (\ref{Einstens}). Inspection of (\ref{Einstens}) shows that these are of two kinds, 
depending on an overall factor of either (A) $e^{-2 \a}$, or (B) $e^{-2\a -2 \n + 2 \j}$. The first set of (A) terms, containing the second derivatives $\b_{rr}, \j_{rr}$ or $\w_{rr}$, may give 
rise to $\d$-functions that would have to be retained if the surface were not at the horizon, but that vanish at the horizon where $e^{-2\a} \to 0$. Since $ 2\n_{rr} = e^{-2 \n} (e^{2\n})_{rr} - 4 \n_r^2$, 
(A) terms involving $\n_{rr}$ (and only those terms) can contribute a $\d$-function surface term comparable to the (B) terms. Note that there are no $\a_{rr}$ terms at all in (\ref{Einstens}).

Thus, terms that may contain a Dirac $\d$-function distribution on the horizon as $R\!\to\,\RH$ can come only from (A) terms containing $\n_{rr}$ in possible combination 
with first derivatives of the discontinuous $\a, \n$ functions, and (B) terms containing $\w_{rr}$, each of which after multiplication by $\exp(\a + \n)$ can be combined to give 
total $r$-derivatives as in (\ref{Gththdens}). Proceeding in this way, we obtain the contributions
\begin{subequations}
\begin{align}
e^{\a +\n} G^t_{~t} & = \sdfrac{\w}{2} e^{- \a - \n +2 \j } \Big\{\w_{rr} - \w_r \big(\a_r + \n_r \!\big)\Big\} + \dots 
= \frac{\w}{2}e^{2\j}\frac{\pa}{\pa r}\left( e^{-\a -\n} \frac{\pa \w }{\pa r} \right)  + \cdots \\
e^{\a +\n} G^t_{~\f} &=-\sdfrac{1}{2}e^{-\a-\n+ 2\j} \Big\{\w_{rr}- \w_r \big(\a_r + \n_r\big)\Big\} + \dots = -\sdfrac{1}{2}\, e^{2\j}\, \frac{\pa }{\pa r} \left(e^{- \a - \n} \frac{\pa \w}{\pa r}\right) + \dots\\
e^{\a +\n} G^\f_{~t} &= \sdfrac{1}{2} e^{- \a + \n} \Big\{- \!2 \w \n_{rr}  + 2 \w \n_r(\a_r \!- \!\n_r)\!\Big\} 
+  \sdfrac{\w^2\!}{\!2} e^{-\a -\n +2 \j}\Big\{ \w_{rr} \! - \!\w_r (\a_r \!+\! \n_r)\!\Big\} + \dots \nn
&\hspace{2mm} =  - \frac{\w}{2} \frac{\pa}{\pa r} \left( e^{-\a-\n} \frac{\pa e^{2\n}}{\pa r}\right) + \frac{\w^2\!}{2}\, e^{2\j}\, \frac{\pa }{\pa r} \left(e^{- \a - \n} \frac{\pa \w}{\pa r}\right) + \dots\\ 
e^{\a +\n} G^\f_{~\f}&= e^{-\a+ \n} \Big\{\n_{rr} -\a_r\n_r + \n_r^2\Big\}-\sdfrac{\w}{2} e^{- \a - \n +2 \j }\Big\{\w_{rr} -\w_r(\a_r+ \n_r)\Big\} + \dots \nn
&=\frac{1}{2}\frac{\pa}{\pa r} \left( e^{-\a - \n}  \frac{\pa e^{2\n}}{\pa r} \right) - \frac{\w}{2} e^{2\j} \frac{\pa}{\pa r} \left(e^{- \a - \n} \frac{\pa \w}{\pa r} \right) + \dots \\
e^{\a +\n}  G^\th_{~\th} & =  e^{-\a+ \n} \Big\{\n_{rr} -\a_r\n_r + \n_r^2\Big\} + \dots = \frac{1}{2}\frac{\pa}{\pa r} \left( e^{-\a - \n}  \frac{\pa e^{2\n}}{\pa r} \right) + \dots \label{nukap}
\end{align}
\label{Eindens}\end{subequations}
where the ellipsis and components not listed contain terms that do not contribute a $\d$-function at the horizon, because they either involve lower numbers of $r$-derivatives
of metric functions, or contain an additional power of $e^{2 \n}\!\to\! 0$ relative to the terms in (\ref{Eindens}), as can be checked explicitly from (\ref{Einstens}).

Since both $e^{-\a-\n} (e^{2\n})_r$ and $e^{-\a-\n}\w_r$ are of the form (\ref{genF}), just as in the nonrotating case, 
applying therefore the general rule (\ref{gendF}) to their derivatives we have
\begin{subequations}
\begin{align}
\frac{\pa}{\pa r}\left( e^{-\a -\n} \frac{\pa \w}{\pa r}\right) &= \left[e^{-\a -\n}\frac{\pa \w}{\pa r}\right] \, \d(r-R) + \dots\\
\frac{\pa}{\pa r}\left( e^{-\a - \n}  \frac{\pa e^{2\n}}{\pa r} \right) &= \left[ e^{-\a-\n} \frac{\pa}{\pa r} e^{2\n}\right]\, \d(r-R) + \dots
\end{align}
\label{derivrdel}\end{subequations}
where the square brackets denote the discontinuity at $r\!=\! R$ of the function enclosed. The coefficients of these $\d(r-R)$ terms 
in $e^{\a +\n}G^\m_{\ \n}$ can be identified with $8\pi G\, ^{(\S)}\!T^\m_{\ \n}\ e^{\a + \n}$ of the surface stress tensor. Note that there are no second derivative terms with 
respect to $r$ or singular terms in the $G^r_{\ r}$ component (\ref{Grr_rot}) which could give rise to singular $\d$-function terms on the surface. 
Thus the stress tensor localized on the surface has only $t,\th,\f$ components and is intrinsic to the surface (\ref{surfmet}) itself.

The two quantities appearing in (\ref{derivrdel}) can be expressed in a geometrically invariant form with a clear physical meaning. One can introduce the null 
vector $\bN$, orthogonal to a two-surface of constant $t$ and $r$, such that
\vspace{-6mm}
\begin{align}
\bN\cdot\bN = 0\,,\qquad ~\bN\cdot \boldsymbol{\pa}_\th = 0\,, \qquad ~\bN\cdot \boldsymbol{\pa}_\f = 0
\label{Nnull}
\end{align}
where $\boldsymbol{\pa}_\th=\pa/\pa\th$ and $\boldsymbol{\pa}_\f=\pa/\pa \f=\bKf$ are tangent to the two-surface. $\bN$ is normalized by
\begin{align}
\bl \cdot \bN = -1
\label{Nnorm}
\end{align}
so that the area element of the two-surface of constant $t$ and $r$ is $l_{[\m} N_{\n]} \, dA $ as in (\ref{2surftr}). On the horizon $\bl$ becomes a null 
vector tangent to the horizon, and $\bN$ becomes the second linearly independent null vector on the horizon (called $\bn$ in Ref.~\cite{BarCarHawk:1973}).

In the $(t,r,\th,\f)$ coordinates of (\ref{axisymstat}), the components of $\bN$ satisfying (\ref{Nnull})-(\ref{Nnorm}) are given explicitly by
\vspace{-7mm}
\begin{align}
N_\m &= (-1, - e^{\a -\n}, 0, 0)\,, \qquad N^\m = (e^{-2\n}, - e^{-\a -\n}, 0, \w e^{-2\n})
\label{Ncomp}
\end{align}
Then one finds
\begin{subequations}
\label{kcj}
\begin{align}
N_\m\,\ell_\n\, \big(\na^\m K^\n_{(t)}\big) &= \k + \w \cJ\\
N_\m\,\ell_\n\, \big(\na^\m K^\n_{(\f)}\big) &= - \cJ\\
N_\m\,\ell_\n \,\big(\na^\m \ell^\n\big) &= \k
\end{align}
\end{subequations}
where
\vspace{-7mm}
\begin{align}
\k = \frac{1}{2} \, e^{-\a-\n} \frac{\pa}{\pa r} \, e^{2\n} 
\label{kaprot}
\end{align}
is the surface gravity in the general case of nonvanishing angular momentum, defined to be positive for an inwardly directed gravitational acceleration.
It reduces to (\ref{surfgrav}) in the nonrotating case. The second invariant scalar quantity
\vspace{-6mm}
\begin{align}
\cJ =-\sdfrac{1}{2} \,e^{2\j} \, e^{-\a-\n}\ \frac{\pa \w}{\pa r} 
\label{cjdef}
\end{align}
will be shown in Sec.~\ref{Sec:Komar} to be proportional to the angular momentum density. 

The surface gravity is the acceleration with respect to the Killing time $t$ of a particle at momentary rest, projected to the normal direction to the constant 
$r$ surface. Its discontinuity at the horizon 
\begin{align}
\big[\k\big] = \frac{1}{2}\left[ e^{-\a-\n} \frac{\pa}{\pa r} e^{2\n}\right]
\label{kapdisc}
\end{align}
is therefore the difference of inward and outward accelerations at the horizon. The discontinuity of $\cJ$ at the horizon 
\vspace{-6mm}
\begin{align}
\big[\cJ\big]=  -\frac{1}{2}\, e^{2\j} \left[ e^{-\a-\n} \frac{\pa\w}{\pa r} \right]
\label{cJdisc}
\end{align}
will be related to the angular momentum carried by the singular horizon surface itself in Sec.~\ref{Sec:Komar}. 

We remark that the derivation of the $\d$-function terms (\ref{derivrdel}) on the rotating horizon in terms of $[\k]$ and $[\cJ]$ did not require any restrictions on the function 
$e^{-\a}$. However, for the discontinuities (\ref{kapdisc})-(\ref{cJdisc}) to be finite, $e^{-\a}$ must tend to zero at the horizon at least as fast as $e^\n$. If it does not, then not 
only would the discontinuities and the resulting surface stress tensor on the horizon be infinite, but the Einstein tensor (\ref{Einstens}) itself would also contain terms that grow 
without bound {\it in the bulk} as $R \to \RH$. Conversely, if $e^{-\a}$ tends to zero at the horizon faster than $e^\n$, then the $\d$-function terms in (\ref{derivrdel}), and 
resulting horizon surface stress tensor will vanish in the limit $R \to \RH$. In both the nonrotating Schwarzschild exterior considered in Sec.~\ref{Sec:NRotat} and 
the rotating Kerr BH vacuum exterior, in fact $e^{-\a}\! \propto\! e^\n \!\to\! 0$ tend to zero the same way as $R \to \RH$. Thus in the most relevant physical applications 
to BH horizons, $e^{-\a-\n}$ and the two discontinuities (\ref{kapdisc})-(\ref{cJdisc}) are well-defined and finite. A finite $e^{-\a-\n}$ as $R \to \RH$ is necessary for a finite 
nonzero $\sqrt{-g} = e^{\a+\n+\b+\j}$ volume measure at the horizon.

From the Einstein equations $G^\m_{\ \n} = 8 \pi G \, T^\m_{\ \n}$, the stress tensor localized as a Dirac $\d$-function distribution on the horizon surface is
\vspace{-1mm}
\begin{align}
^{(\S )}T^i_{\ j}\ e^{\a + \n}  = \cS^i_{\ j} \, \d(r-\RH)
\label{SurfTS}
\end{align}
where from (\ref{Eindens}), (\ref{derivrdel}) and (\ref{kapdisc})-(\ref{cJdisc}), the nonvanishing components of $ \cS^i_{\ j}$ for a general rotating null horizon are given by 
\vspace{-6mm}
\begin{subequations}
\begin{align}
8 \pi G\, \cS^t_{\ t} &= - \wH\big[\cJ\big] \\
8 \pi G\, \cS^t_{\ \f} &= \big[\cJ\big] \\
8 \pi G\, \cS^\f_{\ t} &= -\wH\big[\k\big] - \wH^2\big[\cJ\big] \label{Sphit}\\
8 \pi G\, \cS^\f_{\ \f} &= \big[\k\big] + \wH\big[\cJ\big] \\
8 \pi G\, \cS^\th_{\ \th} &=  \big[\k\big] 
\end{align}
\label{surfTrot}\end{subequations}
obtained directly from the possible $\d$-function contributions to the Einstein tensor. 
The tensor (\ref{SurfTS}) is a well-defined distribution when integrated at fixed $t$ 
against the continuous measure $\int dr\, d\th \,d\f\, e^{\b + \j} = \int dr\, dA$, where $dA$ is the area element (\ref{dArea}) of the induced metric on the two-surface of fixed 
$t$ and $r$. The surface stress tensor $\cS^{i}_{\ j}$ differs in general from the horizon limit of the surface energy tensors $S^{i}_{\ j}$ 
defined in~\cite{Sen:1924,Lanczos:1924,Israel:1966a,BarrabesIsrael:1991}; its relation to them is discussed in Sec.~\ref{Sec:Junc}.

As a check on our result~\eqref{surfTrot}, we also obtained the $\d$-function terms in the Einstein tensor density $\sqrt{-g} \, G^\m_{\ \n}$ on the rotating horizon by transforming all 
second derivatives of the metric in $\sqrt{-g} \, G^\m_{\ \n}$ into derivatives of the form $(\sqrt{-g} \, g_{\m\n,\a})_{,\b}$, and taking the limit $R \to \RH$ 
assuming a finite $\sqrt{-g}$ and continuity of the Killing vectors and the induced metric on the horizon. This gives the coefficients $\cG^\m_{\ \n}$ of the $\d$-function 
Einstein tensor density $e^{\a+\n} G^\m_{\ \n} \!=\! \cG^\m_{\ \n} \d(r-\RH)$ concentrated at the null horizon. Then Einstein's equations relate these $\d$-function terms 
to those of the stress-energy tensor through $\cG^\m_{\ \n} \!=\! 8 \pi G \cS^\m_{\ \n}$, which leads to $\cS^{r}_{\ r} \!=\! \cS^{i}_{\ r}\!=\!\cS^{r}_{\ i} \!=\!0$ and the $\cS^i_{\ j}$ 
components of the surface stress tensor in (\ref{surfTrot}).
 
All the surface stress tensor components (\ref{SurfTS}), when taken to the horizon, can be written in terms of $[\k]$ and $[\cJ]$.
This generalizes (\ref{surfTsph}) to the case of a rotating horizon. In the limit where both $\wH\! \to \!0$ and $[\cJ] \!\to \!0$, the previous result 
(\ref{Snonrot}) for the spherically symmetric horizon surface stress is recovered. Equations.~(\ref{SurfTS}) and (\ref{surfTrot}) together with (\ref{kcj})-(\ref{cJdisc})
are a principal result of this paper.

Since 
\vspace{-6mm}
\begin{align}
\cS^i_{\ t} + \wH\cS^i_{\ \f} = 0\,, \qquad i = t, \th, \f
\label{Seqn}
\end{align}
the surface stress tensor (\ref{surfTrot}) is particularly simple in the corotating local orthonormal tangent frame basis of (\ref{tetradbasis})-(\ref{dualbasis})
\vspace{-5mm}
\begin{align}
\cS^a_{\ b} = e^a_{\ \m}\, \cS^\m_{\ \n}\, \y^\n_{\ b}
\label{Sab}
\end{align}
in which it has components
\begin{subequations}
\begin{align}
\cS^0_{\ 0} &= \cS^t_{\ t} + \wH \cS^t_{\ \f} = 0 \\
\cS^0_{\ 3} &= e^{\n-\j} \cS^t_{\ \f} = e^{\n-\j}\frac{\big[\cJ\big]}{8\pi G}  \label{S03}\\
\cS^3_{\ 0} &= e^{\j-\n} \Big\{ \cS^\f_{\ t} + \wH \cS^\f_{\ \f}  -\wH \big(\cS^t_{\ t} + \wH \cS^t_{\ \f} \big)\Big\} = 0\\
\cS^3_{\ 3} &= \cS^\f_{\ \f}  - \wH \cS^t_{\ \f} = \frac{\big[\k\big]}{8 \pi G}\\
\cS^2_{\ 2} &=  \cS^\th_{\ \th} = \frac{\big[\k\big]}{8 \pi G}  
\end{align}
\label{surfSrot}\end{subequations}
where although (\ref{S03}) goes to zero on the horizon, we retain its explicit form with $e^\n$ small but finite in order for the transformation (\ref{Sab}) to be strictly 
invertible back to the coordinate basis (\ref{surfTrot}). Similar simplifications occur if the components of the Einstein tensor 
in the bulk (\ref{Einstens}) are expressed in the corotating orthonormal tangent frame basis. 

The surface stress tensor in  (\ref{surfTrot}) has the null vector $\bl$ as eigenvector with zero eigenvalue, i.e., $\cS^\m_{\ \n} \, \ell^\n = 0$, or 
\begin{align}
\cS^i_{\ j} \, \ell^j = 0
\end{align}
since both $\bl$ and $\cS$ are tangential to the horizon (this is the geometrical meaning of Eq.~(\ref{Seqn}), and the considerations in this paragraph can be carried out equally well in the tangent space of the spacetime or of the horizon). For $\big[\k\big]\ne0$, $\cS^i_{\ j}$ has two additional independent eigenvectors orthogonal to $\bl$ and tangent to the null horizon, with degenerate eigenvalue $\big[\k\big]$. They can be chosen to be the vectors $\bhth=e^{-\b} \boldsymbol{\pa}_\th$ and $\bhvf=e^{-\j}\big(\boldsymbol{\pa}_\f + \bl \, \big[\cJ\big]/\big[\k\big]\big)$, which are spacelike unit vectors orthogonal to each other. The surface stress tensor then decomposes as 
\begin{align}
\cS^i_{\ j} = \big[\k\big]  \, ( \hth^{\mkern2mu i} \, \hth_{j} + \hvf^{\mkern2mu i} \, \hvf_{j})
\label{Sij_decomposed}
\end{align}
where the index $j$ in $\hth_{j}$ and $\hvf_{j}$ has been lowered with the induced metric, resulting in $\hth_{j} \, dx^j = e^{\b} \, d\th$ and $\hvf_{j} \, dx^j = e^{\j} \, (d\f - \wH dt)$. The decomposition (\ref{Sij_decomposed}) of the surface stress tensor $\cS^i_{\ j} $ shows that it is locally isotropic in the frame $(\bl,\bhth,\bhvf)$ on the rotating null horizon.

The equations allow for $\big[\k\big]=0$ and $\big[\cJ\big]\ne0$, in which case the surface stress tensor $\cS^i_{\ j} $ is traceless and has only one other eigenvector besides $\bl$, namely $\bhth=e^{-\b} \boldsymbol{\pa}_\th$, with eigenvalue zero. It decomposes as 
\begin{align}
\cS^i_{\ j} = e^{-\j} \, \big[\cJ\big] \, \ell^i \, \hvf_j
\label{Sij_decomposed_rad}
\end{align}
and so its structure agrees with the algebraic type of the energy-momentum tensor of a null Maxwell field (pure radiation) \cite{Stephani:2003tm}. Thus $\cS^i_{\ j} $ with $\big[\k\big]=0$ and $\big[\cJ\big]\ne0$ may represent  a null azimuthal flow on the null horizon. Finally, and trivially, for $\big[\k\big]=\big[\cJ\big]=0$, $\cS^i_{\ j}=0$. 

The decompositions~(\ref{Sij_decomposed}) and~(\ref{Sij_decomposed_rad}) of the surface stress tensor $\cS^i_{\ j}$ support the interpretation that the anisotropy between the azimuthal and meridional stresses $\cS^\f_{\ \f}$ and $\cS^\th_{\ \th}$, and the azimuthal ``flow'' indicated by the time components $\cS^t_{\ t}$, $\cS^t_{\ \f}$, and $\cS^\f_{\ t}$ in (\ref{surfTrot}), are kinematical effects due to the rotation of the horizon.
The kinematical coordinate transformation is particularly simple for constant  $\big[\cJ\big]/\big[\k\big]$ with $\big[\k\big]\ne0$,
\begin{align}
\bar{t} = t -  \sdfrac{\big[\cJ\big]}{\big[\k\big]} \left(\f - \wH t\right) ,
\qquad
\bar{\th}=\th, 
\qquad
\bar{\f} = \f - \wH t
\end{align}
It is easy to verify that $\cS^{\bar{i}}_{\ \bar{j}} = \cS^i_{\ j} \, (\pa x^{\bar{i}}/\pa x^i) \, (\pa x^{j}/\pa x^{\bar{j}})$ has only two nonzero components, 
\begin{align}
\cS^{\bar{\th}}_{\ \bar{\th}} = \cS^{\bar{\f}}_{\ \bar{\f}} = \big[\k\big]
\label{Sij_bar}
\end{align}
To hypothetical observers at fixed coordinates $(\bar{\th},\bar{\f})$ on the horizon, which circle the horizon at fixed $\th$ with angular velocity $\wH$, the surface stresses appear as a stationary isotropic surface tension $\big[\k\big]$. 

In Appendix \ref{Sec:Weyl} we show from the form of the Weyl tensor that the surface stress tensor (\ref{surfTrot}) or (\ref{surfSrot}) describes a simple infinitely thin
stationary matter shell, and does not contain any propagating spin-$2$ gravitational shock wave component \cite{BarrabesHogan:2003}. 

\vspace{-1mm}
\section{Komar Mass and Angular Momentum of Rotating Geometry}
\label{Sec:Komar}
\vspace{-1mm}

Since the general stationary axisymmetric geometry is invariantly characterized by the existence of the two Killing vectors (\ref{Killvecs})
of time translation and azimuthal axial symmetry, one can define the Komar mass-energy and angular momentum in terms of these 
Killing vectors \cite{Komar:1959}. In addition to their fundamental coordinate invariant significance for geometries possessing Killing
symmetries, the expressions for the Komar mass and angular momentum are also useful for relating the surface stress tensor on 
singular null hypersurfaces to Einstein tensor densities such as (\ref{Gththdens}).

\vspace{-3mm}
\subsection{Integral form}
\vspace{-1mm}

Taking a covariant derivative of Killing's Eq.~(\ref{Killeq}) and using the commutation of covariant derivatives gives the fundamental local differential relation
\begin{align}
\na_\n \na^\m K^\n =  \big[\na_\n, \na^\m\big]\,  K^\n = R^\m_{\ \n} \,K^\n  = 8 \pi G \left( T^\m_{\ \n} - \sdfrac{1}{2} T \d^\m_{\ \n}\right) K^\n
\label{fundKill}
\end{align}
where the last equality follows by Einstein's equations, and $T= T^\m_{\ \m}$ is the trace of $T^\m_{\ \n}$. Eq.~(\ref{fundKill}) can be integrated 
over a three-surface $V$ with boundary $\pa V$. Referring to Appendix \ref{Sec:Tetrad}, since (\ref{Stokes3to2}) applies to any antisymmetric two-tensor, 
it can be applied to $F^{\m\n} = \na^\m K^\n = - \na^\n K^\m$ for $K^\m$ any Killing vector. 
This immediately yields
\begin{align}
\int_{\pa V} \na^\m K^\n d^2\S_{\m\n}  = \int_V \na_\n \na^\m K^\n d^3\S_\m =  8\pi G \int_V \left( T^\m_{\ \n} - \sdfrac{1}{2} T \d^\m_{\ \n}\right) \,K^\n d^3\S_\m  
\label{Stokes32}
\end{align}
Rearranging slightly yields the fundamental integral relation
\begin{align}
\frac{1}{4 \pi G} \int_{\pa V} \na^\n K^\m d^2\S_{\m\n} =- \int_V  \big(2\, T^\m_{\ \n} - T \d^\m_{\ \n}\big) K^\n d^3\S_\m\,. 
\label{KomarStokes}
\end{align}
If applied to the Killing vector of time translation $K^\m_{(t)}$, and in the situation where the boundary of the three-volume $V$ is composed of outer
and inner surfaces $\pa V_+$ and $\pa V_-$ respectively, with oppositely directed normals, (\ref{KomarStokes}) then gives the Komar 
mass-energy~\cite{Komar:1959,WaldBook,MazurMottola:2015}  
\begin{align}
M_K =  \frac{1}{4 \pi G} \int_{\pa V_+}\na^\n K_{(t)}^\m d^2\S_{\m\n} = 
- \int_V \big(2\, T^\m_{\ \n} - T \d^\m_{\ \n}\big) K_{(t)}^\n d^3\S_\m + \frac{1}{4 \pi G} \int_{\pa V_-}\na^\n K_{(t)}^\m d^2\S_{\m\n}
\label{Komarmass}
\end{align}
for any time independent (static or stationary) geometry admitting a time translation Killing vector $K_{(t)}^\m$. This shows that if $T^\m_{\ \n} =0$, the volume
integral vanishes and $M_K$ is independent of the bounding two-surface. Conversely, if the geometry is nonsingular and the volume includes all of space within
$\pa V_+$, then the contribution of the inner boundary $\pa V_-$ vanishes.

In a similar way the Komar angular momentum $J_K$ is given by the two-surface integral of the covariant derivative of the Killing vector of azimuthal 
rotational symmetry $K^\m_{(\f)}$ 
\begin{align}
J_K= -\frac{1}{8 \pi G} \int_{\pa V_+}\na^\n K_{(\f)}^\m d^2\S_{\m\n} = \int_V \left(T^\m_{\ \n} - \sdfrac{1}{2} T \d^\m_{\ \n}\right) K_{(\f)}^\n d^3\S_\m 
- \frac{1}{8 \pi G} \int_{\pa V_-}\na^\n K_{(\f)}^\m d^2\S_{\m\n}
\label{KomarJ}
\end{align}
where the factor of $-1/8\pi G$ can be verified by evaluating the surface integral in the asymptotically flat region on the outer boundary $\pa V_+$. As in the 
case of the mass, (\ref{KomarJ}) implies that in the absence of any matter, the three-volume integral vanishes, and the two-surface integral of $\na^\n K^\m$ 
is independent of the surface chosen. 

Expressions (\ref{Komarmass}) and (\ref{KomarJ}) are coordinate invariant, and may be evaluated in any set of coordinates. The most convenient coordinates 
in which to evaluate them are those adapted to the Killing symmetries (\ref{Killvecs}), which are exactly those of the general stationary axisymmetric line metric 
(\ref{axisymstat}). In these coordinates where (\ref{Kill2}) applies we have
\begin{align}
\na^\n K^\m_{(\g)}= g^{\n\b}\, \G^\m_{\ \b\g} = -\sdfrac{1}{2} \, g^{\m\a} g^{\n\b} \big( \pa_\a g_{\b \g} -\pa_\b g_{\a \g}\big) 
\label{naKill}
\end{align}
for $\g = t,\f$.
If the three-volumes $V$ in (\ref{Komarmass}) and (\ref{KomarJ}) are at fixed $t$ enclosed by two-surfaces at $r_-$ and $r_+$ in coordinates (\ref{axisymstat}),
we may make use of (\ref{2surftr}) for the two-surface element $d^2\S_{\m\n}$ to obtain
\begin{subequations}
\begin{align}
\na^\n K^\m_{(t)} d^2\S_{\m\n} &= \sdfrac{1}{2} \sqrt{-g} \,g^{rr} \,\big(g^{tt} \pa_r g_{tt} + g^{t\f} \pa_r g_{t\f}\big)\,d\th\,d\f = N_\m\,\ell_\n\, \big(\na^\m K^\n_{(t)}\big)\, dA 
= (\k + \w \cJ)\,  dA \label{massint}\\
\na^\n K^\m_{(\f)} d^2\S_{\m\n} &= \sdfrac{1}{2} \sqrt{-g} \, g^{rr} \,\big(g^{tt} \pa_r g_{t\f} + g^{t\f} \pa_r g_{\f\f}\big)\,d\th\,d\f = N_\m\,\ell_\n\, \big(\na^\m K^\n_{(\f)}\big)\,dA
= - \cJ\, dA
\label{angmomint}
\end{align}
\end{subequations}
respectively, since $g_{rt} = g_{r\f} =0$ in coordinates (\ref{axisymstat}), and we have also made use of (\ref{dArea}) and (\ref{kcj}). 

Therefore the mass integral (\ref{Komarmass}) takes the form
\begin{align}
M_K &=  \frac{1}{4 \pi G}\int_{\pa V_+}\!  (\k + \w \cJ)\, dA\nn
&=\int_V \sqrt{-g}\  \Big(\! \!-T^t_{\ t} + T^r_{\ r}  + T^\th_{\ \th}  + T^\f_{\ \f}  \Big)\, dr\,d\th\,d\f\, +\, \frac{1}{4 \pi G} \int_{\pa V_-} \! (\k + \w \cJ)\, dA 
\label{KomarMass}
\end{align}
and likewise the angular momentum integral is
\begin{align}
J_K =  \frac{1}{8 \pi G} \int_{\pa V_+}\!  \! \cJ \, dA = \int_V  \sqrt{-g}\ T^t_{\ \f} \,dr\,d\th\,d\f +\frac{1}{8 \pi G} \int_{\pa V_-}\!  \!\cJ \, dA 
\label{KomarAngmom}
\end{align}
for the three-volume $V$ at fixed $t$ enclosed by two-surfaces at $r_-$ and $r_+$ in coordinates (\ref{axisymstat}). Thus $\cJ$ is 
$8 \pi G$ times the angular momentum areal flux density and $\w\cJ$ the angular momentum contribution to the Komar mass-energy respectively,
of the two-surface at constant $t$ and $r$.

\vspace{-2mm}
\subsection{Local form of mass and angular momentum flux}
\vspace{-1mm}

Since $\na_\n \na^\m K^\n = - \na_\n \na^\n K^\m = -\sq K^\m$, from (\ref{fundKill}) we have
\begin{align}
-\sq K^\m_{(\g)}  =  -\frac{1}{\sqrt{-g}} \pa_\n\, \Big(\sqrt{-g}\, \na^\n K^\m_{(\g)}  \Big) = 4 \pi G\,\Big(\,2\,T^\m_{\ \n} - T \,\d^\m_{\ \n} \Big) K^\n_{(\g)} 
\label{LaplK}
\end{align}
for the two Killing vectors (\ref{Killvecs}) of (\ref{axisymstat}), or using (\ref{naKill}) 
\begin{align}
\frac{\pa}{\pa x^\n} \Big\{ \sqrt{-g}\, g^{\m\a} g^{\n\b} \big( \pa_\a g_{\b \g} -\pa_\b g_{\a \g}\big)\Big\}= 8 \pi G\,\sqrt{-g}\,  \Big(\,2\,T^\m_{\ \g} - T \,\d^\m_{\ \g} \Big) 
\label{Komarlocal}
\end{align}
which gives the local form of the conservation laws for the Komar mass flux and angular momentum flux in coordinates (\ref{axisymstat}) when $\g = t,\f$ respectively. 

For either value of $\g$, since in (\ref{axisymstat}) all the metric functions depend upon only $(r,\th)$, the index $\n$ in (\ref{Komarlocal}) ranges 
over these two values only. Thus for $\g=t$ we have
\begin{align}
\frac{\pa}{\pa r} \Big\{ \sqrt{-g}\, g^{\m\a} g^{rr} \big( -\pa_r g_{\a t}\big)\Big\}
+ \frac{\pa}{\pa \th} \Big\{ \sqrt{-g}\, g^{\m\a} g^{\th \th} \big( -\pa_\th g_{\a t}\big)\Big\}
= 8 \pi G\,\sqrt{-g}\,  \Big(2\,T^\m_{\ t} - T \,\d^\m_{\ t} \Big)
\end{align}
since $\b=r$ and $\b = \th$ are the only nonzero terms respectively in each of the terms on the left side and $g_{tr} = g_{t\th} = 0$. 
The index $\a$ now ranges over $t,\f$ only and so we obtain
\begin{align}
&\frac{\pa}{\pa r} \left\{ \sqrt{-g}\, g^{rr} \left(g^{\m t}\, \frac{\pa g_{tt}}{\pa r} + g^{\m \f}\, \frac{\pa g_{t\f}}{\pa r}\right)\right\} 
+ \frac{\pa}{\pa \th} \left\{ \sqrt{-g}\, g^{\th \th}\left( g^{\m t} \, \frac{\pa g_{tt}}{\pa \th} +  g^{\m\f}\,\frac{\pa  g_{t\f}}{\pa \th}\right)\right\}\nn
&\hspace{2cm} = - 8 \pi G\,\sqrt{-g}\,  \Big( 2\,T^\m_{\ t} - T \,\d^\m_{\ t} \Big)
\label{localmass}
\end{align}
as the local form of the mass-energy flux. 

Repeating these steps for the Killing vector of azimuthal symmetry and $\g= \f$ in (\ref{Komarlocal}) gives
\begin{align}
&\frac{\pa}{\pa r} \left\{ \sqrt{-g}\, g^{rr}\left( g^{\m t}\,  \frac{\pa g_{t\f}}{\pa r} + g^{\m\f}\, \frac{\pa g_{\f\f}}{\pa r}\right)\right\}
+\frac{\pa}{\pa \th} \left\{ \sqrt{-g}\, g^{\th \th} \left(g^{\m t}\, \frac{\pa g_{t\f}}{\pa \th} + g^{\m\f}\, \frac{\pa g_{\f\f}}{\pa r}\right)\right\}\nn
&\hspace{2cm} = -8 \pi G\,\sqrt{-g}\,  \Big(2\,T^\m_{\ \f} - T \,\d^\m_{\ \f} \Big)
\label{localangmom}
\end{align}
for the local form of the angular momentum flux. 

\vspace{-2mm}
\subsection{Discontinuities on a singular surface at fixed $r$}
\vspace{-1mm}

Each of the two local relations (\ref{localmass}) and (\ref{localangmom}) gives two relations for $\m= t, \f$ respectively. 
Making use of (\ref{metric}) and (\ref{detg}), the four quantities under $\pa/\pa r$ derivatives that appear on the left sides 
of these four relations are given explicitly by
\begin{subequations}
\begin{align}
\sqrt{-g}\, g^{rr} \left(g^{tt} \frac{\pa g_{tt}}{\pa r}  + g^{t\f} \frac{\pa g_{t\f}}{\pa r} \right) &=
e^{\b + \j} \ e^{-\a - \n}\left( \frac{\pa e^{2\n}}{\pa r} -\w\, e^{2\j}\, \frac{\pa \w}{\pa r} \right) = 2\, e^{\b + \j}\,(\k + \w \cJ)  \\
\sqrt{-g}\, g^{rr} \left(g^{t\f}  \frac{\pa g_{tt}}{\pa r}  + g^{\f\f} \frac{\pa g_{t\f}}{\pa r}\right) &=
\w\, e^{\b + \j} \ e^{-\a - \n}\left( \frac{\pa e^{2\n}}{\pa r} -\w\, e^{2\j}\, \frac{\pa \w}{\pa r} \right) 
- e^{\b - \j} \ e^{-\a + \n}\, \frac{\pa (\w e^{2\j})}{\pa r}\nn
&=  2\, e^{\b + \j} \w\left( \k + \w \cJ\right) - e^{\b - \j} \ e^{-\a + \n}\, \frac{\pa (\w e^{2\j})}{\pa r} \label{gamt}\\
\sqrt{-g}\, g^{rr} \left(g^{tt} \frac{\pa g_{t\f}}{\pa r}  + g^{t\f} \frac{\pa g_{\f\f}}{\pa r} \right)
&= e^{\b + 3\j} \ e^{-\a - \n}\, \frac{\pa \w}{\pa r} = - 2\, e^{\b + \j} \,\cJ  \label{secquant}\\
\sqrt{-g}\, g^{rr} \left(g^{t\f} \frac{\pa g_{t\f}}{\pa r} +  g^{\f\f}\frac{\pa g_{\f\f}}{\pa r}\right)&= 
\w\,e^{\b + 3 \j}\ e^{-\a - \n} \, \frac{\pa \w}{\pa r} +  e^{\b - \j} \ e^{-\a + \n}\, \frac{\pa e^{2\j}}{\pa r}\nn
&=  - 2\, e^{\b + \j} \,\w \cJ +  e^{\b - \j} \ e^{-\a + \n}\, \frac{\pa e^{2\j}}{\pa r} \label{gamphi}
\end{align}
\label{localaxisym}\end{subequations}
where $\k$ and $\cJ$ are defined by (\ref{kaprot}) and (\ref{cjdef}) respectively.

If each of the four relations (\ref{localmass})-(\ref{localangmom}) are integrated with respect to $r$ over an infinitesimally small range $r \in [r_-, r_+]$, 
where the metric derivatives in $r$ are discontinuous at a surface, but the $\th$ derivatives are not discontinuous, then the $\pa_\th$ terms 
do not contribute. Furthermore, if the surface is located at the horizon where $e^{2\n} = 0$, then the last terms in (\ref{gamt}) and (\ref{gamphi})
involving derivatives of $\w e^{2 \j}$ or $e^{2\j}$ respectively do not contribute. In that case we obtain the four relations
\begin{subequations}
\begin{align}
&\sdfrac{1}{2} \left[\sqrt{-g}\, g^{rr} \left(g^{tt}  \frac{\pa g_{tt}}{\pa r} +  g^{t\f} \frac{\pa g_{t\f}}{\pa r}\right) \right]_{r_-}^{r_+} = e^{\b + \j} \Big(\big[\k\big] + \w \big[\cJ\big] \Big)
=  4 \pi G\int_{r_-}^{r_+}\!dr \sqrt{-g}\  \Big( \!-\!T^t_{\ t} + T^\th_{\ \th} +T^\f_{\ \f}  \Big)\label{massjump1}\\
&\sdfrac{1}{2} \left[\sqrt{-g}\, g^{rr} \left(g^{t\f}  \frac{\pa g_{tt}}{\pa r}  + g^{\f\f} \frac{\pa g_{t\f}}{\pa r}\right) \right]_{r_-}^{r_+} = e^{\b + \j} \, \w \Big(\big[\k\big] + \w \big[\cJ\big] \Big)
= - 8 \pi G \int_{r_-}^{r_+}\! dr \sqrt{-g}\ T^\f_{\ t}\label{massjump2} \\
&\sdfrac{1}{2} \left[\sqrt{-g}\, g^{rr} \left(g^{tt} \frac{\pa g_{t\f}}{\pa r} + g^{t\f} \frac{\pa g_{\f\f}}{\pa r}\right)\right]_{r_-}^{r_+}= -e^{\b + \j} \, \big[\cJ\big] 
= -8 \pi G\int_{r_-}^{r_+} \!dr\sqrt{-g}\ T^t_{\ \f} \label{angmomjump1}\\
&\sdfrac{1}{2} \left[\sqrt{-g}\, g^{rr} \left(g^{t\f} \frac{\pa g_{t\f}}{\pa r} +  g^{\f\f}\frac{\pa g_{\f\f}}{\pa r}\right)\right]_{r_-}^{r_+} = -e^{\b + \j} \, \w\big[\cJ\big] 
= 4 \pi G\int_{r_-}^{r_+}\! dr \sqrt{-g}\,  \Big(T^t_{\ t} +T^\th_{\ \th}- T^\f_{\ \f} \Big)\label{angmomjump2}
\end{align}
\label{discMJ}\end{subequations}
where we have dropped the $T^r_{\ r}$ term since it has no singular surface component at the two-surface of fixed $r$ and $t$.
If the surface at fixed $r$ is not at the horizon, then the additional terms in (\ref{localaxisym}) would have to be retained.

Substituting $\sqrt{-g}$ from (\ref{detg}) we now observe that the integrands on the right side of these relations involve the
singular surface contributions (\ref{SurfTS}) so that
\begin{align}
\int_{r_-}^{r_+}  dr\, \sqrt{-g} \ ^{(\S )\!}T^i_{\ j}= \int_{r_-}^{r_+}  dr\, e^{\n+ \j + \a + \b}\,  e^{-\a - \n}\,  \cS^i_{\ j} \ \d(r-R) = e^{\b + \j}\, \cS^i_{\ j} 
\label{intTS}
\end{align}
since both $\b$ and $\j$ are continuous on the surface, as is $\w$. Making use of this relation the continuous factors $e^{\b+\j}$ cancel
from (\ref{discMJ}) and we obtain the four conditions
\begin{subequations}
\begin{align}
[\k] +  \wH \,[\cJ] &= 4 \pi G\, \Big(\! -\cS^t_{\ t} + \cS^\th_{\ \th} +\cS^\f_{\ \f}  \Big)\\
\wH \, [\k] + \wH^2\, [\cJ] &=  -8 \pi G\,  \cS^\f_{\ t}\\
[\cJ] &= 8 \pi G\,  \cS^t_{\ \f}\\
- \wH \,[\cJ] &= 4 \pi G \, \Big(\cS^t_{\ t} + \cS^\th_{\ \th} -\cS^\f_{\ \f}  \Big)
\end{align}
\label{Komardisc}\end{subequations}
on the null horizon surface. Referring to the results for the horizon surface stress tensor (\ref{surfTrot}) shows that all four relations (\ref{Komardisc}) are
satisfied as identities. Thus the local form of the Komar energy and angular momentum fluxes provide a consistency check on the calculation of the
singular stress tensor from the Einstein tensor in Sec.~\ref{Sec:Rotate} on a rotating null horizon surface. 

\vspace{-1mm}
\section{Modified Junction Conditions and Barrab{\`e}s-Israel Formalism}
\label{Sec:Junc}
\vspace{-1mm}

The first treatments of junction conditions and surface stress tensors generally assumed implicitly or explicitly the existence of {\it admissible} coordinates 
where all metric components are continuous on $\S$ \cite{Lanczos:1924,OBrSyn:1952,Lichnerowicz:1955}. This condition is relaxed in the
approach of Israel \cite{Israel:1966a}, but was still restricted to the cases that the singular hypersurface is entirely timelike or spacelike, so 
that its normal vector $\bn$ can be normalized to $\bn\cdot \bn = \pm 1$ respectively. With this restriction, 
the conventionally defined surface energy tensor $S_{ij}$ of an entirely spacelike or timelike hypersurface can be expressed
in terms of the discontinuities of the extrinsic curvature tensor 
\begin{align}
{\bK}_{ij} \equiv \be^\m_{(i)}\,  \be^\n_{(j)} \,  (\na_\n n_\m)  =   - n_\m \,\be^\n_{(j)} \na_\n \be^\m_{(i)}
\label{extrK}
\end{align}
at the hypersurface as
\begin{align}
8 \pi G\, S_{ij} =- \big[K_{ij}\big] + h_{ij} \, \big[K^l_{\ l}\big]
\label{Israeljunc}
\end{align}
where $K_{ij}$ is ${\bK}_{ij} $ with a unit normal $\bn$ and the indices are raised and lowered with the induced metric $h_{ij}$ \cite{Israel:1966a}. Here $\be_i$ are the three vectors tangent to the hypersurface with components 
\begin{align}
\be^\m_{(i)} = \frac{\pa \bar x^\m (\x)\!\!}{\!\!\pa \x^i}
\end{align}
where $x^\m = \bar x^\m (\x)$ are the parametric equations specifying the surface in the embedding geometry in surface coordinates $\x^i$, and $\bn$ is normal to
the surface so that $\bn \cdot \be_{(i)}= 0$. The conventional definition of the surface energy tensor $S_{ij}$ for a timelike hypersurface is such that 
$S_{ij}= \lim\limits_{\d\to0} \int_0^\d T_{ij} \, dl_{\rm n}$, where $l_{\rm n}$ is the proper distance measured in a direction normal to the hypersurface 
(the normal Gaussian coordinate of the hypersurface, see e.g.~\cite{Lichnerowicz:1955}).

Since it is not possible to normalize a null vector to $\pm 1$, the algorithm of \cite{Israel:1966a} as first formulated is not applicable to null horizon surfaces where $\bn\cdot \bn=0$. 
Barrab{\`e}s and Israel (BI) cited the convenience of a unified formalism to treat both cases of null and nonnull hypersurfaces, and to that end 
introduced a transverse null vector $\bN$, satisfying $\bN\cdot \bN =0$, but $\bn\cdot \bN = \h^{-1}\neq 0$, and an ``oblique” or ``transverse” extrinsic curvature 
and surface energy tensor based on it \cite{BarrabesIsrael:1991}. The components of the transverse vector $N^\m$ and normal $n_\m$ are assumed to be 
continuous across the null hypersurface in this BI approach.

In this section we show that the original Israel formalism applies even to the case of rotating null horizon hypersurfaces if the normalization of $\bn$ 
is allowed to approach zero
\begin{align}
\bn \cdot \bn \equiv \e = e^{2\n} \to 0
\label{normn}
\end{align}
as the horizon is approached, provided also that the discontinuity junction conditions are specified for a modified contravariant/covariant extrinsic curvature tensor 
${\bf K}^i_{\ j}$ with  the nonunit normal $\bn$ in (\ref{normn}) and one index raised compared to (\ref{extrK}). On  a non-null hypersurface, ${\bf K}^i_{\ j} = e^\n K^i_{\ j}$, where $K_{ij}$ is the usual extrinsic curvature tensor defined with a unit normal and the index is raised with the induced metric. The junction conditions determined in this way do not require any oblique extrinsic curvature 
based on $\bN$, and yield a finite limit as $R\!\to\!\RH$ that agrees with the Einstein tensor analysis of Sec.~\ref{Sec:Rotate}.  

\vspace{-2mm}
\subsection{Spherically symmetric nonrotating case}
\label{Sec:BInonrot}
\vspace{-1mm}

To see what the problems are with earlier general formulations of junction conditions on null hypersurfaces, and how to cure them, consider again the spherically symmetric case
analyzed directly from the Einstein tensor in Sec.~\ref{Sec:NRotat}. Specifying the surface to be at fixed $r= R > \RH$ {\it away} from the horizon in the 
coordinates (\ref{sphstat}) by $\F \equiv r-R =0$, the normal $\bn$ is given by
\begin{align}
n_\m &= \sdfrac{1}{\z}\, \pa_\m \F = \sdfrac{1}{\z}\, \d_{\ \m}^r
\label{zdef}
\end{align}
where $\z$ is an arbitrary normalization factor (called $\a$ in the notation of \cite{BarrabesIsrael:1991}) that may depend on $R$.  The normal vector $\bn$
is spacelike for $R > \RH$ and has norm
\begin{align}
\bn \cdot \bn = \frac{h(R)}{\z^2}
\label{nznorm}
\end{align}
in the metric (\ref{sphstat}). If we were to set this normalization to $+1$ by choosing $\z = \sqrt{h}$, a short calculation of (\ref{extrK}) shows that  $2\,K_{ij} = \sqrt{h} \, \pa_r g_{ij}$ 
for the surface coordinates $i,j = t,\th,\f$ \cite{MazurMottola:2015}. Thus all components of $K_{ij} \sim \sqrt{h}$ go to zero as $R \to \RH$, and the usual junction conditions 
give a vanishing discontinuity and vanishing surface energy tensor $S_{ij}$. On the other hand if one raises one of the indices $i$ or $j$, the $K^t_{\ t}$ component is 
$\frac{1}{2} \sqrt{h}\,\pa_r f/f$ which diverges when  $R\to \RH$ since $f\to 0$ with $h/f\to\textrm{const}$, and hence junction conditions with mixed indices would give 
a divergent energy tensor $S^i_{\ j}$ on the null horizon. This divergence stems from the simple kinematical fact that the spacelike normalization $\bn\cdot\bn = 1$ requires 
going to the rest frame of the two-surface, but this involves an infinite Lorentz boost as $R \to \RH$ where the hypersurface is null, 
and no rest frame for the null hypersurface exists.

On the other hand, if in (\ref{zdef}) and (\ref{nznorm}) we set
\begin{align}
\z^{-1} = \sqrt{\frac{f}{h}}\,,\qquad {\rm so \ that} \qquad \bn \cdot \bn = f(R) = \e \to 0
\label{zfix}
\end{align}
in the continuous normalization (\ref{normn}) with $\e =\exp(2\n)= f(R)$ in the spherically symmetric case, and we compute the discontinuity in
\begin{align}
{\bf K}^i_{\ j} = - g^{ik} n^\m \G_{\m jk} = \sdfrac{1}{2} g^{ik} g^{rr} \sqrt{\sdfrac{f}{h}}\frac{\pa g_{jk}}{\pa r}
\label{upK}
\end{align}
with one index raised, for this $\bn$ instead of that of (\ref{extrK}), we find
\begin{subequations}
\begin{align}
{\bf K}^t_{\ t} &= \frac{1}{2} \sqrt{\frac{h}{f}} \frac{df}{dr} = \k\\
{\bf K}^\th_{\ \th} &= {\bf K}^\f_{\ \f} = \frac{\sqrt{fh}}{r} \to 0
\end{align}
\end{subequations}
and 
\vspace{-6mm}
\begin{subequations}
\begin{align}
\big[{\bf K}^t_{\ t}\big] & = [\k]\\
\big[{\bf K}^\th_{\ \th} \big]&= \big[{\bf K}^\f_{\ \f}\big] = 0
\end{align}
\label{discK}\end{subequations}
in the null horizon limit $R\to \RH, \ \e = f(R)\to 0$. Substituting these results into the usual form (\ref{Israeljunc}) of the Israel junction conditions but with one index raised 
and the unit-normal extrinsic curvature $K^i_{\ j}$ replaced by ${\bf K}^i_{\ j}$, gives exactly the surface stress tensor $\cS^i_{\ j}$ in (\ref{surfTsph}) in place of $S^{i}_{\ j}$,
\begin{align}
8 \pi G\, \cS^i_{\ j} =- \big[{\bf K}^i_{\ j}\big] + \d^i_{\ j} \, \big[{\bf K}^l_{\ l}\big]
\label{modjunc}
\end{align}
where the trace $\big[{\bf K}^l_{\ l}\big] = [\k]$, in agreement with the direct calculation from the Einstein 
tensor density (\ref{discGthth}) and (\ref{Snonrot}) \cite{MazurMottola:2015}.

What has happened is that although the discontinuity in $K_{ij}$ defined by (\ref{extrK}) vanishes as $\sqrt{\e}$ as $\e \to 0$ with
the unit normalization of $\bn$, and the modified $\bn$ of (\ref{zfix}) contributes another factor of $\sqrt{\e}$, raising one index in (\ref{upK}) 
contributes $1/\e$ so that (\ref{discK}) is {\it finite} in the limit $R \to \RH$, $\e \to 0$. The contravariant/covariant tensor (\ref{upK}) removes 
the singular coordinate dependent effects in going to the null horizon hypersurface, is physically motivated by the Komar formulae and gives a well-defined 
$\d$-function distribution for the surface stress tensor with the proper integration measure. 

To find the relation between these new or modified junction conditions and the original prescriptions of Refs.~\cite{Israel:1966a, BarrabesIsrael:1991}, note that
the stress-energy tensor $T^\m_{\ \n}$ is determined through Einstein's equations by the Einstein tensor, which is independent of the normalization chosen for
the normal $\bn$. In~\cite{BarrabesIsrael:1991}, 
\vspace{-2mm}
\begin{align}
^{(\S)}T^\a_{\ \,\b} = \z \, \SBI^\a_{\ \,\b} \, \d (\F)
\label{eq:TBI}
\end{align}
and the BI surface tensor $S^\a_{\ \b}$ depends upon the normalization $\z$ of (\ref{zdef}) in such a way that at given $\Phi$
\vspace{-6mm}
\begin{align}
\z_1 \, ^{(1)}\!S^\a_{\ \,\b} = \z_2 \, ^{(2)}\!S^\a_{\,\ \b}
\end{align}
where $^{(1)}\!S^\a_{\ \,\b}$ is computed from (\ref{upK}) with $\bn$ normalized by (\ref{nznorm}) with $\z=\z_1$, and $^{(2)}\!S^\a_{\ \,\b}$ is computed from the same
formula (\ref{upK}), but with $\bn$ normalized by (\ref{nznorm}) with $\z=\z_2$. Therefore the surface stress tensor computed by the original Israel
junction conditions given in \cite{Israel:1966a} with $\bn\cdot\bn =1$ and $\z_1= \sqrt{h}$ is related to the surface stress tensor $^{(2)}\!S^\a_{\,\ \b} = \cS^\a_{\,\ \b}$
computed by our modified junction conditions with $\bn\cdot\bn =\ve$ and $\z_2 = \sqrt{h/f}$  by
\vspace{-2mm}
\begin{align}
\big(\cS^\a_{\ \,\b}\big)_{\rm here}  = \frac{\z_1}{\z_2}  \, \big(S^\a_{\ \,\b}\big)_{\rm Israel} = \sqrt{f (R)} \, \big(S^\a_{\ \,\b}\big)_{\rm Israel}
\label{Isrmod}
\end{align}
which was found previously in \cite{MazurMottola:2015}. When $\e$ is finite and both the metric and $\sqrt{-g}$ are nonsingular and continuous, 
the redshift factor $\sqrt{f (R)}$ is finite and can easily be kept track of, causing no problems. However when $R\to \RH$ and $f(R) \to 0$,
the Israel junction conditions give an infinite or ambiguous result, while the modified junction conditions (\ref{modjunc}) with (\ref{zfix}) work equally 
well for singular null horizon surfaces, even if expressed in singular Schwarzschild coordinates and $\sqrt{-g}$ is discontinuous,
as in the gravastar case of \cite{MazurMottola:2015} and Sec.~\ref{Sec:NRotat}. The modified junction conditions given here also correspond
to the natural integration measure $\sqrt{-g}$ of the Komar mass and angular momentum formulae of Sec.~\ref{Sec:Komar}.

\vspace{-2mm}
\subsection{Comparison to the Barrab{\`e}s-Israel prescription}
\label{Sec:BINonRot}
\vspace{-1mm}

In the BI method of \cite{BarrabesIsrael:1991}, a second transverse vector $\bN$ is introduced to satisfy the condition $\bn\cdot \bN = \h^{-1}$, 
where $\h$ can take on any nonzero value, and may conveniently be taken to be $-1$. This one condition on $4$ components leaves
$3$ components of $\bN$ undetermined, as the only other requirement BI impose on $\bN$ is that its projection onto the surface must be continuous.
If one also requires $\bN$ to be null, and normal to the two-surface of constant $t$ and $r$, satisfying (\ref{Nnull}) and (\ref{Nnorm}) as in 
Refs.~\cite{Poisson:2002, Poisson:2004,MazurMottola:2015}, then $\bN$ is fixed (up to a sign of $N^r$), and may be chosen
to coincide with $\bN$ introduced in Sec.~\ref{Sec:Rotate}.

Specializing (\ref{Ncomp}) to the spherically symmetric case first, the vector $\bN$ has components
\vspace{-2mm}
\begin{align}
N^\m &= \frac{1}{f} \,\d^\m_{\ t} - \sqrt{\frac{h}{f}} \,\d^\m_{\ r}\,, \qquad N_\m = g_{\m\n} N^\n = - \d_{\ \m}^t  - \frac{1}{\!\!\sqrt{fh}}\, \d^r_{\ \m} 
\label{Ndef}
\end{align}
so that $N_\m$ projected onto the surface of constant $r$ has $(t,\th,\f)$ components $(-1,0,0)$ which are all finite and continuous across the surface
for any $r$, while neither $N_r$ nor $n_r$ are continuous across the horizon if $\sqrt{h/f}$ is not, and $N^t$ is singular at $f=0$.

The authors of \cite{BarrabesIsrael:1991} define the ``oblique extrinsic curvature”
\vspace{-2mm}
\begin{align}
\cK_{ij} \equiv - N_\m \,\be^\n_{(j)} \na_\n \be^\m_{(i)} = - N^\m \G_{\m ij} 
\label{oblique}
\end{align}
and the discontinuity on the null hypersurface 
\vspace{-2mm}
\begin{align}
\g_{ij} \equiv 2\, [\cK_{ij}] = -2\, [N^\m \G_{\m ij}] = -2 \big[N^r\G_{rij}\big] =  - \left[ \sqrt{\frac{h}{f}} \frac{d g_{ij}}{dr} \right]
\label{gamK} 
\end{align}
giving
\vspace{-6mm}
\begin{align}
\g_{tt} = \left[ \sqrt{\frac{h}{f}} \frac{d f}{dr} \right] = 2\,[\k]
\end{align}
with the other components vanishing. Then Eqs.~(17)-(18) of Ref.~\cite{BarrabesIsrael:1991} with (\ref{zdef}) and $\g = g^{ij}\g_{ij}$ give 
\begin{align}
-16\pi G\, \cS^\th_{\ \th} =  -16\pi G\, \cS^\f_{\ \f}  = \e \g = -\g_{tt} =- 2\,[\k] =- \frac{2}{\RH\!\!}  = - \frac{1}{GM}
\label{BIsphsym}
\end{align}
with all other components projected onto the surface vanishing. Thus (\ref{surfTsph}) is recovered by the BI method, with the particular (singular) choice (\ref{Ndef}) of $\bN$,
provided one keeps the $\e \g$ term in $\cS^i_{\ j}$, since $\g = g^{tt} \g_{tt} = -\g_{tt}/\e$ is singular, and takes the limit $R\!\to\! \RH, \e\!\to\!0$ which yields the finite
result (\ref{BIsphsym}).

At this point one may ask if the result (\ref{BIsphsym}) is independent of the conditions (\ref{Nnull}) and (\ref{Nnorm}) on $\bN$. The authors 
of \cite{BarrabesIsrael:1991} remark that although the oblique tensor $\cK_{ij}$ is not independent of the choice of transverse vector $\bN$, transforming
under the change
\vspace{-1mm}
\begin{align}
\bN \to \bN + \l^k(\x) \, \be_{(k)} \,,\qquad {\rm as} \qquad \cK_{ij} \to \cK_{ij} - \l^k\,\G_{kij}
\label{lamN}
\end{align}
the discontinuity $[\cK_{ij}]$ is independent of $\l^a$, on condition that the tangential $\G_{kij}$ is continuous at the surface. The continuity of $\bN$ 
is also implicitly assumed in Ref.~\cite{BarrabesIsrael:1991}. However, inspection of (\ref{Ndef}) shows that $N^\m$ and in particular $N^r$ which enters (\ref{gamK}) 
is neither finite, nor continuous at the horizon in the case of the gravastar, {\it cf.}\,(\ref{fhdeS})-(\ref{fhSchw}). Thus although the result (\ref{BIsphsym}) 
is correct in this case, with the particular choice (\ref{Ndef}), as it can be verified independently from the Einstein tensor in Sec.~\ref{Sec:NRotat}, 
the argument given in Ref.~\cite{BarrabesIsrael:1991} for the prescription being independent of the choice of $\bN$ does not apply. For this reason the BI prescription 
may lead to ambiguous or incorrect results for the surface stress if it is applied in the cases that its underlying assumption of continuity of the oblique vector $\bN$ 
does not hold. In Appendix \ref{Sec:BIrot}, we give an explicit example in the case of the rotating horizon, where applying the BI algorithm -
or rather {\it mis}-applying it where $\bN$ is discontinuous and independence of the choice of $\bN$ cannot be assumed - leads to incorrect results 
with the choice (\ref{Ncomp}). 

\vspace{-1mm}
\section{Modified Israel Junction Conditions for a Rotating Null Horizon}
\label{Sec:Krot}
\vspace{-1mm}

We may apply the proposed modification of the original Israel junction conditions (\ref{modjunc}) with (\ref{upK}) to the rotating horizon as well. 
Define the (nonunit) normal vector $\bn$ to the surface at constant $r=R$ to be defined by (\ref{zdef}) and fix 
\vspace{-4mm}
\begin{align}
\z = e^{-\a-\n}\qquad{\rm so\ that}\qquad \bn \cdot \bn = e^{2\n} \equiv \e\to 0
\label{neps}
\end{align}
as $R \to \RH$ approaches the horizon. The normal $\bn$ so defined has components
\vspace{-1mm}
\begin{align}
n_\m = \d^r_{\ \m} \, e^{\a + \n} \,,\hspace{1.2cm} n^\m = \d^\m_{\ r} \, e^{-\a + \n} 
\label{nrot}
\end{align}
and as $R \to \RH$ the generators of the future and past horizons are given respectively by the null vectors $(\bl \pm \bn)/2$.
The extrinsic curvature tensor (\ref{upK}) defined with one index raised for this $\bn$\\
is \vspace{-5mm}
\begin{align}
{\bf K}^i_{\ j} = -g^{ik}n^\m \G_{\m jk}  =\sdfrac{1}{2} e^{-\a + \n} g^{ik} \left(\frac{\pa g_{jk}}{\pa r} - \frac{\pa g_{rk}}{\pa x^j}  - \frac{\pa g_{rj}}{\pa x^k}\right)\,,\qquad i, j= t,\th,\f
\label{Kdef}
\end{align}
which has the nonvanishing components
\begin{subequations}
\begin{align}
&\hspace{3cm}{\bf K}^t_{\ t} = \sdfrac{1}{2} \, e^{-\a-\n} \,\frac{\pa}{\pa r} e^{2\n}-  \sdfrac{1}{2}\, \w\, e^{2\j}\, e^{-a-\n} \,\frac{\pa \w}{\pa r} = \k + \w\, \cJ\\
&\hspace{3cm}{\bf K}^t_{\ \f} = \sdfrac{1}{2}\, e^{2\j}\, e^{-a-\n} \,\frac{\pa \w}{\pa r} = -\cJ\\
{\bf K}^\f_{\ t} &= \sdfrac{1}{2} \,\w\, e^{-\a-\n} \,\frac{\pa}{\pa r} e^{2\n} - \sdfrac{1}{2}\, \w^2 e^{2\j}\, e^{-a-\n} \,\frac{\pa \w}{\pa r} - \sdfrac{1}{2}\, e^{-2\j}\, e^{-a+\n} \,
\frac{\pa}{\pa r}\,\big(\w e^{2\j}\big) = \w\,\k + \w^2\,\cJ + \cO (\e) \\
&\hspace{3cm}{\bf K}^\th_{\ \th} = \sdfrac{1}{2}\, e^{-\a+\n-2\b} \,\frac{\pa}{\pa r}e^{2\b}= \cO (\e) \\
&\hspace{1.5cm}{\bf K}^\f_{\ \f} = \sdfrac{1}{2}\,\w\, e^{2\j}\, e^{-a-\n} \,\frac{\pa \w}{\pa r} + \sdfrac{1}{2}\,e^{-2\j}\, e^{-a+\n} \,\frac{\pa}{\pa r}e^{2\j} = -\w\,\cJ + \cO (\e)
\end{align}
\label{extKrot}\end{subequations}
with the nonlisted components vanishing. The discontinuities of these components on the null horizon hypersurface where $\e \to 0$ are
\begin{subequations}
\begin{align}
\left[{\bf K}^t_{\ t}\right] &= \sdfrac{1}{2} \left[e^{-\a-\n} \,\frac{\pa}{\pa r} e^{2\n}\right]-  \sdfrac{1}{2}\, \wH e^{2\j} \left[e^{-\a-\n} \,\frac{\pa \w}{\pa r}\right]
= [\k] + \wH [\cJ]\\
\left[{\bf K}^t_{\ \f}\right] &= \sdfrac{1}{2}\, e^{2\j} \left[e^{-\a-\n} \,\frac{\pa \w}{\pa r}\right]= - [\cJ]\\
\left[{\bf K}^\f_{\ t} \right]&= \sdfrac{1}{2} \,\wH\left[e^{-\a-\n} \,\frac{\pa}{\pa r} e^{2\n}\right]- \sdfrac{1}{2}\, \wH^2 e^{2\j} \left[e^{-\a-\n} \,\frac{\pa \w}{\pa r}\right]
=  \wH[\k] + \wH^2\,[\cJ] \\
\left[{\bf K}^\th_{\ \th}\right] &= 0 \\
\left[{\bf K}^\f_{\ \f}\right]&= \sdfrac{1}{2}\,\wH e^{2\j}\left[e^{-\a-\n} \,\frac{\pa \w}{\pa r}\right]= -\,\wH[\cJ]
\end{align}
\label{extKrotdisc}\end{subequations}
where we have used $\e \!=\! e^{2\n} \!\to \!0$ and $\w \!=\!\wH$ on the horizon. If the singular surface is not located on the horizon, all terms in (\ref{extKrot})
would have to be retained. Finally computing the surface stress tensor through the junction conditions (\ref{modjunc}), we find a result that coincides
with (\ref{surfTrot}) computed directly from the Einstein tensor, which verifies the modified prescription for rotating null horizons.

The discussion in the previous Sec.~\ref{Sec:BInonrot} and the relation (\ref{Isrmod}) for the nonrotating case carries over to the rotating case as well, and
we see that the modified junction conditions given here can be obtained from the original Israel junction conditions of \cite{Israel:1966a} by having one index raised
and the other lowered and multiplying the latter by $e^\n$ before taking the horizon limit $e^\n \to 0$. This then corresponds to (\ref{Kdef}) with (\ref{neps}) and
yields the finite result (\ref{modjunc}) for the rotating null horizon stress tensor, upon using (\ref{extKrotdisc}). This simple modification of the junction conditions works perfectly well for lightlike or timelike surfaces, 
and eliminates the need for a special formalism for null surfaces, involving the oblique extrinsic curvature introduced in Ref.~\cite{BarrabesIsrael:1991}.

\vspace{-1mm}
\section{Summary and Conclusions}
\label{Sec:Concl}
\vspace{-1mm}

In this paper we have given the junction conditions and surface stress tensor of a rotating null surface, stationary in time, and in particular
for the horizon of a rotating ``black hole,”  where the region interior to the Kerr horizon generally may differ from the analytic continuation of
the exterior Kerr solution. Relations (\ref{SurfTS}) and (\ref{surfTrot}) are the principal result of this analysis, derived directly from the
singular contributions to the Einstein tensor density (\ref{Eindens}) for a null surface at constant $r$ in the general axisymmetric, stationary
coordinates of (\ref{axisymstat}). These results and surface stress tensor are applicable in particular to the matching of a regular or nonsingular
solution to Einstein's equations in a rotating black hole interior to the Kerr solution external to its horizon.

The approach to the junction conditions and stress tensor of a singular null surface followed in this paper differs from previous work
in focusing on the tensor density $\sqrt{-g}\, G^\m_{\ \n}$ with one covariant and one contravariant index. It is this Einstein tensor density 
and corresponding stress-energy density that appear in expressions (\ref{KomarMass}) and (\ref{KomarAngmom}) for the Komar mass 
and angular momentum, and which are free of coordinate artifacts. By identifying all the terms in this tensor density that are singular 
and can be expressed as total derivatives with respect to $r$ in (\ref{Eindens}) we have identified those terms and only those terms 
that can give rise to a well-defined $\d$-function distribution localized on a singular null surface. The two quantities $[\k]$ and $[\cJ]$ determining 
these surface contributions (\ref{kaprot}) and (\ref{cjdef}) having a coordinate invariant physical meaning, are well-defined and finite, 
even if the horizon is defined in the most familiar singular Schwarzschild or Boyer-Lindquist coordinates of BH exteriors. 

In deriving the junction condition for a rotating null horizon based on the Komar expressions and the tensor density $\sqrt{-g}\, T^\m_{\ \n}$,
we observed that difficulties which the earlier Israel formalism encounters when applied to null surfaces are easily circumvented. 
The modified junction algorithm is the following:
\begin{enumerate}
\vspace{-2mm}
\item The normalization  $\bn \cdot \bn = \e = e^{2\n} \to 0$ of the normal $\bn$ to the surface should be continuously defined and
taken to zero for the null surface only in the final step;
\vspace{-3mm}
\item The extrinsic tensor ${\bf K}^i_{\ j}$ (\ref{extrK}) of the surface is calculated using this $\bn$ but with one contravariant and one
covariant index (\ref{upK}), which is well-defined for $\e \neq 0$ and has a finite limit as $\e\to 0$.
\vspace{-3mm}
\item The surface stress tensor $\cS^i_{\ j}$ is calculated from the discontinuities $[{\bf K}^i_{\ j}]$ at the surface according
to (\ref{modjunc}), just as one would ordinarily in the Israel approach;
\vspace{-3mm}
\item The limit $\e \!\to\! 0$ is taken to obtain the finite stress tensor $\cS^i_{\ j}$ on a null horizon hypersurface.
\end{enumerate}
\vspace{-2mm}

In the case that $\e \neq 0$ and the metric at the surface is nonsingular, this algorithm is equivalent to the original one,
related by (\ref{Isrmod}), or a factor of $e^\n$ in the general rotating case. However, for joining on null hypersurfaces
when $\e\to 0$ and the original Israel junction conditions fail, the simple modification above gives a finite result 
for the surface stress tensor. This is in contrast to the original prescription in terms of ${K}_{ij}$ which vanishes 
as $\e \to 0$ and thus would lead to an indeterminate result for the surface stress tensor. By keeping $\e$ finite and 
raising the index before  taking the $\e \to 0$ limit, as is also motivated by the Komar mass and angular momentum, 
one obtains finite, coordinate invariant results for the surface stress on a singular null hypersurface, which coincide 
with the direct derivation from the Einstein tensor density.

With this modified Israel prescription for the junction conditions and surface stress there is no need to introduce a transversal
$\bN$ vector or oblique extrinsic curvature tensor as in the approach of Ref.~\cite{BarrabesIsrael:1991}, and indeed that approach
should not be applied uncritically to the cases in which the normal $\bn$, transversal $\bN$ or integration measure $\sqrt{-g}$ in the density are
themselves singular or discontinuous, where independence of the results from the choice of $\bN$ cannot be assumed.
These are the cases of most direct physical interest for BH horizons and interiors, where Schwarzschild or Boyer-Lindquist 
coordinates are most convenient.  Misapplication of the BI approach to these singular cases may lead to incorrect results, 
as the example of Appendix \ref{Sec:BIrot} shows, while the direct method of derivation of the singular null surface stress
tensor from the Einstein tensor density in Sec.~\ref{Sec:Rotate} does not suffer from this drawback.
The application of the junction conditions for a rotating gravastar in the slow rotation approximation of Hartle and Thorne 
is given in a second paper accompanying this one.

\vspace{3mm}
\centerline{\bf Acknowledgement}
\vspace{1mm}

E. M. acknowledges support from LANL LDRD grant No. 20200661ER, and several useful conversations with P. O. Mazur, which brought to the authors' 
attention several of references cited on axisymmetric geometries. P.G. acknowledges partial support from NSF grant No. PHY-2014075, and is very grateful to Prof.\ Masahide Yamaguchi for his generous support under JSPS Grant-in-Aid for Scientific Research Number JP18K18764 at the Tokyo Institute of Technology.

\vspace{-5mm}
\bibliographystyle{apsrev4-1}
\bibliography{posadaeqs}

\begin{thebibliography}{49}%
\makeatletter
\providecommand \@ifxundefined [1]{%
 \@ifx{#1\undefined}
}%
\providecommand \@ifnum [1]{%
 \ifnum #1\expandafter \@firstoftwo
 \else \expandafter \@secondoftwo
 \fi
}%
\providecommand \@ifx [1]{%
 \ifx #1\expandafter \@firstoftwo
 \else \expandafter \@secondoftwo
 \fi
}%
\providecommand \natexlab [1]{#1}%
\providecommand \enquote  [1]{``#1''}%
\providecommand \bibnamefont  [1]{#1}%
\providecommand \bibfnamefont [1]{#1}%
\providecommand \citenamefont [1]{#1}%
\providecommand \href@noop [0]{\@secondoftwo}%
\providecommand \href [0]{\begingroup \@sanitize@url \@href}%
\providecommand \@href[1]{\@@startlink{#1}\@@href}%
\providecommand \@@href[1]{\endgroup#1\@@endlink}%
\providecommand \@sanitize@url [0]{\catcode `\\12\catcode `\$12\catcode
  `\&12\catcode `\#12\catcode `\^12\catcode `\_12\catcode `\%12\relax}%
\providecommand \@@startlink[1]{}%
\providecommand \@@endlink[0]{}%
\providecommand \url  [0]{\begingroup\@sanitize@url \@url }%
\providecommand \@url [1]{\endgroup\@href {#1}{\urlprefix }}%
\providecommand \urlprefix  [0]{URL }%
\providecommand \Eprint [0]{\href }%
\providecommand \doibase [0]{http://dx.doi.org/}%
\providecommand \selectlanguage [0]{\@gobble}%
\providecommand \bibinfo  [0]{\@secondoftwo}%
\providecommand \bibfield  [0]{\@secondoftwo}%
\providecommand \translation [1]{[#1]}%
\providecommand \BibitemOpen [0]{}%
\providecommand \bibitemStop [0]{}%
\providecommand \bibitemNoStop [0]{.\EOS\space}%
\providecommand \EOS [0]{\spacefactor3000\relax}%
\providecommand \BibitemShut  [1]{\csname bibitem#1\endcsname}%
\let\auto@bib@innerbib\@empty
%</preamble>
\bibitem [{\citenamefont {Lanczos}(1922)}]{Lanczos:1922}%
  \BibitemOpen
  \bibfield  {author} {\bibinfo {author} {\bibfnamefont {K.}~\bibnamefont
  {Lanczos}},\ }\href@noop {} {\bibfield  {journal} {\bibinfo  {journal}
  {Physik Z.}\ }\textbf {\bibinfo {volume} {23}},\ \bibinfo {pages} {537}
  (\bibinfo {year} {1922})}\BibitemShut {NoStop}%
\bibitem [{\citenamefont {Sen}(1924)}]{Sen:1924}%
  \BibitemOpen
  \bibfield  {author} {\bibinfo {author} {\bibfnamefont {N.}~\bibnamefont
  {Sen}},\ }\href {https://doi.org/10.1002/andp.19243780505} {\bibfield
  {journal} {\bibinfo  {journal} {Ann. Physik}\ }\textbf {\bibinfo {volume}
  {378}},\ \bibinfo {pages} {365} (\bibinfo {year} {1924})}\BibitemShut
  {NoStop}%
\bibitem [{\citenamefont {Lanczos}(1924)}]{Lanczos:1924}%
  \BibitemOpen
  \bibfield  {author} {\bibinfo {author} {\bibfnamefont {K.}~\bibnamefont
  {Lanczos}},\ }\href {\doibase https://doi.org/10.1002/andp.19243791403}
  {\bibfield  {journal} {\bibinfo  {journal} {Ann. Physik}\ }\textbf {\bibinfo
  {volume} {379}},\ \bibinfo {pages} {518} (\bibinfo {year}
  {1924})}\BibitemShut {NoStop}%
\bibitem [{\citenamefont {O'Brien}\ and\ \citenamefont
  {Synge}(1952)}]{OBrSyn:1952}%
  \BibitemOpen
  \bibfield  {author} {\bibinfo {author} {\bibfnamefont {S.}~\bibnamefont
  {O'Brien}}\ and\ \bibinfo {author} {\bibfnamefont {J.}~\bibnamefont
  {Synge}},\ }\href@noop {} {\bibfield  {journal} {\bibinfo  {journal} {Dublin
  Inst. Adv. Stud. Ser. A}\ }\textbf {\bibinfo {volume} {9}} (\bibinfo {year}
  {1952})}\BibitemShut {NoStop}%
\bibitem [{\citenamefont {Lichnerowicz}(1955)}]{Lichnerowicz:1955}%
  \BibitemOpen
  \bibfield  {author} {\bibinfo {author} {\bibfnamefont {A.}~\bibnamefont
  {Lichnerowicz}},\ }\href@noop {} {\emph {\bibinfo {title} {Th{\'e}ories
  relativistes de la gravitation et de l'{\'e}lectromagn{\'e}tisme:
  relativit{\'e} g{\'e}n{\'e}rale et th{\'e}ories unitaires}}}\ (\bibinfo
  {publisher} {Masson},\ \bibinfo {year} {1955})\BibitemShut {NoStop}%
\bibitem [{\citenamefont {Synge}(1957)}]{Synge:1957}%
  \BibitemOpen
  \bibfield  {author} {\bibinfo {author} {\bibfnamefont {J.~L.}\ \bibnamefont
  {Synge}},\ }\href@noop {} {\bibfield  {journal} {\bibinfo  {journal} {Proc.
  R. Irish Acad. Section A: Math. Phys. Sci.}\ }\textbf {\bibinfo {volume}
  {59}},\ \bibinfo {pages} {1} (\bibinfo {year} {1957})}\BibitemShut {NoStop}%
\bibitem [{\citenamefont {Synge}(1960)}]{Synge:1960}%
  \BibitemOpen
  \bibfield  {author} {\bibinfo {author} {\bibfnamefont {J.~L.}\ \bibnamefont
  {Synge}},\ }\href@noop {} {\emph {\bibinfo {title} {Relativity: The General
  Theory}}}\ (\bibinfo  {publisher} {Interscience Publishers},\ \bibinfo {year}
  {1960})\BibitemShut {NoStop}%
\bibitem [{\citenamefont {Dautcourt}(1964)}]{Dautcourt:1964}%
  \BibitemOpen
  \bibfield  {author} {\bibinfo {author} {\bibfnamefont {G.}~\bibnamefont
  {Dautcourt}},\ }\href {\doibase https://doi.org/10.1002/mana.19640270504}
  {\bibfield  {journal} {\bibinfo  {journal} {Math. Nachr.}\ }\textbf {\bibinfo
  {volume} {27}},\ \bibinfo {pages} {277} (\bibinfo {year} {1964})}\BibitemShut
  {NoStop}%
\bibitem [{\citenamefont {Israel}(1966)}]{Israel:1966a}%
  \BibitemOpen
  \bibfield  {author} {\bibinfo {author} {\bibfnamefont {W.}~\bibnamefont
  {Israel}},\ }\href {\doibase 10.1007/BF02710419} {\bibfield  {journal}
  {\bibinfo  {journal} {N. Cimento B}\ }\textbf {\bibinfo {volume} {44}},\
  \bibinfo {pages} {1} (\bibinfo {year} {1966})}\BibitemShut {NoStop}%
\bibitem [{\citenamefont {Taub}(1980)}]{Taub:1980}%
  \BibitemOpen
  \bibfield  {author} {\bibinfo {author} {\bibfnamefont {A.~H.}\ \bibnamefont
  {Taub}},\ }\href {\doibase 10.1063/1.524568} {\bibfield  {journal} {\bibinfo
  {journal} {Jour. Math. Phys.}\ }\textbf {\bibinfo {volume} {21}},\ \bibinfo
  {pages} {1423} (\bibinfo {year} {1980})}\BibitemShut {NoStop}%
\bibitem [{\citenamefont {Clarke}\ and\ \citenamefont
  {Dray}(1987)}]{Clarke:1987}%
  \BibitemOpen
  \bibfield  {author} {\bibinfo {author} {\bibfnamefont {C.~J.~S.}\
  \bibnamefont {Clarke}}\ and\ \bibinfo {author} {\bibfnamefont
  {T.}~\bibnamefont {Dray}},\ }\href {\doibase 10.1088/0264-9381/4/2/010}
  {\bibfield  {journal} {\bibinfo  {journal} {Class. Quant. Grav.}\ }\textbf
  {\bibinfo {volume} {4}},\ \bibinfo {pages} {265} (\bibinfo {year}
  {1987})}\BibitemShut {NoStop}%
\bibitem [{\citenamefont {Barrab{\`e}s}(1989)}]{Barrabes:1989}%
  \BibitemOpen
  \bibfield  {author} {\bibinfo {author} {\bibfnamefont {C.}~\bibnamefont
  {Barrab{\`e}s}},\ }\href {\doibase 10.1088/0264-9381/6/5/003} {\bibfield
  {journal} {\bibinfo  {journal} {Class. Quant. Grav.}\ }\textbf {\bibinfo
  {volume} {6}},\ \bibinfo {pages} {581} (\bibinfo {year} {1989})}\BibitemShut
  {NoStop}%
\bibitem [{\citenamefont {Barrab\`es}\ and\ \citenamefont
  {Israel}(1991)}]{BarrabesIsrael:1991}%
  \BibitemOpen
  \bibfield  {author} {\bibinfo {author} {\bibfnamefont {C.}~\bibnamefont
  {Barrab\`es}}\ and\ \bibinfo {author} {\bibfnamefont {W.}~\bibnamefont
  {Israel}},\ }\href {\doibase 10.1103/PhysRevD.43.1129} {\bibfield  {journal}
  {\bibinfo  {journal} {Phys. Rev. D}\ }\textbf {\bibinfo {volume} {43}},\
  \bibinfo {pages} {1129} (\bibinfo {year} {1991})}\BibitemShut {NoStop}%
\bibitem [{\citenamefont {Poisson}(2002)}]{Poisson:2002}%
  \BibitemOpen
  \bibfield  {author} {\bibinfo {author} {\bibfnamefont {E.}~\bibnamefont
  {Poisson}},\ }\href@noop {} {\emph {\bibinfo {title} {An Advanced Course in
  General Relativity}}}\ (\bibinfo  {publisher} {Cambridge Univ. Press},\
  \bibinfo {year} {2002})\BibitemShut {NoStop}%
\bibitem [{\citenamefont {Berezin}\ \emph {et~al.}(1983)\citenamefont
  {Berezin}, \citenamefont {Kuzmin},\ and\ \citenamefont
  {Tkachev}}]{BerKuzTka:1983}%
  \BibitemOpen
  \bibfield  {author} {\bibinfo {author} {\bibfnamefont {V.}~\bibnamefont
  {Berezin}}, \bibinfo {author} {\bibfnamefont {V.}~\bibnamefont {Kuzmin}}, \
  and\ \bibinfo {author} {\bibfnamefont {I.}~\bibnamefont {Tkachev}},\ }\href
  {\doibase https://doi.org/10.1016/0370-2693(83)90630-5} {\bibfield  {journal}
  {\bibinfo  {journal} {Phys. Lett. B}\ }\textbf {\bibinfo {volume} {120}},\
  \bibinfo {pages} {91 } (\bibinfo {year} {1983})}\BibitemShut {NoStop}%
\bibitem [{\citenamefont {Berezin}\ \emph {et~al.}(1987)\citenamefont
  {Berezin}, \citenamefont {Kuzmin},\ and\ \citenamefont
  {Tkachev}}]{BerKuzTka:1987}%
  \BibitemOpen
  \bibfield  {author} {\bibinfo {author} {\bibfnamefont {V.~A.}\ \bibnamefont
  {Berezin}}, \bibinfo {author} {\bibfnamefont {V.~A.}\ \bibnamefont {Kuzmin}},
  \ and\ \bibinfo {author} {\bibfnamefont {I.~I.}\ \bibnamefont {Tkachev}},\
  }\href {\doibase 10.1103/PhysRevD.36.2919} {\bibfield  {journal} {\bibinfo
  {journal} {Phys. Rev. D}\ }\textbf {\bibinfo {volume} {36}},\ \bibinfo
  {pages} {2919} (\bibinfo {year} {1987})}\BibitemShut {NoStop}%
\bibitem [{\citenamefont {Blau}\ \emph {et~al.}(1987)\citenamefont {Blau},
  \citenamefont {Guendelman},\ and\ \citenamefont {Guth}}]{BlauGueGuth:1987}%
  \BibitemOpen
  \bibfield  {author} {\bibinfo {author} {\bibfnamefont {S.~K.}\ \bibnamefont
  {Blau}}, \bibinfo {author} {\bibfnamefont {E.~I.}\ \bibnamefont
  {Guendelman}}, \ and\ \bibinfo {author} {\bibfnamefont {A.~H.}\ \bibnamefont
  {Guth}},\ }\href {\doibase 10.1103/PhysRevD.35.1747} {\bibfield  {journal}
  {\bibinfo  {journal} {Phys. Rev. D}\ }\textbf {\bibinfo {volume} {35}},\
  \bibinfo {pages} {1747} (\bibinfo {year} {1987})}\BibitemShut {NoStop}%
\bibitem [{\citenamefont {Aurilia}\ \emph {et~al.}(1987)\citenamefont
  {Aurilia}, \citenamefont {Kissack}, \citenamefont {Mann},\ and\ \citenamefont
  {Spallucci}}]{AurKisManSpa:1987}%
  \BibitemOpen
  \bibfield  {author} {\bibinfo {author} {\bibfnamefont {A.}~\bibnamefont
  {Aurilia}}, \bibinfo {author} {\bibfnamefont {R.~S.}\ \bibnamefont
  {Kissack}}, \bibinfo {author} {\bibfnamefont {R.}~\bibnamefont {Mann}}, \
  and\ \bibinfo {author} {\bibfnamefont {E.}~\bibnamefont {Spallucci}},\ }\href
  {\doibase 10.1103/PhysRevD.35.2961} {\bibfield  {journal} {\bibinfo
  {journal} {Phys. Rev. D}\ }\textbf {\bibinfo {volume} {35}},\ \bibinfo
  {pages} {2961} (\bibinfo {year} {1987})}\BibitemShut {NoStop}%
\bibitem [{\citenamefont {Misner}\ \emph {et~al.}(1973)\citenamefont {Misner},
  \citenamefont {Thorne},\ and\ \citenamefont {Wheeler}}]{MTW}%
  \BibitemOpen
  \bibfield  {author} {\bibinfo {author} {\bibfnamefont {C.}~\bibnamefont
  {Misner}}, \bibinfo {author} {\bibfnamefont {K.}~\bibnamefont {Thorne}}, \
  and\ \bibinfo {author} {\bibfnamefont {J.}~\bibnamefont {Wheeler}},\
  }\href@noop {} {\emph {\bibinfo {title} {Gravitation}}}\ (\bibinfo
  {publisher} {W. H. Freeman},\ \bibinfo {year} {1973})\BibitemShut {NoStop}%
\bibitem [{\citenamefont {Hawking}\ and\ \citenamefont
  {Ellis}(1973)}]{HawkingEllis:1973}%
  \BibitemOpen
  \bibfield  {author} {\bibinfo {author} {\bibfnamefont {S.~W.}\ \bibnamefont
  {Hawking}}\ and\ \bibinfo {author} {\bibfnamefont {G.~F.~R.}\ \bibnamefont
  {Ellis}},\ }\href@noop {} {\emph {\bibinfo {title} {{The Large Scale
  Structure of Space-Time}}}}\ (\bibinfo  {publisher} {Cambridge Univ. Press},\
  \bibinfo {year} {1973})\BibitemShut {NoStop}%
\bibitem [{\citenamefont {Hawking}(1976)}]{Hawking:1976}%
  \BibitemOpen
  \bibfield  {author} {\bibinfo {author} {\bibfnamefont {S.~W.}\ \bibnamefont
  {Hawking}},\ }\href {\doibase 10.1103/PhysRevD.14.2460} {\bibfield  {journal}
  {\bibinfo  {journal} {Phys. Rev. D}\ }\textbf {\bibinfo {volume} {14}},\
  \bibinfo {pages} {2460} (\bibinfo {year} {1976})}\BibitemShut {NoStop}%
\bibitem [{\citenamefont {{'t Hooft}}(1995)}]{tHooft:1995}%
  \BibitemOpen
  \bibfield  {author} {\bibinfo {author} {\bibfnamefont {G.}~\bibnamefont {{'t
  Hooft}}},\ }\href {\doibase https://doi.org/10.1016/0920-5632(95)00444-E}
  {\bibfield  {journal} {\bibinfo  {journal} {Nucl. Phys. B - Proc. Suppl.}\
  }\textbf {\bibinfo {volume} {43}},\ \bibinfo {pages} {1 } (\bibinfo {year}
  {1995})}\BibitemShut {NoStop}%
\bibitem [{\citenamefont {Unruh}\ and\ \citenamefont
  {Wald}(2017)}]{UnruhWald:2017}%
  \BibitemOpen
  \bibfield  {author} {\bibinfo {author} {\bibfnamefont {W.~G.}\ \bibnamefont
  {Unruh}}\ and\ \bibinfo {author} {\bibfnamefont {R.~M.}\ \bibnamefont
  {Wald}},\ }\href {\doibase 10.1088/1361-6633/aa778e} {\bibfield  {journal}
  {\bibinfo  {journal} {Rep. Prog. Phys.}\ }\textbf {\bibinfo {volume} {80}},\
  \bibinfo {pages} {092002} (\bibinfo {year} {2017})}\BibitemShut {NoStop}%
\bibitem [{\citenamefont {Mathur}(2005)}]{Mathur:2015}%
  \BibitemOpen
  \bibfield  {author} {\bibinfo {author} {\bibfnamefont {S.}~\bibnamefont
  {Mathur}},\ }\href {\doibase https://doi.org/10.1002/prop.200410203}
  {\bibfield  {journal} {\bibinfo  {journal} {Fortsch. Phys.}\ }\textbf
  {\bibinfo {volume} {53}},\ \bibinfo {pages} {793} (\bibinfo {year}
  {2005})}\BibitemShut {NoStop}%
\bibitem [{\citenamefont {Mazur}\ and\ \citenamefont
  {Mottola}(2015)}]{MazurMottola:2015}%
  \BibitemOpen
  \bibfield  {author} {\bibinfo {author} {\bibfnamefont {P.~O.}\ \bibnamefont
  {Mazur}}\ and\ \bibinfo {author} {\bibfnamefont {E.}~\bibnamefont
  {Mottola}},\ }\href {http://stacks.iop.org/0264-9381/32/i=21/a=215024}
  {\bibfield  {journal} {\bibinfo  {journal} {Class. Quant. Grav.}\ }\textbf
  {\bibinfo {volume} {32}},\ \bibinfo {pages} {215024} (\bibinfo {year}
  {2015})}\BibitemShut {NoStop}%
\bibitem [{\citenamefont {Buchdahl}(1959)}]{Buchdahl:1959}%
  \BibitemOpen
  \bibfield  {author} {\bibinfo {author} {\bibfnamefont {H.~A.}\ \bibnamefont
  {Buchdahl}},\ }\href {\doibase 10.1103/PhysRev.116.1027} {\bibfield
  {journal} {\bibinfo  {journal} {Phys. Rev.}\ }\textbf {\bibinfo {volume}
  {116}},\ \bibinfo {pages} {1027} (\bibinfo {year} {1959})}\BibitemShut
  {NoStop}%
\bibitem [{\citenamefont {Mazur}\ and\ \citenamefont
  {Mottola}(2001)}]{MazurMottola:2001}%
  \BibitemOpen
  \bibfield  {author} {\bibinfo {author} {\bibfnamefont {P.~O.}\ \bibnamefont
  {Mazur}}\ and\ \bibinfo {author} {\bibfnamefont {E.}~\bibnamefont
  {Mottola}},\ }\href@noop {} {\enquote {\bibinfo {title} {{Gravitational
  condensate stars: An alternative to black holes}},}\ } (\bibinfo {year}
  {2001}),\ \Eprint {http://arxiv.org/abs/gr-qc/0109035} {arXiv:gr-qc/0109035}
  \BibitemShut {NoStop}%
\bibitem [{\citenamefont {Mazur}\ and\ \citenamefont
  {Mottola}(2004)}]{MazurMottola:2004}%
  \BibitemOpen
  \bibfield  {author} {\bibinfo {author} {\bibfnamefont {P.~O.}\ \bibnamefont
  {Mazur}}\ and\ \bibinfo {author} {\bibfnamefont {E.}~\bibnamefont
  {Mottola}},\ }\href {\doibase 10.1073/pnas.0402717101} {\bibfield  {journal}
  {\bibinfo  {journal} {Proc. Nat. Acad. Sci.}\ }\textbf {\bibinfo {volume}
  {101}},\ \bibinfo {pages} {9545} (\bibinfo {year} {2004})}\BibitemShut
  {NoStop}%
\bibitem [{\citenamefont {Chapline}\ \emph {et~al.}(2001)\citenamefont
  {Chapline}, \citenamefont {Hohlfeld}, \citenamefont {Laughlin},\ and\
  \citenamefont {Santiago}}]{ChapHohlLaughSant:2001}%
  \BibitemOpen
  \bibfield  {author} {\bibinfo {author} {\bibfnamefont {G.}~\bibnamefont
  {Chapline}}, \bibinfo {author} {\bibfnamefont {E.}~\bibnamefont {Hohlfeld}},
  \bibinfo {author} {\bibfnamefont {R.~B.}\ \bibnamefont {Laughlin}}, \ and\
  \bibinfo {author} {\bibfnamefont {D.~I.}\ \bibnamefont {Santiago}},\ }\href
  {\doibase 10.1080/13642810108221981} {\bibfield  {journal} {\bibinfo
  {journal} {Phil. Mag. B}\ }\textbf {\bibinfo {volume} {81}},\ \bibinfo
  {pages} {235} (\bibinfo {year} {2001})}\BibitemShut {NoStop}%
\bibitem [{\citenamefont {Mottola}(2010)}]{Mottola:2010}%
  \BibitemOpen
  \bibfield  {author} {\bibinfo {author} {\bibfnamefont {E.}~\bibnamefont
  {Mottola}},\ }\href {https://www.actaphys.uj.edu.pl/R/41/9/2031/pdf}
  {\bibfield  {journal} {\bibinfo  {journal} {Acta Phys. Polon. B}\ }\textbf
  {\bibinfo {volume} {41}},\ \bibinfo {pages} {2031} (\bibinfo {year}
  {2010})},\ \Eprint {http://arxiv.org/abs/1008.5006} {arXiv:1008.5006 [gr-qc]}
  \BibitemShut {NoStop}%
\bibitem [{\citenamefont {Abbott~{\it et al.}}(2019)}]{LIGO:2019}%
  \BibitemOpen
  \bibfield  {author} {\bibinfo {author} {\bibfnamefont {B.~P.}\ \bibnamefont
  {Abbott~{\it et al.}}} (\bibinfo {collaboration} {LIGO and Virgo Scientific
  Collaborations}),\ }\href
  {https://link.aps.org/doi/10.1103/PhysRevD.100.104036} {\bibfield  {journal}
  {\bibinfo  {journal} {Phys. Rev. D}\ }\textbf {\bibinfo {volume} {100}},\
  \bibinfo {pages} {104036} (\bibinfo {year} {2019})}\BibitemShut {NoStop}%
\bibitem [{\citenamefont {Cardoso}\ and\ \citenamefont
  {Pani}(2019)}]{CardosoPani:2019}%
  \BibitemOpen
  \bibfield  {author} {\bibinfo {author} {\bibfnamefont {V.}~\bibnamefont
  {Cardoso}}\ and\ \bibinfo {author} {\bibfnamefont {P.}~\bibnamefont {Pani}},\
  }\href {\doibase 10.1007/s41114-019-0020-4} {\bibfield  {journal} {\bibinfo
  {journal} {Living Rev. Relativity}\ }\textbf {\bibinfo {volume} {22}},\
  \bibinfo {pages} {4} (\bibinfo {year} {2019})}\BibitemShut {NoStop}%
\bibitem [{\citenamefont {Barrabes}\ and\ \citenamefont
  {Hogan}(2003)}]{BarrabesHogan:2003}%
  \BibitemOpen
  \bibfield  {author} {\bibinfo {author} {\bibfnamefont {C.}~\bibnamefont
  {Barrabes}}\ and\ \bibinfo {author} {\bibfnamefont {P.}~\bibnamefont
  {Hogan}},\ }\href@noop {} {\emph {\bibinfo {title} {Singular Null
  Hypersurfaces in General Relativity: Light-like Signals from Violent
  Astrophysical Events}}}\ (\bibinfo  {publisher} {World Scientific},\ \bibinfo
  {year} {2003})\BibitemShut {NoStop}%
\bibitem [{\citenamefont {Carter}(1970)}]{Carter:1970}%
  \BibitemOpen
  \bibfield  {author} {\bibinfo {author} {\bibfnamefont {B.}~\bibnamefont
  {Carter}},\ }\href {\doibase 10.1007/BF01647092} {\bibfield  {journal}
  {\bibinfo  {journal} {Comm. Math. Phys.}\ }\textbf {\bibinfo {volume} {17}},\
  \bibinfo {pages} {233} (\bibinfo {year} {1970})}\BibitemShut {NoStop}%
\bibitem [{\citenamefont {Stephani}\ \emph {et~al.}(2003)\citenamefont
  {Stephani}, \citenamefont {Kramer}, \citenamefont {MacCallum}, \citenamefont
  {Hoenselaers},\ and\ \citenamefont {Herlt}}]{Stephani:2003tm}%
  \BibitemOpen
  \bibfield  {author} {\bibinfo {author} {\bibfnamefont {H.}~\bibnamefont
  {Stephani}}, \bibinfo {author} {\bibfnamefont {D.}~\bibnamefont {Kramer}},
  \bibinfo {author} {\bibfnamefont {M.~A.~H.}\ \bibnamefont {MacCallum}},
  \bibinfo {author} {\bibfnamefont {C.}~\bibnamefont {Hoenselaers}}, \ and\
  \bibinfo {author} {\bibfnamefont {E.}~\bibnamefont {Herlt}},\ }\href@noop {}
  {\emph {\bibinfo {title} {{Exact solutions of Einstein's field
  equations}}}},\ Cambridge Monographs on Mathematical Physics\ (\bibinfo
  {publisher} {Cambridge Univ. Press},\ \bibinfo {year} {2003})\BibitemShut
  {NoStop}%
\bibitem [{\citenamefont {Gourgoulhon}(2011)}]{Gourgoulhon:2010}%
  \BibitemOpen
  \bibfield  {author} {\bibinfo {author} {\bibfnamefont {E.}~\bibnamefont
  {Gourgoulhon}},\ }\href@noop {} {\enquote {\bibinfo {title} {An introduction
  to the theory of rotating relativistic stars},}\ } (\bibinfo {year} {2011}),\
  \Eprint {http://arxiv.org/abs/1003.5015} {arXiv:1003.5015} \BibitemShut
  {NoStop}%
\bibitem [{\citenamefont {Lewis}\ and\ \citenamefont
  {Schott}(1932)}]{Lewis:1932}%
  \BibitemOpen
  \bibfield  {author} {\bibinfo {author} {\bibfnamefont {T.}~\bibnamefont
  {Lewis}}\ and\ \bibinfo {author} {\bibfnamefont {G.~A.}\ \bibnamefont
  {Schott}},\ }\href {\doibase 10.1098/rspa.1932.0073} {\bibfield  {journal}
  {\bibinfo  {journal} {Proc. R. Soc. Lond. A}\ }\textbf {\bibinfo {volume}
  {136}},\ \bibinfo {pages} {176} (\bibinfo {year} {1932})}\BibitemShut
  {NoStop}%
\bibitem [{\citenamefont {Papapetrou}(1966)}]{Papapetrou:1966}%
  \BibitemOpen
  \bibfield  {author} {\bibinfo {author} {\bibfnamefont {A.}~\bibnamefont
  {Papapetrou}},\ }in\ \href@noop {} {\emph {\bibinfo {booktitle} {Annales de
  l'IHP Physique th{\'e}orique}}},\ Vol.~\bibinfo {volume} {4}\ (\bibinfo
  {year} {1966})\ pp.\ \bibinfo {pages} {83--105}\BibitemShut {NoStop}%
\bibitem [{\citenamefont {{Hartle}}\ and\ \citenamefont
  {{Sharp}}(1967)}]{HartleSharp:1967}%
  \BibitemOpen
  \bibfield  {author} {\bibinfo {author} {\bibfnamefont {J.~B.}\ \bibnamefont
  {{Hartle}}}\ and\ \bibinfo {author} {\bibfnamefont {D.~H.}\ \bibnamefont
  {{Sharp}}},\ }\href {\doibase 10.1086/149002} {\bibfield  {journal} {\bibinfo
   {journal} {\apj}\ }\textbf {\bibinfo {volume} {147}},\ \bibinfo {pages}
  {317} (\bibinfo {year} {1967})}\BibitemShut {NoStop}%
\bibitem [{\citenamefont {{Bardeen}}(1970)}]{Bardeen:1970}%
  \BibitemOpen
  \bibfield  {author} {\bibinfo {author} {\bibfnamefont {J.~M.}\ \bibnamefont
  {{Bardeen}}},\ }\href {\doibase 10.1086/150635} {\bibfield  {journal}
  {\bibinfo  {journal} {Astrophys. Jour.}\ }\textbf {\bibinfo {volume} {162}},\
  \bibinfo {pages} {71} (\bibinfo {year} {1970})}\BibitemShut {NoStop}%
\bibitem [{\citenamefont {{Chandrasekhar}}\ and\ \citenamefont
  {{Friedman}}(1972)}]{FriedmanChandra:1972}%
  \BibitemOpen
  \bibfield  {author} {\bibinfo {author} {\bibfnamefont {S.}~\bibnamefont
  {{Chandrasekhar}}}\ and\ \bibinfo {author} {\bibfnamefont {J.~L.}\
  \bibnamefont {{Friedman}}},\ }\href {\doibase 10.1086/151566} {\bibfield
  {journal} {\bibinfo  {journal} {Astrophys. Jour.}\ }\textbf {\bibinfo
  {volume} {175}},\ \bibinfo {pages} {379} (\bibinfo {year}
  {1972})}\BibitemShut {NoStop}%
\bibitem [{\citenamefont {Chandrasekhar}(1983)}]{Chandrasekhar:1984}%
  \BibitemOpen
  \bibfield  {author} {\bibinfo {author} {\bibfnamefont {S.}~\bibnamefont
  {Chandrasekhar}},\ }\href@noop {} {\emph {\bibinfo {title} {The Mathematical
  Theory of Black Holes}}}\ (\bibinfo  {publisher} {Oxford Univ. Press},\
  \bibinfo {year} {1983})\BibitemShut {NoStop}%
\bibitem [{\citenamefont {Thorne}\ \emph {et~al.}(1986)\citenamefont {Thorne},
  \citenamefont {Thorne}, \citenamefont {Price},\ and\ \citenamefont
  {MacDonald}}]{ThorneMembrane}%
  \BibitemOpen
  \bibfield  {author} {\bibinfo {author} {\bibfnamefont {K.}~\bibnamefont
  {Thorne}}, \bibinfo {author} {\bibfnamefont {K.}~\bibnamefont {Thorne}},
  \bibinfo {author} {\bibfnamefont {R.}~\bibnamefont {Price}}, \ and\ \bibinfo
  {author} {\bibfnamefont {D.}~\bibnamefont {MacDonald}},\ }\href@noop {}
  {\emph {\bibinfo {title} {Black Holes: The Membrane Paradigm}}}\ (\bibinfo
  {publisher} {Yale Univ. Press},\ \bibinfo {year} {1986})\BibitemShut
  {NoStop}%
\bibitem [{\citenamefont {Misner}\ and\ \citenamefont
  {Sharp}(1964)}]{MisnerSharp:1964}%
  \BibitemOpen
  \bibfield  {author} {\bibinfo {author} {\bibfnamefont {C.~W.}\ \bibnamefont
  {Misner}}\ and\ \bibinfo {author} {\bibfnamefont {D.~H.}\ \bibnamefont
  {Sharp}},\ }\href {\doibase 10.1103/PhysRev.136.B571} {\bibfield  {journal}
  {\bibinfo  {journal} {Phys. Rev.}\ }\textbf {\bibinfo {volume} {136}},\
  \bibinfo {pages} {B571} (\bibinfo {year} {1964})}\BibitemShut {NoStop}%
\bibitem [{\citenamefont {Bardeen}\ \emph {et~al.}(1973)\citenamefont
  {Bardeen}, \citenamefont {Carter},\ and\ \citenamefont
  {Hawking}}]{BarCarHawk:1973}%
  \BibitemOpen
  \bibfield  {author} {\bibinfo {author} {\bibfnamefont {J.~M.}\ \bibnamefont
  {Bardeen}}, \bibinfo {author} {\bibfnamefont {B.}~\bibnamefont {Carter}}, \
  and\ \bibinfo {author} {\bibfnamefont {S.~W.}\ \bibnamefont {Hawking}},\
  }\href {\doibase 10.1007/BF01645742} {\bibfield  {journal} {\bibinfo
  {journal} {Comm. Math. Phys.}\ }\textbf {\bibinfo {volume} {31}},\ \bibinfo
  {pages} {161} (\bibinfo {year} {1973})}\BibitemShut {NoStop}%
\bibitem [{\citenamefont {Komar}(1959)}]{Komar:1959}%
  \BibitemOpen
  \bibfield  {author} {\bibinfo {author} {\bibfnamefont {A.}~\bibnamefont
  {Komar}},\ }\href {\doibase 10.1103/PhysRev.113.934} {\bibfield  {journal}
  {\bibinfo  {journal} {Phys. Rev.}\ }\textbf {\bibinfo {volume} {113}},\
  \bibinfo {pages} {934} (\bibinfo {year} {1959})}\BibitemShut {NoStop}%
\bibitem [{\citenamefont {Wald}(2010)}]{WaldBook}%
  \BibitemOpen
  \bibfield  {author} {\bibinfo {author} {\bibfnamefont {R.~M.}\ \bibnamefont
  {Wald}},\ }\href@noop {} {\emph {\bibinfo {title} {General Relativity}}}\
  (\bibinfo  {publisher} {Univ. of Chicago Press},\ \bibinfo {year}
  {2010})\BibitemShut {NoStop}%
\bibitem [{\citenamefont {Poisson}(2004)}]{Poisson:2004}%
  \BibitemOpen
  \bibfield  {author} {\bibinfo {author} {\bibfnamefont {E.}~\bibnamefont
  {Poisson}},\ }\href@noop {} {\emph {\bibinfo {title} {A relativist's toolkit:
  the mathematics of black-hole mechanics}}}\ (\bibinfo  {publisher} {Cambridge
  Univ. Press},\ \bibinfo {year} {2004})\BibitemShut {NoStop}%
\bibitem [{\citenamefont {Guendelman}\ \emph {et~al.}(2010)\citenamefont
  {Guendelman}, \citenamefont {Kaganovich}, \citenamefont {Nissimov},\ and\
  \citenamefont {Pacheva}}]{GuenKagNis:2010}%
  \BibitemOpen
  \bibfield  {author} {\bibinfo {author} {\bibfnamefont {E.}~\bibnamefont
  {Guendelman}}, \bibinfo {author} {\bibfnamefont {A.}~\bibnamefont
  {Kaganovich}}, \bibinfo {author} {\bibfnamefont {E.}~\bibnamefont
  {Nissimov}}, \ and\ \bibinfo {author} {\bibfnamefont {S.}~\bibnamefont
  {Pacheva}},\ }\href {\doibase 10.1142/S0217751X10047762} {\bibfield
  {journal} {\bibinfo  {journal} {Int. Jour. Mod Phys. A}\ }\textbf {\bibinfo
  {volume} {25}},\ \bibinfo {pages} {1405} (\bibinfo {year}
  {2010})}\BibitemShut {NoStop}%
\end{thebibliography}%

\appendix

\vspace{-2mm}
\section{Tetrads, Surfaces, Integration and Stokes' Theorem}
\label{Sec:Tetrad}
\vspace{-1mm}

The metric and curvature conventions used in this paper are those of Misner, Thorne and Wheeler \cite{MTW}.
Greek indices are four-dimensional and $x^\m$ ranges over $(t,r,\th,\f)$ in the metric of (\ref{axisymstat}).
Orthonormal tangent space indices are indicated by $a,b,c,\dots$, while coordinates of a three- or two-dimensional hypersurface
are labeled $\x^i, \x^j$ {\it etc.} In the tetrad or vierbein formalism the line element is written 
\begin{align}
ds^2 = g_{\m\n} dx^\m dx^\n = w^a \h_{ab} w^b
\label{genmet}
\end{align}
with $\h_{ab} =$ diag $(-1,1,1,1)$ the flat spacetime Minkowski metric in the tangent space and $w^a$ are the one-forms 
\vspace{-6mm}
\begin{align}
w^a = e^a_{\ \m}\, dx^\m \qquad {\rm with} \qquad e^a_{\ \m}\, e^b_{\ \n}\, \h_{ab} = g_{\m\n}\,,\quad g^{\m\n} \,e^a_{\ \m }\,e^b_{\ \n} = \h^{ab}
\label{tetrad}
\end{align}
for $a,b=0,\dots,3$, the orthonormal tangent space indices. In the case of (\ref{axisymstat}) a basis of such one-forms is
\vspace{-3mm}
\begin{subequations}
\begin{align}
w^{0} &= e^0_{\ t} \,dt = \exp (\n) \,dt\\
w^{1} & = e^1_{\ r} \, dr = \exp(\a)\, dr\\
w^{2} & = e^2_{\ \th}\, d\th = \exp(\b)\, d\th \\
w^{3} & = e^3_{\ \f}\, d\f + e^3_{\ t}\, dt = \exp (\j)\, (d\f - \w \,dt)
\end{align}
\label{tetradbasis}\end{subequations}
with all other components of the vierbein $e^a_{\ \m}$ not shown vanishing \cite{Chandrasekhar:1984}. The dual basis of vectors $\by_a$ 
\begin{subequations}
\begin{align}
\by_0 &= \exp(-\n)\, \Big(1, 0, 0, \w\Big)  = \exp(-\n) \left(\frac{\pa}{\pa t} + \w  \frac{\pa}{\pa \f}\right) = e^{-\n} \,  \bl \label{ups0} \\
\by_1 &= \exp(-\a)\, \Big(0, 1, 0, 0\Big)= \exp(-\a)\, \frac{\pa}{\pa r} \\
\by_2 &= \exp(-\b)\, \Big(0, 0, 1, 0\Big)= \exp(-\b)\, \frac{\pa}{\pa \th} \\
\by_3 &= \exp(-\j)\, \Big(0, 0, 0, 1\Big)= \exp(-\j)\, \frac{\pa}{\pa \f}
\end{align}
\label{dualbasis}\end{subequations}
in $(t,r,\th,\f)$ coordinates, satisfy $e^a_{\ \m}  \y^\m_{\ b} \!= \!\d^a_{ \ b},\  e^a_{\ \n}  \y^\m_{\ a} \!=\! \d^\m_{ \ \n}$ and $\y^\m_{\ a}  \y^\n_{\ b}\, g_{\m\n} \!=\! \h_{ab},\ 
\y^\m_{\ a}  \y^\n_{\ b} \h^{ab} \!=\! g^{\m\n}$. The vector $\by_0= e^{-\n} \,\bl$ is the four-velocity of a zero angular momentum observer (ZAMO) \cite{Bardeen:1970,ThorneMembrane}. 
Note that the orthonormal tangent space basis (\ref{dualbasis}) becomes singular as $e^\n\to 0$,which limit we take only at the very end.

The tetrad formalism is useful for defining the integration measures for surfaces embedded in the geometry, and for Stokes' theorem.
The four-volume integration measure relies on the Hodge star dual of the constant scalar function $f(x)=1$, which we denote by $\mathds{1}$, namely
\begin{align}
^*\!\mathds{1} &=  \sdfrac{1}{\,4!} \,\e_{abcd}\, w^{a} \wedge w^{b} \wedge w^{c} \wedge w^{d} =  
\frac{1}{\,4!}\, \e_{abcd}\,e^a_{\,\m}\, e^b_{\, \n}\, e^c_{\ \l}\, e^d_{\ \r}\ dx^\m \wedge dx^\n\wedge dx^\l \wedge dx^\r\nn
&=  \e_{abcd}\,e^a_{\ t}\, e^b_{\ r}\, e^c_{\ \th}\, e^d_{\ \f}\ dt\wedge dr\wedge d\th\wedge d\f = \sqrt{-g} \ dt\wedge dr\wedge d\th\wedge d\f
\label{Volform}
\end{align}
where $\e_{abcd}$ is the totally antisymmetric Levi-Civita symbol of four indices, with orientation fixed by $\e_{0123} = +1$.
It is convenient also to define the metric dependent anti-symmetric Levi-Civita tensor density
\vspace{-6mm}
\begin{align}
\ve_{\m\n\l\r} \equiv  \e_{abcd}\,e^a_{\,\m}\, e^b_{\, \n}\, e^c_{\ \l}\, e^d_{\ \r} 
\end{align}
in the coordinate basis. The integral of the four-form (\ref{Volform}) over the four-volume $\W$ is defined to be
\begin{align}
\int\! ^*\!\mathds{1}  \equiv  \int_{\W} \e_{abcd}\,e^a_{\ t}\, e^b_{\ r}\, e^c_{\ \th}\, e^d_{\ \f} \, dt\,dr\,d\th\,d\f =  \int_{\W} {\rm det}(e^a_{\ \m} ) \,d^4x = \int_{\W} \!\sqrt{-g} \,d^4x
\label{intvol}
\end{align}
since viewing $e^a_{\ \m}$ as a $4 \times 4$ matrix, the determinant
\begin{align}
{\rm det} (e^a_{\ \m}) = \e_{abcd}\, e^a_{\ t} e^b_{\ r}e^c_{\ \th}e^d_{\ \f} = \sqrt{-g} = \exp \,(\n + \j + \a + \b)
\label{detg}
\end{align}
is the covariant volume element measure factor of the metric in (\ref{axisymstat}). 

A three-dimensional hypersurface $\S_3$ embedded in the four-geometry is defined by the $4$ coordinate functions
$x^\m = \bar x^\m (\x^1, \x^2,\x^3)$ of the $3$ variables $(\x^1,\x^2,\x^3)$ on the hypersurface. 
The Hodge star dual of a $1$-form $j \!= \!j_\m\, dx^\m$ is the three-form
\vspace{-1mm}
\begin{align}
 \,^*\!j = ^*\!\!(j_\m\, dx^\m) =\sdfrac{1}{\,3!}\,  j^\m \ve_{\m\n\l\r}\ dx^\n\wedge dx^\l \wedge dx^\r 
\label{starJ}
\end{align}
The integral of the three-form (\ref{starJ}) over the three-surface $\S_3$ is by definition
\begin{align}
\int_{\S_3}\,^*\!j = \int_{\S_3} j^\m\  d^3\S_\m 
\label{threesurfint}
\end{align}
where the integrand is evaluated at $x^\m = \bar x^\m (\x^1, \x^2,\x^3)$ and the three-surface element 
\begin{align}
d^3\S_\m = \ve_{\m\n\l\r}\,
\left(\frac{\pa \bar x^\n}{\pa \x^1}\right)  \left(\frac{\pa \bar x^\l}{\pa \x^2}\right)  \left(\frac{\pa \bar x^\r}{\pa \x^3}\right)  d^3\x
\label{threesurf}
\end{align}
follows from the pullback of $\,^*\!j$ onto the three-surface $\S_3$
\begin{align}
&\sdfrac{1}{\,3!}\, \ve_{\m\n\l\r}\, d\bar x^\n\wedge d\bar x^\l \wedge d\bar x^\r 
=  \sdfrac{1}{\,3!}\, \ve_{\m\n\l\r}\, 
\left(\frac{\pa \bar x^\n}{\pa \x^i}\right)  \left(\frac{\pa \bar x^\l}{\pa \x^j}\right) \left(\frac{\pa \bar x^\r}{\pa \x^k}\right)d\x^i\wedge d\x^j \wedge d\x^k\nn
&= \ve_{\m\n\l\r}\, 
\left(\frac{\pa \bar x^\n}{\pa \x^1}\right)  \left(\frac{\pa \bar x^\l}{\pa \x^2}\right)  \left(\frac{\pa \bar x^\r}{\pa \x^3}\right) d\x^1\wedge d\x^2 \wedge d\x^3\,.
\label{threeform}
\end{align}
Integrals of the kind (\ref{threesurfint}) on closed three-surfaces are related to a four-volume integral by Stokes' theorem
\vspace{-6mm}
\begin{align}
\int_\W d \,^*\!j = \int_{\pa\W} \,^*\!j 
\label{Stokes}
\vspace{-3mm}
\end{align}
expressed in the language of differential forms, where $\pa \W = \S_3$ is the three-surface boundary of the four-volume $\W$ and $d$ is the exterior derivative. Now
\begin{align}
d \,^*\!j = \frac{\pa}{\pa x^\a}\left( \sdfrac{j^\m}{3!}\,\ve_{\m\n\l\r} \right) dx^\a \wedge dx^\n\wedge dx^\l \wedge dx^\r 
= \pa_\m\left(j^\m \sqrt{-g} \right) dt \wedge dr\wedge d\th \wedge d\f
\end{align}
as can be verified by writing out the four terms for $\a = t,r,\th,\f$ explicitly and observing that $\a, \n, \l ,\r$ must all be different by the antisymmetry of $\ve_{abcd}$, 
so that $\m =\a$ necessarily. Since the four-form at right is $^*{\mathds 1} \,(\na_\m j^\m)$, Stokes' theorem (\ref{Stokes}) may be written in this case as
\begin{align}
\int_\W \na_\m j^\m \sqrt{-g}\, d^4x =  \int_{\pa\W}  j^\m d^3\S_\m
\label{43Stokes}
\end{align}
with the three-surface element given by (\ref{threesurf}).
 
These steps may be repeated in applying Stokes' theorem again to a vector $j^\m = \na_\n F^{\m\n}$ which is the covariant derivative of an antisymmetric tensor field
$F^{\m\n} = - F^{\n\m}$. In this case the two-form
\begin{align}
\sdfrac{1}{\,2!}\, \ve_{\m\n\l\r}\ dx^\l \wedge dx^\r =  
\ve_{\m\n\l\r}\left(\frac{\pa \bar x^\l}{\pa \x^1}\right) 
\left(\frac{\pa \bar x^\r}{\pa \x^2}\right) d\x^1\wedge d\x^2
\end{align}
characterizes the two-dimensional surface parametrized by $x^\m = \bar x^\m (\x^1,\x^2)$. Contracting this with $F^{\m\n}/2!$ gives
\vspace{-5mm}
\begin{align}
\left(\sdfrac{1}{\,2!}\right)^2 F^{\m\n} \ve_{\m\n\l\r}\ dx^\l \wedge dx^\r = \,^*\!F
\end{align}
which is the Hodge star dual of the two-form
\begin{align}
F &\equiv  \sdfrac{1}{\,2!}\, F_{\m\n}\ dx^\m \wedge dx^\n\,.
\end{align}
Thus applying Stokes' theorem to this case, 
\begin{align}
\int_{V} d\,^*\!F = \int_{\pa V} \,^*\!F 
\label{32Stokes} 
\end{align}
where on the left side we have the integral of the three-form
\begin{align}
d \,^*\!F &= \frac{\pa}{\pa x^\a} \left(\sdfrac{1}{(2!)^2}\, F^{\m\n} \ve_{\m\n\l\r}\right) dx^\a \wedge dx^\l \wedge dx^\r =\pa_\n (\sqrt{-g} F^{t \n})\,dr \wedge d\th \wedge d\f \nn
&-\pa_\n (\sqrt{-g} F^{r\n})\,dt \wedge d\th \wedge d\f + \pa_\n (\sqrt{-g} F^{\th\n})\,dt \wedge dr \wedge d\f - \pa_\n (\sqrt{-g} F^{\f\n})\,dt \wedge dr \wedge d\th\nn
&= \na_\n F^{\m\n} \sdfrac{1}{\,3!}\,\ve_{\m\n\l\r}\ dx^\n\wedge dx^\l \wedge dx^\r  = \,^*\!j
\label{dFJ}
\end{align}
over the three-volume, relabeled $V$ here instead of $\S_3$. 

The three-volume integral on the left side of (\ref{32Stokes}) can be written exactly as in the right side of (\ref{43Stokes}) for this $j^\m$ with (\ref{threesurf}).
At the same time the right side of (\ref{32Stokes}) can be written in terms of the element of directed $2$-surface area
\begin{align}
d^2\S_{\m\n} =  \sdfrac{1}{2} \ve_{\m\n\l\r}\left(\frac{\pa \bar x^\l}{\pa \x^1}\right)  \left(\frac{\pa \bar x^\r}{\pa \x^2}\right) d^2\x 
\label{2surf}
\end{align}
with one combinatoric factor of $1/2!$ cancelled by specifying the order of $\x^1,\x^2$, while the second factor of $1/2!$ remains to account for the
$2$ equal terms in the sum over $\m,\n$.  Thus finally (\ref{32Stokes}) can be written
\vspace{-5mm}
\begin{align}
\int_{V}  \na_\n F^{\m\n} d^3\S_\m =  \int_{\pa V} F^{\m\n} d^2\S_{\m\n}  
\label{Stokes3to2}
\end{align}
in component form, with $d^3\S_\m$ and $d^2\S_{\m\n}$ given by (\ref{threesurf}) and (\ref{2surf}) respectively.

The surface elements simplify if the surfaces are at fixed values of the coordinates of the embedding spacetime.
In the case of interest for the Komar mass and angular momentum considered in the text, the three-surface is the spacelike constant $t$ slice 
of (\ref{axisymstat}), with coordinates $\x = (r,\th,\f)$ the same as the embedding space coordinates. In that case (\ref{threesurf}) becomes
\begin{align}
d^3\S_\m = \d^t_{\ \m}\, \sqrt{-g}\ dr\,d\th\,d\f =  \d^t_{\ \m}\ e^{\a +\n + \b + \j}\ dr\,d\th\,d\f 
\label{3surft}
\end{align}
with only a future directed $\m =t$ component, and where (\ref{detg}) has been used. Likewise, if the two-surface in (\ref{2surf}) is 
at fixed $t$ and $r$ in the same coordinates
\begin{align}
d^2\S_{\m\n} =  \sdfrac{1}{2} \e_{abcd}\,e^a_{\, \m}\, e^b_{\, \n}\, e^c_{\, \th}\, e^d_{\, \f} \, d\th\,d\f =  \d^t_{\ [\m}\, \d^r_{\ \n]}  \,\sqrt{-g} \,d\th\,d\f 
=\ell_{[\m} N_{\n]} \,dA 
\label{2surftr}
\end{align}
in terms of the vectors $\bl$ and $\bN$ of (\ref{Nnull})-(\ref{Ncomp}) orthogonal to the two-surface, and where
\begin{align}
dA = e^{\b + \j} \, d\th\,d\f
\label{dArea}
\end{align}
is the area element on the metric induced from (\ref{axisymstat}) on the two-surface of constant $t$ and $r$. Since $\bl = \bn + e^{2\n}\, \bN$, with $\bn$
given by (\ref{nrot}), the area element (\ref{2surftr}) is also $n_{[\m} N_{\n]} \,dA$.

\vspace{-2mm}
\section{Axisymmetric Geometry and Weyl Tensor on the Singular Horizon Surface}
\label{Sec:Weyl}
\vspace{-1mm}

The Riemann tensor corresponding to the general stationary axisymmetric metric (\ref{axisymstat}) has exactly $12$ nonzero and {\it a priori} independent components, 
of a possible $20$ in the most general case of no symmetry~\cite{Chandrasekhar:1984}. These $12$ components can be decomposed into $6$ for the 
nontrivial traceless Weyl tensor, and $6$ for the nonzero components of the Ricci tensor in the trace parts. Following from this, there are a total of $6$ nontrivial Einstein 
equations for the remaining $6$ components of the Ricci and Einstein tensors, which are given explicitly by (\ref{Einstens}), and correspondingly $6$ generally nonzero 
components of the stress-energy tensor,  {\it viz.} $T_{tt}, T_{t\f},T_{\f\f},T_{rr},T_{\th\th},T_{r\th}$, each of which are functions of just $r$ and $\th$. 
The remaining $4$ components of the Einstein (and Ricci) tensor, {\it viz.} $G_{tr},G_{t\th},G_{\f r},G_{\f\th}$ vanish identically, by stationarity and axial 
symmetry, and hence so must the corresponding stress-energy tensor components $T_{tr},T_{t\th},T_{\f r},T_{\f\th}$. 

This has the immediate consequence that the covariant conservation equations $\na_\n T^\n_{\ t} = 0= \na_\n T^\n_{\ \f}$ are also satisfied identically. However, the 
two other (first order) covariant conservation equations $\na_\n T^\n_{\ r} = 0$ and $\na_\n T^\n_{\ \th} = 0$ are in general nontrivial and independent of each other, 
so that they can be substituted for $2$ of the $6$ Einstein equations if desired. Finally, in either description, there are $6$ partial differential equations in $(r,\th)$ for $10$ 
 functions, {\it i.e.} $4$ metric functions plus $6$ stress-energy components, in the general stationary axisymmetric geometry. 

Without further restrictions, this system is underdetermined and the Segre class is also the most general, with the Ricci tensor possessing in general $4$ distinct eigenvalues 
at each point. If the stress-energy tensor vanishes or has the specific form of a cosmological constant, the $2$ covariant conservation equations also 
become vacuous, and there are just $4$ Einstein equations for $4$ metric functions of $r,\th$. The solution of these is then unique up to integration constants depending
upon boundary conditions. The Kerr solution is an example of a such a special vacuum solution which is asymptotically flat.

The Weyl conformal tensor, defined by \cite{MTW}
\vspace{-2mm}
\begin{align}
C^{\a\b}_{~~~\m\n} = R^{\a\b}_{~~\,\m\n} - 2\, \d^{[\a}_{~~[\m} R^{\b]}_{~~\n]} + \sdfrac{1}{3} \d^{[\a}\,_{\!\![\m} \d^{\b]}\,_{\!\n]} R
\label{Weyldef}
\end{align}
is antisymmetric under exchange of $\m$ and $\n$ and of $\a$ and $\b$, symmetric under exchange of the pairs $\m\n$ and $\a\b$,
and completely traceless, $C^{\a\b}_{~~\ \a\n} \!=\! 0$. The orthonormal tangent frame components of the Weyl tensor 
are defined via the tetrads and dual basis vectors of Appendix \ref{Sec:Tetrad} by
\begin{align}
C^{ab}_{\ \ \,cd} = e^a_{\ \a}\, e^b_{\ \b}\,  C^{\a\b}_{\ \ \ \m\n}\,\y^\m_{\ c}\, \y^\n_{\ d} 
\label{Weylorth}
\end{align}
with orthonormal indices $0,\dots,3$ raised and lowered by the flat metric $\h_{ab}= \mathrm{diag(-+++)}$.

Because of the symmetries obeyed by the Weyl tensor, each pair of indices $(ab)$ or $(cd)$ can be restricted to just one of six possible pair 
values $(ab) = (01), (02),(03),(23),(31),(12)$, and $C^{ab}_{\ \ cd}$ can be expressed in the form of a symmetric $6\times6$ traceless matrix
with the $3 \times 3$ block structure
\vspace{-2mm}
\begin{align}
C^{ab}_{\ \ \,cd} = 
\begin{array}{l} (01)\\[-8pt](02)\\[-8pt](03)\\[-8pt](23)\\[-8pt](31)\\[-8pt](12) \end{array} 
\left(\begin{tabular}{c|c} 
&\\[-12pt]
\hbox to 2em{\hfil$E$\hfil} & \hbox to 2em{\hfil$H$\hfil} \\
&\\[-12pt]
\hline
&\\[-12pt]
$\!\!\!-H$ & $E$ \\[-12pt]
& 
\end{tabular} \right)
\label{eq:WeylCmatrix}
\end{align}
\vspace{-1mm}\noindent
where the $(ab)$ rows are labeled by six index pair values as shown, and the columns $(cd)$ are labeled similarly. The information in the matrix \eqref{eq:WeylCmatrix} can be expressed 
still more compactly in the form of the complex $3 \times 3$ matrix
\vspace{-7mm}
\begin{align}
Q = E+iH 
\end{align}
with relabeled indices ranging over $i= 1,2,3$ for $(0i) \to i$, and $(jk) \to i$ for $i,j,k$ a cyclic permutation of $1, 2, 3$. Thus
\vspace{-7mm}
\begin{subequations}
\begin{align}
E_{ij} &= C^{0i}_{\ \ 0j} = \sdfrac{1}{4} \,\e_{ikl}\,\e_{jmn} \,C^{kl}_{\ \ mn} =  C^{kj}_{\ \ \,ik} + \sdfrac{1}{2}\,\d_{ij}\, C^{kl}_{\ \ kl} \label{Eij}\\
H_{ij} &= \sdfrac{1}{2} \,\e_{jkl}\, C^{0i}_{\ \ kl} = -\sdfrac{1}{2}\, \e_{jkl}\,C^{kl}_{\ \ 0i}\label{Hij}
\end{align}
\label{EHij}\end{subequations}
with the real and imaginary parts of $Q$ the ``electric” and ``magnetic” parts respectively of the Weyl tensor~\cite{Stephani:2003tm}. Note that in fact 
$C^{kl}_{\ \ kl} \!=\!0\!=\!\e_{ijk}C^{0i}_{\ \ jk}$. The matrix $Q$ inherits both its transpose symmetry $Q_{ij} = Q_{ji}$, and traceless property $Q_{ii} = 0$ 
from those of $E_{ij}$ and $H_{ij}$ separately.

The nonzero components of the matrices $E_{ij}$ and $H_{ij}$ for the stationary, axisymmetric metric of \eqref{axisymstat} are given explicitly by
\begin{subequations}
\begin{align}
E_{11}&=  \sdfrac{1}{6} e^{-2 \a } \Big\{\b _{rr}+\j _{rr}-2 \n_{rr} -(\b_r + \j_r -2\n_r) (\a_r-\n_r) + (\b_r-\j_r)^2\Big\}
+\sdfrac{1}{3} e^{-2 \a -2 \n +2 \j }\,\w _r^2 \nn
&+ \sdfrac{1}{6} e^{-2 \b } \Big\{\a_{\th \th } +\n _{\th \th } -2 \j _{\th \th}  -(\a_\th + \n_\th -2\j_\th) (\b_\th-\j_\th) + (\a_\th-\n_\th)^2\Big\}  
-\sdfrac{1}{6}  e^{-2 \b -2 \n +2 \j} \, \w _{\th}^2\\
E_{12}&=E_{21} = \sdfrac{1}{2} e^{-\a -\b } \Big\{\j _{r\th}-\n _{r\th} + \a _{\th } \left(\n _r-\j _r\right)+\n _{\th } \left(\b _r-\n _r\right)+\j_{\th } \left(\j _r-\b _r\right)
+ e^{2 \j -2 \n } \,\w_\th\,\w_r\Big\} \\
E_{22}&=\sdfrac{1}{6} e^{-2 \a } \Big\{\b _{rr}+\n _{rr}-2 \j_{rr} -(\b_r + \n_r -2\j_r) (\a_r-\j_r) + (\b_r-\n_r)^2\Big\}
-\sdfrac{1}{6} e^{-2 \a -2 \n +2 \j }\,\w _r^2 \nn
&+\sdfrac{1}{6} e^{-2 \b} \Big\{\a_{\th \th }+\j _{\th \th }-2 \n _{\th \th } -(\a_\th + \j_\th -2\n_\th) (\b_\th-\n_\th) + (\a_\th-\j_\th)^2\Big\}   
+ \sdfrac{1}{3}  e^{-2 \b -2 \n +2\j }\,\w _{\th }^2\\
E_{33}&=\sdfrac{1}{6} e^{-2 \a } \Big\{\j _{rr}+\n _{rr}-2 \b _{rr} -(\j_r + \n_r -2\b_r) (\a_r-\b_r) + (\n_r-\j_r)^2\Big\}
-\sdfrac{1}{6} e^{-2 \a -2 \n +2 \j }\, \w _r^2\nn
&+ \sdfrac{1}{6} e^{-2 \b } \Big\{\j _{\th \th } + \n_{\th \th } -2\a _{\th \th } -(\j_\th + \n_\th -2\a_\th) (\b_\th-\a_\th) + (\j_\th-\n_\th)^2 \Big\} 
-\sdfrac{1}{6} e^{-2 \b -2 \n +2 \j }\, \w _{\th }^2 \\
H_{11}&=\sdfrac{1}{2} e^{-\a -\b -\n +\j } \Big\{\w _{r\th} - \w _r \left(\a _{\th }+\n _{\th }-2 \j _{\th }\right)-\w _{\th } \left(\b _r-\j _r\right)\!\Big\}\\
&\hspace{-7mm}H_{12}=\!H_{21} \!=\! \sdfrac{1}{4} e^{\j-\n} \Big\{\!e^{-2 \b}\big\{\w _{\th \th } \!-\!\w _{\th } \left(\a _{\th }+\b _{\th }+\n _{\th }\!-\!3 \j _{\th}\right)\!\big\}
\!-\! e^{-2 \a} \big\{\w _{rr}\!-\!\w _r \left(\a _r+\b_r+\n _r\!-\!3 \j _r\right)\!\big\}\!\Big\}\\
H_{22}& =\sdfrac{1}{2} e^{-\a -\b -\n +\j } \Big\{-\w _{r\th} - \w _r \left(\j _{\th }-\a _{\th }\right)+\w _{\th } \left(\b_r+\n _r-2 \j _r\right)\!\Big\}\\
H_{33}& =\sdfrac{1}{2} e^{-\a -\b -\n +\j } \Big\{\w _r \left(\n _{\th }-\j _{\th }\right)+\w _{\th } \left(\j _r-\n_r\right)\!\Big\}
\end{align}
\label{EHweyl}\end{subequations}
with all other components not listed vanishing.

It follows that for a stationary, axisymmetric spacetime and for the choice of tetrad in Eq.~(\ref{tetradbasis}), the complex matrix $Q$ has the following block structure
\begin{align}
Q = \begin{pmatrix}
Q_{11} & Q_{12} & 0 \\
Q_{12} & Q_{22} & 0 \\
0 & 0 & Q_{33}\\
\end{pmatrix} \qquad {\rm with} \qquad Q_{11} + Q_{22} +  Q_{33} = 0
\label{eq:Q_BlockStructure}
\end{align}
and hence contains just $3$ independent complex components (the Newman-Penrose coefficients $\Psi_0,\Psi_2,\Psi_4$), accounting for the $6$ nonzero 
real components of the Weyl tensor in this geometry. The three eigenvalues of $Q$ are
\begin{align}
\sdfrac{1}{2} \Big\{Q_{11}+Q_{22} \pm \sqrt{(Q_{22}-Q_{11})^2+4Q_{12}^2}\Big\} \qquad {\rm and} \qquad  Q_{33} = - \left( Q_{11}+Q_{22} \right)
\label{eigenQ}
\end{align}
If these are all different, which is the generic case, then the geometry is Petrov Type {\bf I}. If any two of the eigenvalues are equal (but not equal to the third),
at all points, then the geometry is Petrov Type {\bf D}. This is an algebraically special case, of which the vacuum Kerr solution is a well-known example.

The symmetric matrix $Q$ can be brought to a diagonal form $Q_D$ by a complex $SO(2,\mathbb{C})$ rotation 
\begin{align}
Q_{_{\! D}} = M Q M^{T}
\end{align} 
where the matrix $M$ also has the $2\times 2$ block form 
\begin{align}
M = \begin{pmatrix} 
\cos(\x+i\h) & -\sin(\x+i\h) & 0 \\ 
\sin(\x+i\h) & \cos(\x+i\h) & 0 \\ 
0 & 0 & 1 
\end{pmatrix} 
\end{align}
and  
\vspace{-5mm}
\begin{align}
\tan \big\{2(\x + i \h)\big\} = \frac{2\, Q_{12}}{Q_{22}-Q_{11}}
\end{align}
in terms of two real functions $(\x,\h)$ of $(r,\th)$. This corresponds to the local Lorentz transformation $w^{\prim a} = \L^{a}_{~b} w^b$ 
of the orthonormal tangent frame basis one-forms (\ref{tetradbasis}) to
\begin{subequations}
\begin{align}
& w^{\prim 0} = \cosh\h  \, w^{0} + \sinh\h \, w^{1} \\
& w^{\prim 1} = \sinh\h  \, w^{0} + \cosh\h \, w^{1} \\
& w^{\prim 2} = \cos\x  \, w^{2} - \sin\x \, w^{3}\\
& w^{\prim 3} = \sin\x  \, w^{2} + \cos\x \, w^{3} 
\end{align}
\end{subequations}
in which $\x$ is the rotation angle in the $r\th$ plane, and $\h$ is the boost parameter (rapidity) in the $t\f$ plane. Thus $2$ of the $6$ real components of
the Weyl tensor at any point are accounted for by this rotation-boost to the tangent frame basis where $Q$ is diagonal, and the remaining
$4$ real components are accounted for by any two of the unequal complex eigenvalues (\ref{eigenQ}). 

Identifying the null surface singular $\d$-function contributions to the Weyl tensor proceeds in the same fashion as for the Einstein tensor in Sec.~\ref{Sec:Rotate},
namely by multiplying the components (\ref{EHweyl}) by the potentially discontinuous factor $\exp(\a + \n)$ of $\sqrt{-g}$, grouping all second derivative terms 
with respect to $r$ together with the singular $\a_r + \n_r$ terms to form total $r$-derivatives, and finally taking the horizon limit $e^{2\n} \to 0$. In this way
we obtain
\begin{subequations}
\begin{align}
e^{\a + \n}E_{11}&= - \sdfrac{1}{6} \left[e^{- \a-\n }\frac{\pa e^{2\n}}{\pa r}\right] \d (r-R) + \dots = - \sdfrac{1}{3} \big[\k\big] \d (r-R)+ \dots \\
e^{\a + \n}E_{22}&=  \sdfrac{1}{12} \left[e^{- \a-\n }\frac{\pa e^{2\n}}{\pa r}\right] \d (r-R)+ \dots = \sdfrac{1}{6} \big[\k\big] \d (r-R)+ \dots \\
e^{\a + \n}E_{33}&= \sdfrac{1}{12} \left[e^{- \a-\n }\frac{\pa e^{2\n}}{\pa r}\right] \d (r-R)+ \dots = \sdfrac{1}{6} \big[\k\big] \d (r-R)+ \dots \\
e^{\a + \n}H_{12}&\!=\!e^{\a + \n}H_{21} = \sdfrac{1}{4} e^{\j+\n} \left[e^{-\a-\n }\frac{\pa \w}{\pa r}\right] \d (r-R)+ \dots\! = \sdfrac{1}{2} e^{\n-\j}\big[\cJ\big] \d (r-R)+ \dots 
\label{H12surf}
\end{align}
\label{EHsurf}\end{subequations}
where (\ref{nukap}) and the definitions (\ref{kapdisc}), (\ref{cJdisc}) have been used, and with other components not listed giving no null-horizon contributions. 
We have retained the last contribution (\ref{H12surf}), even though like (\ref{S03}) it goes to zero on the horizon, because it is needed if one transforms to the
$(t,r,\th,\f)$ coordinate basis. This transformation involves $e^{-\n}$ through (\ref{ups0}), and leads to finite $[\cJ]$ dependence of $^{(\S)\!}C^{\a\b}_{\ \ \ \m\n}$ on the surface, 
as does (\ref{surfSrot}) if transformed back to the coordinate basis (\ref{surfTrot}).

The singular surface contributions (\ref{EHsurf}) of the Weyl tensor can be related to those of the Einstein tensor and surface stress tensor as follows.
From the definition of the Weyl tensor (\ref{Weyldef}) and Einstein's equations it follows that
\vspace{-2mm}
\begin{align}
  C^{\a\b}_{~~~\m\n} = R^{\a\b}_{~~\,\m\n} - 16\pi G\, \d^{[\a}_{~~[\m} T^{\b]}_{~~\n]} + \sdfrac{16\pi G}{3}\, \d^{[\a}\,_{\!\![\m} \d^{\b]}\,_{\!\n]}\, T
  \label{eq:WeylT}
\end{align}
so that multiplying by $\exp(\a + \n)$, and selecting out only the singular surface contributions proportional to $\d (r-\RH)$ gives
\vspace{-3mm}
\begin{align}
  e^{\a + \n}\ ^{(\S)\!}C^{ab}_{~~\,cd} = e^{\a + \n}\ ^{(\S)\!}R^{ab}_{~~\,cd} + \left( - 16\pi G\, \d^{[a}_{~~[c} \cS^{b]}_{~~d]} 
  + \sdfrac{16\pi G}{3}\, \d^{[a}\,_{\!\![c} \d^{b]}\,_{\!d]} \,\cS^f_{\ f} \right) \, \d(r-\RH)
\label{WeylorthoS}
\end{align}
in the orthonormal tangent frame basis. Using (\ref{EHij}), (\ref{EHsurf}) and (\ref{surfSrot}) one may check that for the six independent components
$C^{31}_{\ \ \,31}\!=\!E_{22},\, C^{03}_{\ \ \,03}\!=\!E_{33},\, C^{01}_{\ \ \,31}\!=\!H_{12}$, and $C^{02}_{\ \ \,31}\!=\!H_{22},\, C^{03}_{\ \ \,12}\!=\!H_{33},\, C^{01}_{\ \ \,02}\!=\! E_{12}$, 
the Weyl tensor $\d$-function surface contributions on the left are precisely given by the surface stress tensor terms on the right of (\ref{WeylorthoS}), the last three with vanishing
surface contributions. In other words, these six components have zero $^{(\S)\!}R^{ab}_{~~\,cd}$ surface terms, and hence the Weyl tensor and $Q$ matrix (\ref{eq:Q_BlockStructure})
on the surface are {\it completely determined} by the surface stress tensor $\cS^a_{\ b}$. Since all the Weyl tensor $\d$-function surface contributions are completely fixed in this way by the 
matter stress tensor on the horizon, there is no independent lightlike singular gravitational shock wave propagating along the horizon \cite{BarrabesHogan:2003}.

By inverting the transformation (\ref{Weylorth}), {\it i.e.}
\begin{align}
C^{\a\b}_{\ \ \ \m\n} = \y^\a_{\ a}\, \y^\b_{\ b}\,   C^{ab}_{\ \ \,cd} \,e^c_{\ \m}\, e^d_{\ \n}\,  \,
\label{Weycoord}
\end{align}
which is well defined and nonsingular for all $e^{\n} \neq 0, e^{-\a} \neq 0$, one can also express the Weyl tensor in the original coordinate basis of (\ref{axisymstat}),
and evaluate the singular surface components in this basis in the horizon limit, $e^{2\n} \to 0$. Defining the $\d$-function contributions to the Weyl tensor at the null horizon as
 \begin{align}
 e^{\a + \n}\ ^{(\S )}C^{\a\b}_{\ \ \m\n}\  \equiv \cC^{\a\b}_{\ \ \m\n} \, \d(r-\RH)
\label{SurfWeyl}
\end{align}
and making use of (\ref{EHsurf}), in this way we obtain
\begin{subequations}
\begin{align}
\cC^{t\th}_{\ \ \,t\th}&=  \cC^{r\f}_{\ \ \,r\f} = \sdfrac{1}{6} \big[\k\big]  + \sdfrac{1}{2} \omega_{_{\!H}} \big[\cJ\big] \\
\cC^{t\f}_{\ \ \,t\f}&=  \cC^{r\th}_{\ \ \,r\th} = \sdfrac{1}{6} \big[\k\big]  \\
\cC^{\th\f}_{\ \ \,\th\f}&=   \cC^{rt}_{\ \ \,rt} = - \sdfrac{1}{3} \big[\k\big]  - \sdfrac{1}{2} \omega_{_{\!H}} \big[\cJ\big] \\
\cC^{t\th}_{\ \ \,\th\f}&=  \cC^{rt}_{\ \ \,r\f} = \sdfrac{1}{2} \big[\cJ\big]  \\
\cC^{\th\f}_{\ \ \,t\th}&=   \cC^{r\f}_{\ \ \,rt} = - \sdfrac{1}{2} \omega_{_{\!H}} \big[\k\big]  - \sdfrac{1}{2} \omega_{_{\!H}}^2 \big[\cJ\big] 
\end{align}
\label{Csurf}\end{subequations}
with all other components not obtained from those listed by the antisymmetry in the first and second pair of indices equal to zero. 

Naturally this same result is also obtained by computing the components of (\ref{Weyldef}) directly in the coordinate basis (\ref{axisymstat}).
As a consistency check, or in case the reader is worried about the use of the orthonormal basis at the horizon where it becomes singular, we have evaluated 
the $\d$-function terms in the Weyl tensor density $\sqrt{-g} \, C^{\a\b}_{\ \ \,\m\n}$ directly in the coordinate basis, transforming all second derivatives of the metric 
$g_{\m\n}$ into derivatives of the form $(\sqrt{-g} \, g_{\m\n,\a})_{,\b}$, and taking the limit $R \to \RH$ assuming a finite $\sqrt{-g}$ at the horizon and continuity 
of the Killing vectors and the induced metric.  We then reach the same conclusion about the absence of a gravitational shock wave at the null horizon working 
in the coordinate basis as follows. The coefficients of the $\d(r-\RH)$ terms in Eq.~(\ref{eq:WeylT}) obey the equation
\begin{align}
  \cC^{\a\b}_{~~\,\m\n} = \cR^{\a\b}_{~~\,\m\n} - 16\pi G\, \d^{[\a}_{~~[\m} \cS^{\b]}_{~~\n]} + \sdfrac{16\pi G}{3}\, \d^{[\a}\,_{\!\![\m} \d^{\b]}\,_{\!\n]} \,\cS^\l_{\ \l} 
\label{WeylCRS}
\end{align}
where $\ e^{\a + \n} \,^{(\S )}R^{\a\b}_{\ \ \,\m\n} = \cR^{\a\b}_{\ \ \,\m\n} \, \d(r-\RH)$.
Substituting the explicit results (\ref{Csurf}) and (\ref{surfTrot}) into (\ref{WeylCRS}), or by direct calculation of $\cR^{\a\b}_{\ \ \,\m\n}$, one can verify that all the 
tangential components of the surface Weyl tensor, i.e., all the components $\cC^{ij}_{\ \ kl}$ in (\ref{SurfWeyl}) with indices $i,j,k,l$ ranging over $t,\th,\f$, receive 
no contribution from the Riemann tensor term in (\ref{WeylCRS}) and are determined completely by the terms in (\ref{WeylCRS}) that contain the surface stress tensor 
$\cS^\m_{\ \n}$, to wit 
\begin{subequations}
\begin{align}
  \cR^{ij}_{\ \ kl}=0, \qquad  \cC^{ij}_{\ \ kl} & = - 16\pi G\, \d^{[i}_{~~[k} \cS^{j]}_{~~l]} + \sdfrac{16\pi G}{3}\, \d^{[i}\,_{\!\![k} \d^{j]}\,_{\!l]}\, \cS^{m}_{\ m}
\end{align}
\label{Ctangential}\end{subequations}
with $ i,j,k,l,m = t,\th,\f$.
All other components of $\cC^{\a\b}_{~~\,\m\n}$, {\it i.e.} those with one or more indices equal to $r$, either vanish or are equal to one of the tangential components $\cC^{ij}_{\ \ kl}$ 
according to the leftmost equal signs in (\ref{Csurf}) and the antisymmetries of the Weyl tensor. Therefore by an analogous accounting as that leading to (\ref{WeylorthoS}),
{\it all} components of the Weyl tensor density $\cC^{\a\b}_{~~\,\m\n}\,  \d(r-\RH)$ at the null horizon are completely determined by the local value of the surface stress tensor 
$\cS^{\m}_{\ \n}$ at the same point on the horizon. There is no impulsive gravitational wave on the null horizon.

\vspace{-2mm}
\section{Barrab\`es-Isreal Formalism for the Null Surface Stress Tensor in the Rotating Case}
\label{Sec:BIrot}
\vspace{-1mm}

The BI transverse null vector $\bN$ satisfying (\ref{Nnull}) and (\ref{Nnorm}) in the general axisymmetric stationary metric of (\ref{axisymstat})
is given by (\ref{Ncomp}). Proceeding with the BI formalism of~\cite{BarrabesIsrael:1991} we compute the oblique curvature
\vspace{-4mm}
\begin{align}
\cK_{ij} = - N_\m \be^\l_{(j)} \na_\l \be^\m_{(i)} = -N_\m \G^\m_{\ ij} = \sdfrac{1}{2} N^\m \Big\{ \pa_\m g_{ij} - \pa_i g_{\m j} - \pa_j g_{\m i}\Big\} 
\end{align}
so that the discontinuity $\g_{ij} = 2\, [\cK_{ij}]$ has components 
\begin{subequations}
\begin{align}
\g_{tt} = \left[N^r\,\frac{\pa g_{tt}}{\pa r}\right] &=  \left[e^{-\a-\n} \frac{\pa}{\pa r} \left(e^{2\n} - \w^2 \, e^{2\j}\right)\right]\\
\g_{t\f} = \left[N^r\,\frac{\pa g_{t\f}}{\pa r}\right] &=  \left[e^{-\a-\n} \frac{\pa}{\pa r} \left(\w \, e^{2\j}\right)\right]\\
\g_{\f\f} = \left[N^r\,\frac{\pa  g_{t\f}}{\pa r}\right]&= -  \left[e^{-\a-\n} \frac{\pa}{\pa r} e^{2\j}\right]\\
\g_{\th\th} = \left[N^r\,\frac{\pa  g_{\th\th}}{\pa r}\right]&= -  \left[e^{-\a-\n} \frac{\pa}{\pa r} e^{2\b}\right]\\
\g_{t\th} = -\left[N^t\,\frac{\pa g_{tt}}{\pa \th}\right] -\left[ N^\f\,\frac{\pa g_{t\f}}{\pa \th} \right] &=  
\left[e^{-2\n} \frac{\pa}{\pa \th} e^{2\n}\right]- \left[\w\, e^{-2\n + 2\j} \frac{\pa \w}{\pa \th}\right]\\
\g_{\th\f}  = -\left[N^t\,\frac{\pa  g_{t\f}}{\pa \th}\right]&-\left[N^\f\,\frac{\pa g_{\f\f}}{\pa \th} \right]= \left[e^{-2\n + 2\j} \, \frac{\pa \w}{\pa \th}\right]
\end{align}
\label{gsurf}\end{subequations}
Note that all of these $\g_{ab}$ are {\it finite} on the horizon. 
The first four entries in (\ref{gsurf}) depend only upon $N^r$ and are clearly independent of the other components, which characterize
displacements tangential to the surface described by BI Eq.~(12): $\bN \to \bN + \l^k \be_{(k)}$ and which would change $N^t, N^\th, N^\f$.
The only possibly problematic nonzero entry in (\ref{gsurf}) is the fifth one, $\g_{t\th} \neq 0$, which does depend upon the tangential
components, $N^t$ and $N^\f$, and causes a problem below.

We then apply BI Eqs.~(17)-(18) for the surface stress tensor $\cS^i_{\ j}$, with $\h\!=\! -1$. Noting that the normal $\bn$ as given by (\ref{zdef}) has no $(t,\th,\f)$
components, so that $\g^\m\! =\!0 \!= \!\g^\dag$, this is equivalent to BI Eq.~(22) or
\vspace{-6mm}
\begin{align}
- 16 \pi G\, \cS^i_{\ j} = - \e\,  \big(g^{ik} \g_{jk} - \d^i_{\ j} g^{kl}\g_{kl}\big)
\end{align}
which is well defined and finite away from the horizon where $\e > 0$. Making use of (\ref{gsurf}) to evaluate these components and then taking $\e = e^{2\n} \to 0$ 
on the null horizon gives
\begin{subequations}
\begin{align}
16 \pi G\, \cS^t_{\ t}&= e^{2\n}\big(g^{tc}\g_{tc} - g^{cd}\g_{cd}\big) \to \w \g_{t\f} + \w^2 \g_{\f\f} = \w\, e^{2\j} \left[e^{-\a-\n} \, \frac{\pa \w}{\pa r}\right]\\
16 \pi G\, \cS^t_{\ \f}&= e^{2\n}\big(g^{tt}\g_{t\f} - g^{t\f}\g_{\f\f}\big) = -\g_{t\f} -\w\, \g_{\f\f} = - e^{2 \j}\left[e^{-\a-\n} \, \frac{\pa\w}{\pa r}\right]\\
16 \pi G\, \cS^\f_{\ t}&=  e^{2\n} \big(g^{t\f}\g_{tt} - g^{\f\f}\g_{t\f}\big)\to  -\w \g_{tt} - \w^2\g_{t\f} = -\w\left[e^{-\a-\n} \frac{\pa}{\pa r} e^{2\n}\right] + \w^2 e^{2\j} \left[e^{-\a-\n}\, \frac{\pa \w}{\pa r}\right]\\
16 \pi G\, \cS^\f_{\ \f}&= e^{2\n}\big(g^{\f c}\g_{\f c} - g^{cd}\g_{cd}\big) \to \g_{tt} + \w \,\g_{t\f} = \left[e^{-\a-\n}\frac{\pa}{\pa r} e^{2\n}\right] - \w\, e^{2\j}  \, 
\left[e^{-\a-\n}\frac{d \w}{dr}\right]\label{Sff}\\
16 \pi G\, \cS^\th_{\ \th}&= e^{2\n}\big(g^{\th c}\g_{\th c} - g^{cd}\g_{cd}\big)\to \g_{tt} + 2\w\, \g_{t\f} + \w^2 \g_{\f\f} = \left[e^{-\a-\n}\frac{\pa}{\pa r} e^{2\n}\right]\label{Sthth}\\
16 \pi G\, \cS^t_{\ \th}&= \e\,  (g^{tc} \g_{c \th}) = e^{2\n} g^{tt}\g_{t \th}= -\left[e^{-2\n} \frac{\pa}{\pa \th} e^{2\n}\right]\qquad {\rm (incorrect)}\\
16 \pi G\, \cS^\th_{\ t}&= \e\,  (g^{\th c} \g_{c t})  = e^{2\n} g^{\th\th}\g_{t \th}=e^{2\n -2\b}  \left[e^{-2\n} \frac{\pa}{\pa \th} e^{2\n}\right]\qquad {\rm (incorrect)} \\
16 \pi G\, \cS^\th_{\ \f}&= \e\,  (g^{\th c} \g_{c \f}) =  e^{2\n} g^{\th\th}\g_{\th \f}= 0\\
16 \pi G\, \cS^\f_{\ \th}&= \e\,  (g^{\f c} \g_{c \th})  = e^{2\n} g^{\f t}\g_{t \th}= -\w \left[e^{-2\n} \frac{\pa}{\pa \th} e^{2\n}\right] \qquad {\rm (incorrect)}
\end{align}
\label{Stensor}\end{subequations}
where we have put to zero all terms which vanish as $\e = e^{2\n} \to 0$ on the null horizon surface. We note from (\ref{Stensor}) that in the case of $\w\!=\!0$ 
and no $\th$ dependence, only $\cS^\f_{\ \f}$ and $\cS^\th_{\ \th}$ in (\ref{Sff}) and (\ref{Sthth}) are nonzero, proportional to $\g_{tt}$ and equal to each other, 
reducing to the spherically symmetric case already considered in Sec.~\ref{Sec:BInonrot}. Note that although some of these components agree with those derived 
directly from the Einstein tensor in Sec.~\ref{Sec:Rotate}, the nonzero results for $\cS^t_{\ \th}, \cS^\th_{\ t}$ and $\cS^\f_{\ \th}$ {\it do not agree}. These components cannot 
possibly appear in the true answer since the relevant components of the Einstein tensor vanish {\it identically} by symmetry. This discrepancy is a result of the 
discontinuous oblique $\bN$ vector, in violation of the assumptions made in \cite{BarrabesIsrael:1991} to argue for the independence of the result for the 
discontinuity of $\cK_{ij}$ and the surface stress tensor on the choice of $\bN$, which makes the BI formalism inapplicable to this case. Rotating lightlike
horizon surface ``branes” were also considered in \cite{GuenKagNis:2010}.

\end{document}